\documentclass[twocolumn]{aastex631}
\usepackage{amsmath} 
\usepackage{soul} 
\usepackage{multirow} 
\usepackage{bm} 
\usepackage{amsmath} 
\usepackage{xspace} 
\usepackage{gensymb} 
\usepackage{float}
\usepackage{customcommands} 
\usepackage{newunicodechar,graphicx}
\PassOptionsToPackage{hyphens}{url}\usepackage{hyperref}

\submitjournal{The Astronomical Journal}
\received{Mar. 22, 2021}
\revised {Aug. 20, 2021}
\accepted{Sept. 17, 2021}

\begin{document}
\title{Another super-dense sub-Neptune in \oneEightTwo b and refined mass measurements for \oneNineNine b and c\footnote{Based on observations obtained at the W. M. Keck Observatory, which is operated jointly by the University of California and the California Institute of Technology.}}

\correspondingauthor{Joseph M. Akana Murphy}
\email{joseph.murphy@ucsc.edu}

\author[0000-0001-8898-8284]{Joseph M. Akana Murphy}
\altaffiliation{NSF Graduate Research Fellow}
\affiliation{Department of Astronomy and Astrophysics, University of California, Santa Cruz, CA 95064, USA}

\author[0000-0002-6115-4359]{Molly R. Kosiarek}
\altaffiliation{NSF Graduate Research Fellow}
\affiliation{Department of Astronomy and Astrophysics, University of California, Santa Cruz, CA 95064, USA}

\author[0000-0002-7030-9519]{Natalie M. Batalha}
\affiliation{Department of Astronomy and Astrophysics, University of California, Santa Cruz, CA 95064, USA}

\author[0000-0002-9329-2190]{Erica J. Gonzales}
\altaffiliation{NSF Graduate Research Fellow}
\affiliation{Department of Astronomy and Astrophysics, University of California, Santa Cruz, CA 95064, USA}

\author[0000-0002-0531-1073]{Howard Isaacson}
\affiliation{{Department of Astronomy,  University of California Berkeley, Berkeley CA 94720, USA}}
\affiliation{Centre for Astrophysics, University of Southern Queensland, Toowoomba, QLD, Australia}

\author[0000-0003-0967-2893]{Erik A Petigura}
\affiliation{Department of Physics \& Astronomy, University of California Los Angeles, Los Angeles, CA 90095, USA}

\author[0000-0002-3725-3058]{Lauren M. Weiss}
\affiliation{Department of Physics, University of Notre Dame, Notre Dame, IN 46556, USA}

\author[0000-0003-4976-9980]{Samuel K. Grunblatt}
\altaffiliation{Kalbfleisch Fellow}
\affiliation{American Museum of Natural History, 200 Central Park West, Manhattan, NY 10024, USA}
\affiliation{Center for Computational Astrophysics, Flatiron Institute, 162 5$^\text{th}$ Avenue, Manhattan, NY 10010, USA}

\author[0000-0002-5741-3047]{David R. Ciardi}
\affiliation{NASA Exoplanet Science Institute-Caltech/IPAC, 1200 E. California Blvd, Pasadena, CA 91125 USA}

\author[0000-0003-3504-5316]{Benjamin Fulton}
\affiliation{NASA Exoplanet Science Institute/Caltech-IPAC, MC 314-6, 1200 E California Blvd, Pasadena, CA 91125, USA}

\author[0000-0001-8058-7443]{Lea A.\ Hirsch}
\affiliation{Kavli Institute for Particle Astrophysics and Cosmology, Stanford University, Stanford, CA 94305, USA}

\author[0000-0003-0012-9093]{Aida Behmard}
\altaffiliation{NSF Graduate Research Fellow}
\affiliation{Division of Geological and Planetary Science, California Institute of Technology, Pasadena, CA 91125, USA}

\author[0000-0001-8391-5182]{Lee J.\ Rosenthal}
\affiliation{Cahill Center for Astronomy $\&$ Astrophysics, California Institute of Technology, Pasadena, CA 91125, USA}

\begin{abstract}
We combine multiple campaigns of \ktwo photometry with precision radial velocity measurements from Keck-HIRES to measure the masses of three sub-Neptune-size planets. We confirm the planetary nature of the massive sub-Neptune \oneEightTwo b (\periodb $= 4.7$ days, \rplanetb $= 2.69$ \rearth) and derive refined parameters for \oneNineNine b and c (\periodb $ = 3.2$ days, \rplanetb $= 1.73$ \rearth, and \periodc $ = 7.4$ days, \rplanetc $ = 2.85$ \rearth). These planets provide valuable data points in the mass-radius plane, especially as \emph{TESS} continues to reveal an increasingly diverse sample of sub-Neptunes. \oneEightTwo (EPIC 211359660) is a moderately bright ($V = 12.0$ mag) early-K dwarf observed during \ktwo campaigns 5 and 18. We find \oneEightTwo b is potentially one of the densest sub-Neptunes known to date (\oneEightTwoMass\xspace \mearth and \oneEightTwoRho\xspace g cm$^{-3}$). \oneNineNine (EPIC 212779596; $V = 12.3$ mag) is a K5V dwarf observed in \ktwo campaigns 6 and 17 which hosts two recently-confirmed planets. We refine the orbital and planetary parameters for \oneNineNine b and c by modeling both campaigns of \ktwo photometry and adding 12 Keck-HIRES measurements to the existing radial velocity data set ($N$ = 33). We find \oneNineNine b is likely rocky, at \oneNineNineBMass\xspace \mearth and \oneNineNineBRho\xspace g cm$^{-3}$. \oneNineNine c has an intermediate density at \oneNineNineCMass\xspace \mearth and \oneNineNineCRho\xspace g cm$^{-3}$. We contextualize these planets on the mass-radius plane, discuss a small but intriguing population of ``super-dense" sub-Neptunes (\rplanet $< 3$ \rearth, \mplanet $> 20$ \mearth), and consider our prospects for the planets' atmospheric characterization.
\end{abstract}


\section{Introduction} \label{sec:intro} 
\emph{Kepler} \citep{borucki10} demonstrated that sub-Neptune-size planets (1.7-4 \rearth) are common in the Milky Way Galaxy \citep{howard12, batalha13, fressin13, petigura13, fulton17, bryson21}. Still, the formation and evolution processes that lead to their diverse bulk compositions are not well-understood. Thanks to more than a decade of discoveries from space and dedicated follow-up on the ground, the field of exoplanets has entered an era where emerging sub-structure in the mass-radius diagram can help inform which physical processes drive the diversity.

While degeneracies in bulk composition limit the conclusions that can be drawn from mass and radius measurements alone \citep{adams08, valencia07, otegi20b}, constraints on atmospheric metallicity can help disambiguate the interior structure of sub-Neptunes \citep{rogersSeager10}. However, to interpret transmission spectra, small planets generally require 5-$\sigma$ mass measurements to break the degeneracy between surface gravity and atmospheric mean molecular weight \citep{batalha19}. Fortunately, radial velocity surveys of bright planet candidate hosts from \tess \citep{ricker14} have identified the sub-Neptune regime as a fruitful balance between the inherent intrigue of small planets and the Doppler amplitudes needed to quickly achieve precise mass measurements. Multi-planet sub-Neptune systems are even more valuable to questions of bulk composition, as they are natural testbeds for theories in planet formation and dynamical evolution. With the launch of the \emph{James Webb Space Telescope} (\jwst) on the horizon, sub-Neptune confirmations continue to serve as the critical first step that will enable future investigations in planetary astrophysics.

In this paper we measure the mass of the unusually dense \oneEightTwo b (\periodb $= 4.7$ days, \rplanetb $=$ \oneEightTwoRad\ \rearth, \mplanetb $=$ \oneEightTwoMass\ \mearth, \rhob $=$ \oneEightTwoRho\ \gcc) and provide refined planet ephemerides and mass measurements for \oneNineNine b (\periodb $= 3.2$ days, \rplanetb $=$ \oneNineNineBRad\ \rearth, \mplanet $=$ \oneNineNineBMass\ \mearth) and \oneNineNine c (\periodc $= 7.4$ days, \rplanetc $=$ \oneNineNineCRad\ \rearth, \mplanetc $=$ \oneNineNineCMass\ \mearth). We also discuss a group of super-dense sub-Neptunes similar to \oneEightTwo b (\rplanet $< 3$ \rearth, \mplanet $> 20$ \mearth) that seems to be emerging in the mass-radius diagram.

The paper is organized as follows: In \S\ref{sec:phot} we extract the \ktwo light curves and simultaneously model stellar variability and planet transits. The photometric analysis is summarized in Figures \ref{fig:k2-182_phot} and \ref{fig:k2-199_phot}. In \S\ref{sec:stellar} we characterize \oneEightTwo and \oneNineNine with high-resolution spectroscopy and high-contrast imaging. Derived stellar parameters are listed in Table \ref{tab:star_planet_combined}. We describe our radial velocity follow-up and analysis in \S\ref{sec:rvs}. \revision{Models of the radial velocities of \oneEightTwo and \oneNineNine are shown in Figures \ref{fig:k2-182_rvs_combined} and \ref{fig:k2-199_rv}, respectively.} A list of measured and derived planetary parameters for \oneEightTwo b and \oneNineNine b and c can also be found in Table \ref{tab:star_planet_combined}. In \S\ref{discuss:comp} we explore possible bulk compositions for \oneEightTwo b and \oneNineNine b and c, \revision{illustrated in Figures \ref{fig:mass_radius}, \ref{fig:fh2o_combined}, and \ref{fig:fhhe_combined}}. In \S\ref{discuss:spur} we discuss a selection of super-dense sub-Neptunes and investigate whether or not their high mass measurements could be artificially inflated by unmitigated signatures of stellar activity in models of radial velocities. \revision{Figure \ref{fig:mass_radius} shows \oneEightTwo b and \oneNineNine b and c among confirmed planets in the mass-radius diagram, with special attention drawn to the super-dense sub-Neptunes.} In \S\ref{discuss:atmos} we return to \oneEightTwo b and \oneNineNine b and c to discuss their prospects for space-based atmospheric characterization. We conclude in \S\ref{sec:conclusion}.

\section{\ktwo photometry and modeling}
\label{sec:phot}

\subsection{\oneEightTwo and \oneNineNine photometry} \label{sub:182_199_phot}

Both \oneEightTwo and \oneNineNine were observed in two \ktwo campaigns and the total baseline for each system is 3 years. \oneEightTwo was observed at long-cadence (29.4-minute exposures) during \ktwo Campaigns 5 (2015 Apr 27—2015 Jul 10) and 18 (2018 May 12—2018 Jul 2)\footnote{C18 is notably shorter than normal (about 50 days instead of 80) because data collection was halted when spacecraft fuel levels became dangerously low in early July 2018. Despite the low fuel levels, C18 spacecraft pointing and roll behavior was consistent with other campaigns.}. The \oneEightTwo raw \revision{Simple Aperture Photometry (SAP)} flux from the two campaigns has a median 6-hour combined differential photometric precision (CDPP; \citealt{christiansen12}) of 288 and 206 parts per million (ppm), respectively. \oneNineNine was observed in long-cadence during Campaigns 6 (2015 Jul 14—2015 Sep 30) and 17 (2018 Mar 1—2018 May 8) with a 6-hour median CDPP of 442 and 452 ppm, respectively. \revision{For reference, the median 6-hour CDPP for the raw SAP fluxes of dwarfs in \ktwo Campaigns 0-7 is approximately 200 ppm at $K_\mathrm{p} = 12$ mag (see Figure 11 in \citealt{everest:luger16}). \oneEightTwo and \oneNineNine are $K_\mathrm{p} = 11.74$ mag and $11.93$ mag, respectively \citep{epic}. While the 6-hour CDPP for the raw SAP fluxes of \oneEightTwo and \oneNineNine is slightly higher than the median level across dwarfs from Campaigns 0-7, these levels are still within the bulk of the distribution and should not be cause for concern \citep{everest:luger16}.}

\revision{While \oneEightTwo b and \oneNineNine b and c have all been robustly identified and validated, the literature still lacks a simultaneous analysis of their multiple \ktwo campaigns, which is critical for refining the planet ephemerides.} \cite{pope16} first identified \oneEightTwo b and \oneNineNine b and c as planet candidates using the \ktwo C5 and C6 photometry, followed by \cite{petigura17}. Soon after, both \cite{mayo18} and \cite{livingston18} (hereafter \mayo and \livingston, respectively) independently identified and statistically validated all three planets from the C5 and C6 photometry using \vespa to calculate false positive probabilities (FPPs; \citealt{morton12} and \citealt{morton15}). \cite{crossfield18} then independently identified \oneNineNine b and c as planet candidates in the C17 photometry. Recently, \cite{wittenmyer20} reported updated semi-major axis and radius measurements for the three planets using stellar properties from observations with the High Efficiency and Resolution Multi-Element Spectrograph (HERMES) on the Anglo-Australian Telescope \citep{simpson16}. In \S\ref{sub:photImprove} we highlight the improvement to the precision of the planet orbital periods and mid-transit times by including both campaigns of photometry in our analysis.

\subsection{Light curve extraction} \label{phot:lightcurve}

We extracted the light curves using the EPIC Variability Extraction and Removal for Exoplanet Science Targets pipeline (\everest\footnote{\url{https://github.com/rodluger/everest}}, \citealt{everest:luger16} and \citealt{everest:luger18}). \everest uses a variant of pixel level decorrelation (PLD; \citealt{deming15}) to remove systematic artifacts related to \ktwo's imprecise pointing and to minimize scatter on 6-hour timescales. By now the community has developed numerous data reduction methods and open-source pipelines to correct for \ktwo's unique systematics in an attempt to push back down to \emph{Kepler}-like precision e.g. \ktwoSFF\footnote{\url{https://www.cfa.harvard.edu/~avanderb/k2.html}} \citep{vanderburgJohnson14}, \texttt{K2SC}\footnote{\url{https://github.com/OxES/k2sc}} \citep{aigrain16}, \texttt{Kadenza}\footnote{\url{https://github.com/KeplerGO/kadenza}} \citep{kadenza:barentsen18}. In general, \everest's PLD method has shown to be slightly more successful at mitigating \ktwo systematics than other popular reduction approaches (e.g. \citealt{hirano18}, \citealt{lillobox20}). For each campaign of \oneEightTwo and \oneNineNine observations we verified that the \everest light curves had smaller median 6-hour CDPP than those from \ktwoSFF. The CDPP of the \oneEightTwo C5 and C18 \everest light curves is 19 and 22 ppm, respectively. For \oneNineNine C6 and C17 the CDPP levels are 20 and 28 ppm.

\revision{We elected to use the default \everest apertures for \oneEightTwo and \oneNineNine, both of which are aperture number 15 from the \ktwoSFF catalog \citep{vanderburg14, vanderburgJohnson14}. \cite{everest:luger16} find that of the 20 apertures provided for each source by \ktwoSFF (which are derived with knowledge of the \emph{Kepler} pixel response function), aperture 15 strikes a favorable balance between including enough pixels to form a good basis set for PLD and avoiding contamination from nearby stars. \cite{everest:luger16} also note that when contamination is not an issue, the choice of aperture has little influence on the \everest algorithm.}

\revision{In the Appendix of \livingston, the authors note that a bright, nearby star ($\Delta K_\mathrm{p} = 5.6$ mag, separation $\approx 8$\arcsec) falls in their \texttt{k2phot}\footnote{\url{https://github.com/petigura/k2phot}} aperture for \oneNineNine. The same star falls in our \everest aperture. Since the neighbor is about two \emph{Kepler} pixels away from \oneNineNine, \livingston used multi-aperture photometry to confirm that the transit signals belong to \oneNineNine. They also find that transit depths are not dependent on the aperture radius. In addition to the results of their multi-aperture analysis, \livingston point out that, a priori, it is more likely to find two sub-Neptune-size planets transiting \oneNineNine than to find multiple stellar-size objects transiting the fainter neighbor. At $\Delta K_\mathrm{p} = 5.6$ mag, dilution from the secondary star in \oneNineNine's \everest aperture could mean that the radii of \oneNineNine b and c are potentially larger than the values we report in Table \ref{tab:star_planet_combined} by $\lesssim 1\%$ (see Equation 7 in \citealt{ciardi15}). However, this level of dilution is negligible compared to other sources of error on the planet radius measurements (e.g. our determination of \oneNineNine's radius, see \S\ref{sec:stellar}). Therefore, even with the faint neighbor in \oneNineNine's \everest aperture, we are not concerned with dilution or ambiguity in the transits' host. There are no bright, contaminating sources in \oneEightTwo's \everest aperture. We comment further on possible dilution scenarios and their implications in our presentation of high-resolution, high-contrast adaptive optics imaging of \oneEightTwo and \oneNineNine in \S\ref{stellar:img}.}

\revision{To remove the short-timescale ($\sim$6 hours) systematic signal in the raw \ktwo SAP photometry related to the spacecraft's pointing, the \everest detrending model solves a generalized least-squares (GLS) problem. The GLS system is shown in Equation 7 and 8 of \cite{everest:luger16}, where the ``design matrix," $\bm{X}$, is constructed using a combination of third-order PLD and principal component analysis (PCA; e.g. \citealt{jolliffe86}).} Before computing the \everest detrending model, we first masked transits using the periods and mid-transit times reported for the three planets by \livingston. This was done to prevent the \everest algorithm from creating systematically shallower transits by smoothing over the planetary signals \revision{during the least-squares minimization.} \revision{The uncertainties in the mid-transit times from \livingston are $\pm 0.0007$, $\pm 0.0027$, and $\pm 0.0017$ days for \oneEightTwo b, \oneNineNine b, and \oneNineNine c, respectively. Propagating these values to the end of the second campaign of \ktwo photometry for each system produces mid-transit time uncertainties of similar order to the transit durations. For this reason when extracting the \everest light curves we masked all data $\pm$3 hours of the nominal mid-transit times as predicted from the \livingston ephemerides (where 6 hours is $>2\times$ longer than the transit duration of any of the planets, see Table \ref{tab:star_planet_combined}.). We visually inspected each transit mask to ensure that even with the uncertainties in the predicted mid-transit times from the \livingston ephemerides (especially during the second campaign of \ktwo photometry) the masks still covered each transit (they did).} 

\revision{Following the transit masking, we then produced the corrected \everest light curves using the detrending model trained on the out-of-transit data. Finally, we removed all data (either in- or out-of-transit) with poor quality flags as well as any out-of-transit data that \everest had identified as outliers. To further convince ourselves that the mid-transit time uncertainties from the \livingston ephemerides did not bias our light curve extraction, for both systems we modeled each campaign of our \everest light curves independently according to the methodology below in \S \ref{sub:phot_model_and_fit} and \S \ref{sub:phot_post_est}. For each system the posteriors of the transit depths for each planet were entirely consistent between models fit to the separate campaigns. If our transit masking was imperfect due to the uncertainties on the \livingston ephemerides, we would have expected the second campaign’s transits to have been smoothed over by the \everest detrending model and appear shallower, but this was not the case.}

\subsection{Simultaneous modeling of stellar variability and planet transits} \label{sub:phot_model_and_fit}

\begin{figure*}
    \centering
    \includegraphics[width=\textwidth]{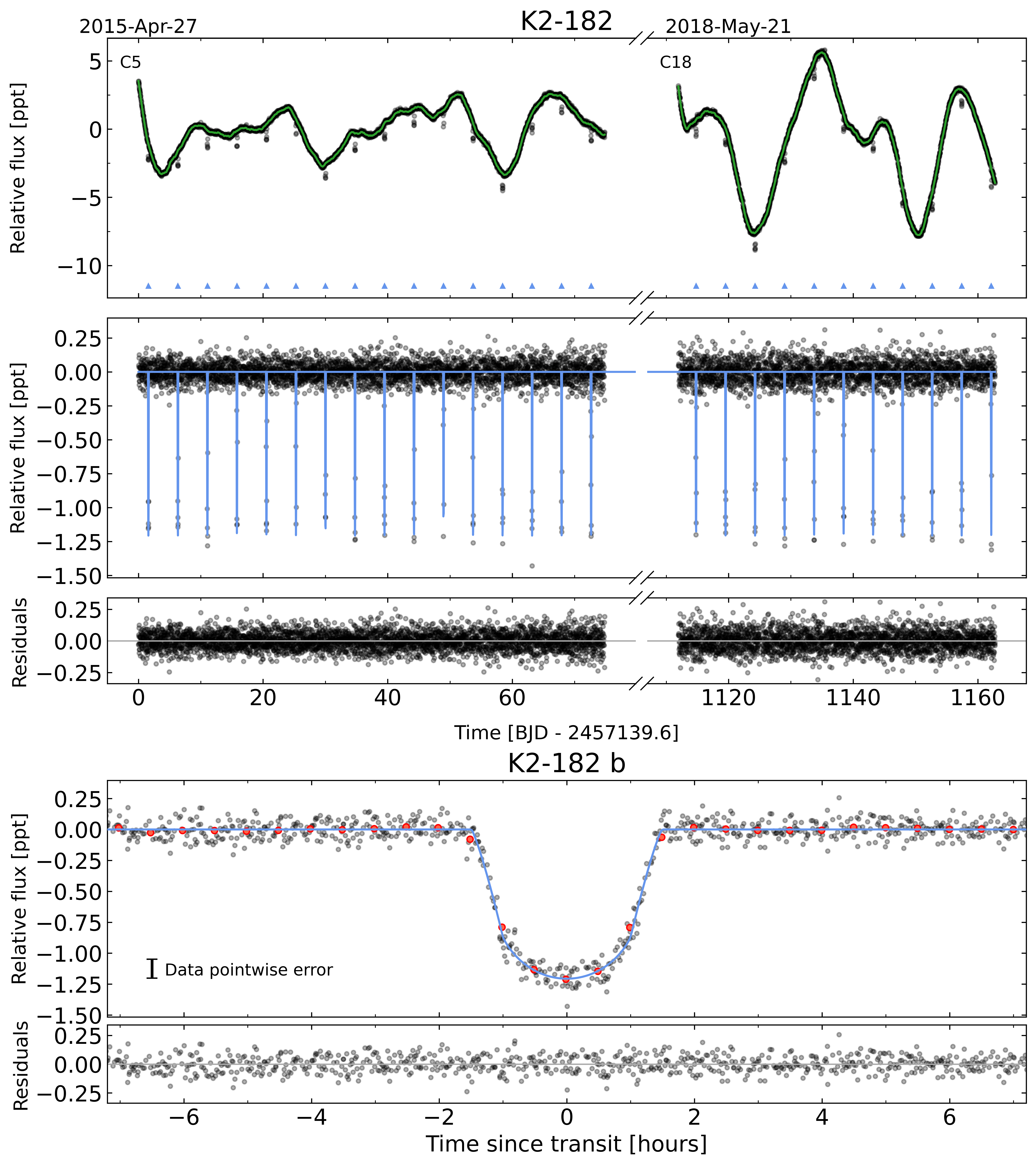}
    \caption{The maximum a posteriori (MAP) solution for the \oneEightTwo photometric model and phase-folded transit of \oneEightTwo b. \emph{Top:} The \everest C5 and C18 light curves are shown as the black points. The Gaussian process ($+$ a constant offset fit to the data) used to \revision{model stellar activity} is overplotted in green. Blue triangles denote \oneEightTwo b transits. \emph{Middle:} Data \revision{minus the GP model of the stellar activity} are shown as the black points with the \oneEightTwo b orbital solution from \texttt{starry} shown in blue. Residuals are shown below. \emph{Bottom:} The phase-folded best-fitting transit model for \oneEightTwo b and residuals. \revision{Data minus the GP model of the stellar activity} are shown in black, and these data binned by 0.5 hours are shown in red. The MAP transit model is shown as the solid blue line.}
    \label{fig:k2-182_phot}
\end{figure*}

\begin{figure*}
    \centering
    \includegraphics[width=\textwidth]{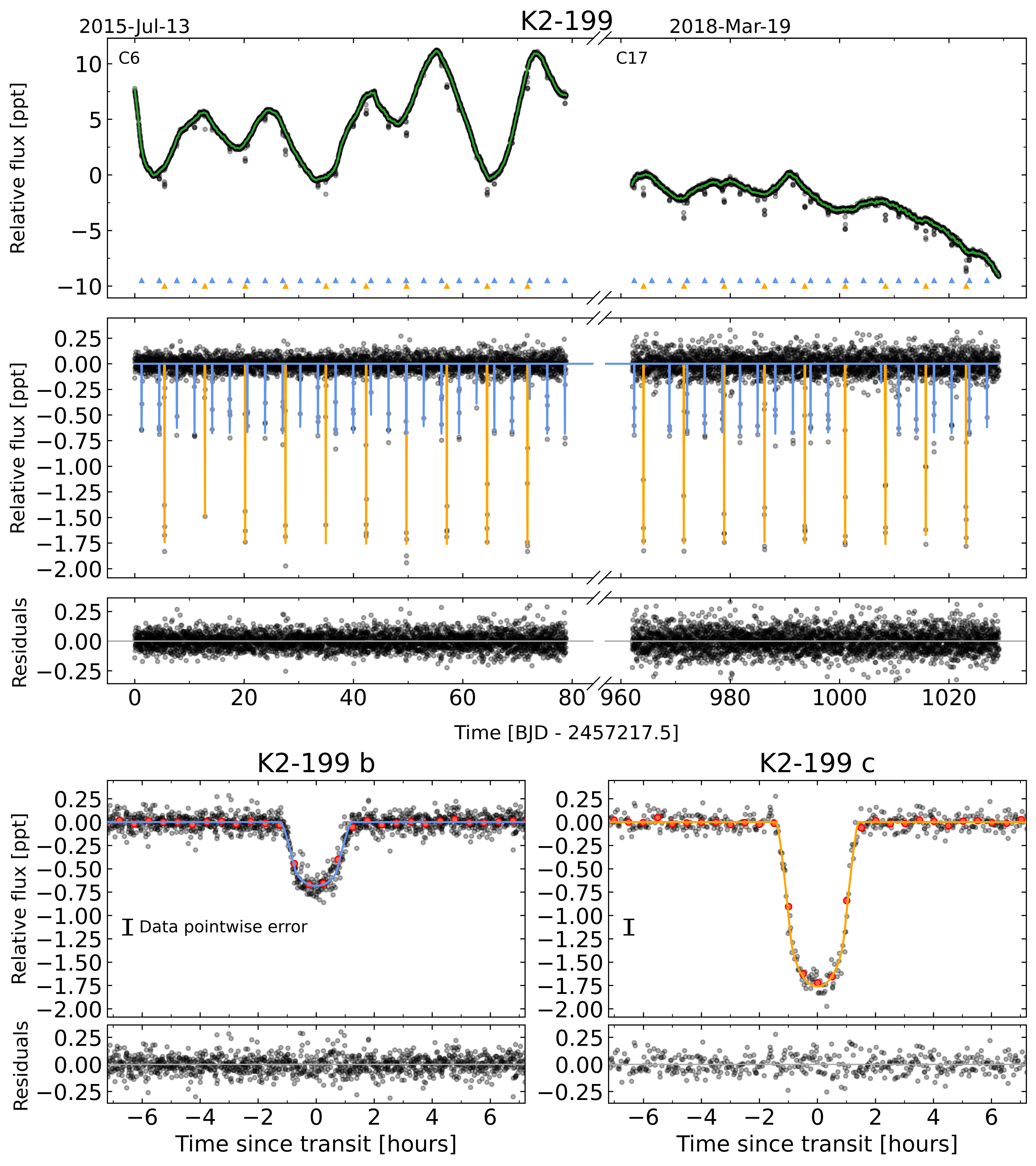}
    \caption{The MAP solution for the \oneNineNine photometric model and phase-folded transits of \oneNineNine b and c. The figure layout is analogous to Figure \ref{fig:k2-182_phot}. \emph{Top:} The \everest C6 and C17 light curves are shown as the black points. The Gaussian process ($+$ a constant offset fit to the data) used to \revision{model stellar activity} is overplotted in green. Blue and orange triangles denote \oneNineNine b and c transits, respectively. \emph{Middle:} Data \revision{minus the GP model of the stellar activity} are shown as the black points with the \oneNineNine b and c orbital solutions from \texttt{starry} shown in blue and orange, respectively. Residuals are shown below. \emph{Bottom:} The phase-folded best-fitting transit models for \oneNineNine b and c with their residuals below. \revision{Data minus the GP model of the stellar activity} are shown in black, and these data binned by 0.5 hours are shown in red. The MAP transit models are shown as the solid blue and orange lines.}
    \label{fig:k2-199_phot}
\end{figure*}

We modeled the \everest photometry in \texttt{exoplanet} \citep{exoplanet:exoplanet} by simultaneously fitting a Gaussian process (GP) to the stellar variability with \celerite \citep{celerite} and modeling transits with a quadratic limb-darkening law \citep{exoplanet:kipping13} via \texttt{starry} \citep{starry}. While in practice GPs may have similar performance in removing low-frequency stellar variability to e.g. a basis spline \citep{vanderburgJohnson14}, they have the added benefit of providing a meaningful propagation of errors, since each GP posterior prediction is itself a Gaussian random variable with an associated variance. Furthermore, GPs have been shown to be useful phenomenological tools for modeling the light curves of spotted stars \citep{angus18}, which can add physical motivation to the hyperparameters of their kernels. (\oneEightTwo and \oneNineNine exhibit rotational modulation on the order of 1\%. See Figures \ref{fig:k2-182_phot} and \ref{fig:k2-199_phot}.)

The kernel of the GP we used to model the low-frequency stellar activity signal in the \everest light curves is a mixture of three terms, each of which has a power spectral density (PSD) in the form of a stochastically-driven, damped harmonic oscillator (SHO). The first term is an overdamped oscillator, meant to describe non-periodic behavior such as spot evolution. The second and third terms are underdamped, with fundamental frequencies corresponding to the stellar rotation period and its first harmonic, respectively. \revision{The PSD of the kernel can be written as
\begin{equation} \label{eqn:activity_kernel}
    S_\mathrm{act}(\omegagp) = S_\mathrm{dec}(\omegagp) + S_{P_\mathrm{rot}}(\omegagp) + S_{P_\mathrm{rot}/2}(\omegagp),
\end{equation}
where \omegagp is the angular frequency at which to evaluate the PSD (not to be confused with the argument of periastron). Each term in Equation \ref{eqn:activity_kernel} is in the form
\begin{equation}
    S(\omegagp) = \sqrt{\frac{2}{\pi}} \frac{S_0 \omega_0^4}{(\omegagp^2 - \omega_0^2)^2 + \omega_0^2 \omegagp^2/Q^2},
\end{equation}
where $S_0$ is the power of the SHO at $\omegagp = 0$, $\omega_0$ is the fundamental angular frequency of the undamped oscillator, and $Q$ is the quality factor of the oscillator. For $S_\mathrm{dec}$ we fix $Q_\mathrm{dec} = \frac{1}{\sqrt{2}}$, since this gives the SHO the same PSD as stellar granulation \citep{harvey85, kallinger14}.}

\revision{
To define the terms describing the rotational modulation, let
\begin{equation}
    S_{P_\mathrm{rot}}(\omegagp) = \sqrt{\frac{2}{\pi}} \frac{S_1 \omega_1^4}{(\omegagp^2 - \omega_1^2)^2 + \omega_1^2 \omegagp^2/Q_1^2}
\end{equation}
and
\begin{equation}
    S_{P_\mathrm{rot}/2}(\omegagp) = \sqrt{\frac{2}{\pi}} \frac{S_2 \omega_2^4}{(\omegagp^2 - \omega_2^2)^2 + \omega_2^2 \omegagp^2/Q_2^2}.
\end{equation}
The hyperparameters of $S_{P_\mathrm{rot}}$ and $S_{P_\mathrm{rot}/2}$ are related via
\begin{align}
    Q_1 & = \frac{1}{2} + Q_0 + \delta Q \\ 
    \omega_1 & = \frac{4 \pi Q_1}{P_\mathrm{rot}\sqrt{4 Q_1^2 - 1}} \\
    S_1 & = \frac{S_{0,\mathrm{rot}}}{(1 + f) \omega_1 Q_1}
\end{align}
and
\begin{align}
    Q_2 & = \frac{1}{2} + Q_0 \\ 
    \omega_2 & = \frac{8 \pi Q_1}{P_\mathrm{rot}\sqrt{4 Q_1^2 - 1}} \\
    S_2 & = \frac{f S_{0,\mathrm{rot}}}{(1 + f) \omega_1 Q_1},
\end{align}
where $S_{0,\mathrm{rot}}$ is the amplitude of $S_{P_\mathrm{rot}} + S_{P_\mathrm{rot}/2}$ relative to $S_\mathrm{dec}$, $Q_0$ is the quality factor minus $\frac{1}{2}$ for the oscillator at $P_\mathrm{rot}/2$, $\delta Q$ is the difference between the quality factors of the oscillators at $P_\mathrm{rot}$ and $P_\mathrm{rot}/2$, $P_\mathrm{rot}$ is the primary period of variability (meant to represent the stellar rotation period), and $f$ is the fractional amplitude of the SHO at $P_\mathrm{rot}/2$ relative to the SHO at $P_\mathrm{rot}$.
}

\revision{
If we define
\begin{equation}
    S_\mathrm{rot}(\omegagp) \equiv S_{P_\mathrm{rot}} + S_{P_\mathrm{rot}/2}
\end{equation}
then the PSD in Equation \ref{eqn:activity_kernel} becomes just the sum of a term describing the exponentially decaying stellar activity and a term describing the rotational modulation:
\begin{equation}\label{eqn:activity_kernel_final}
    S_\mathrm{act}(\omegagp) = S_\mathrm{dec}(\omegagp) + S_\mathrm{rot}(\omegagp).
\end{equation}
Finally, we add a photometric ``jitter" term to the kernel with the PSD in Equation \ref{eqn:activity_kernel_final} to fit additional white noise. In their online tutorials the authors of \exoplanet suggest the kernel we use is a good choice for modeling stellar variability.} We experimented by fitting the model with variants of this kernel (e.g. removing the non-periodic term, removing the first-harmonic term, removing the underdamped oscillators and adding a second overdamped oscillator) and produced similar results.

Since GPs can be sensitive to outliers, before performing an initial fit to the \everest light curves we removed additional data (7 and 24 points for \oneEightTwo and \oneNineNine, respectively) by smoothing the data in bins of 0.3 days with a cubic Savitzky-Golay filter \citep{savitzky64} and iteratively removing out-of-transit, $>$3-$\sigma$ outliers until convergence (4 iterations for both systems). We then fit our initial photometric model and removed 7-$\sigma$ outliers about the maximum a posteriori (MAP) solution. For \oneEightTwo we then re-fit the data, seeding the optimization with the initial MAP solution, to produce the final, best-fitting model. We repeated the fitting and 7-$\sigma$ outlier-removal step a second time for \oneNineNine before fitting the final model.  We found the C17 photometry had a slightly larger scatter than the other data (which is consistent with its larger CDPP). 

The resulting MAP solutions for our models of the \oneEightTwo and \oneNineNine photometry are shown in Figures \ref{fig:k2-182_phot} and \ref{fig:k2-199_phot}, respectively, along with the phase-folded transits for each planet. In each figure the top panel shows the \everest light curve and MAP GP model ($+$ a small constant offset fit to the data) used to \revision{model} the stellar rotation and spot modulation signals. The middle panel shows the light curves \revision{with the stellar activity signal removed} and planet transit models along with the residuals about the full model (GP $+$ offset $+$ transit models). The bottom panels show the phase-folded transits of the planets and their residuals. 

We note that a few transits of \oneNineNine b are missing from the middle panel of Figure \ref{fig:k2-199_phot} (transits near 70, 965, 1000, and 1005 BJD - 2457217.5). These epochs are instances where portions of the in-transit data had poor data quality flags and were removed prior to fitting the model.

\subsection{Posterior estimation} \label{sub:phot_post_est}
Implemented with \texttt{exoplanet} and \texttt{pymc3} \citep{pymc3}, we used No-U-Turn sampling (NUTS; \citealt{nuts:hoffman14}), an adaptive form of Hamiltonian Monte Carlo (HMC; \citealt{hmc:duane87} and \citealt{neal12}), to estimate the posterior distributions of the parameters in our models of the \ktwo photometry. HMC sampling uses the gradient of the posterior to help inform Markov transitions, enabling more efficient exploration of high-dimensional posterior surfaces than brute-force, guess-and-check methods like Metropolis-Hastings (\citealt{metropolis53} and \citealt{hastings70}). For each system, a NUTS sampler ran 20 chains with starting locations randomly drawn from the model prior and with each chain taking 15,000 ``tuning" steps before drawing 10,000 samples. During the tuning stage the NUTS sampler optimizes hyperparameters like step size to meet a targeted sample acceptance rate as it explores the posterior surface. This can help prevent the sampler from taking too large of a step in posterior regions with high curvature. Samples drawn during the tuning period were discarded, similar to how various Markov Chain Monte Carlo (MCMC) methods discard burn-in samples. The chains were concatenated to produce a total of $N$ = 200,000 samples from the marginal posteriors of each model parameter.

Convergence of the HMC sampling was assessed through multiple diagnostic statistics. Recently, \cite{vehtari19} pointed out serious flaws with the standard Gelman-Rubin statistic ($\hat{R}$; \citealt{gelman92}), which is conventionally used to determine convergence for iterative stochastic algorithms like MCMC. Following their prescription, we instead assessed convergence by verifying a sufficiently small ($< 1.001$) \emph{rank-normalized} $\hat{R}$ for each model parameter. In brief, a rank-normalized $\hat{R}$ statistic is computed by calculating $\hat{R}$ on the normalized, rank-transformed chains of the parameter, rather than the values of the parameter itself. To ensure the chains could offer reliable confidence intervals, we also calculated the rank-normalized bulk and tail effective sample sizes from \cite{vehtari19} for each of the marginal posteriors (roughly, the effective sample sizes are the number of ``independent" samples obtained in the bulk and tails of the posterior). \revision{\cite{vehtari19} recommend that the effective sample size should be larger than 400 in both the bulk and the tails of the posterior.} For every parameter we find the minimum between the bulk and tail effective sample sizes was much larger than the recommended minimum threshold ($N_\mathrm{eff} \gtrsim$ 30,000).

To avoid sampling bias, for several physical parameters (e.g. planet orbital period, planet radius, and stellar rotation period) we fit and explored the posterior of the natural logarithm of the parameter of interest. For parameters which are strictly non-negative, a prior's hard bound at 0 can cause the posterior surface to form a ``funnel" geometry, infamous of hierarchical models \citep{neal03}. NUTS sampling can have difficulty exploring the funnel because of its high curvature, which can cause the gradient calculation to diverge \citep{betancourt13}. Fitting the natural logarithm of strictly non-negative parameters rather than the parameter itself can help to avoid the funnel geometry altogether, thereby increasing sampling efficiency and the effective sample size.

The relevant measured and derived physical planet parameters from our photometric analysis are listed in Table \ref{tab:star_planet_combined}. Full lists of the parameters, priors, and posterior median values and 68\% confidence intervals for our photometric models of \oneEightTwo and \oneNineNine are shown in Table \ref{tab:182_phot_model} and Table \ref{tab:199_phot_model}, respectively. In general, we used conservatively broad priors for all physical parameters save for stellar mass and radius, which had informed Gaussian priors stemming from our stellar characterization described in \S\ref{sec:stellar}. We also used the mixture distribution from \cite{exoplanet:vaneylen19} to place a prior on planet eccentricity and marginalized over its hyperparameters.

\revision{While some of the GP hyperparameters that define the stellar activity kernel in Equation \ref{eqn:activity_kernel_final} are not necessarily of immediate physical interest (e.g. $f$, the mixture fraction between the SHO at $P_\mathrm{rot}$ and $P_\mathrm{rot}/2$), it can still be a useful sanity check to compare hyperparameter posteriors to physical expectations. We used Equation 11 from \cite{kawaler89} to obtain a rough estimate of the expected stellar rotation period for \oneEightTwo and \oneNineNine based on the (poorly-constrained) stellar ages we determined via high-resolution spectroscopy and isochrone modeling (see \S \ref{sec:stellar}, also Table \ref{tab:star_planet_combined}) and their $B - V$ colors from the EPIC catalog \citep{epic}. For \oneEightTwo, the relation from \cite{kawaler89} estimates $P_\mathrm{rot} = 21 \pm 2$ days. The GP model of the \oneEightTwo stellar variability seems to agree reasonably well, finding $P_\mathrm{rot} = 24.8 \pm 1.1$ days. For \oneNineNine, the \cite{kawaler89} relation implies $P_\mathrm{rot} = 37 \pm 4$ days. However, the GP included to model the \oneNineNine light curves initially had difficulty inferring $P_\mathrm{rot}$ because the posterior was multimodal, perhaps due to the stark change in spot behavior between the two campaigns of photometry (see the top panel of Figure \ref{fig:k2-199_phot}). We placed a tight prior on $P_\mathrm{rot}$ (really, on $\log P_\mathrm{rot}$) for the \oneNineNine GP to avoid the multimodality, so $P_\mathrm{rot}$ for this system should not be thought of as an inference of the stellar rotation period, but simply as a model parameter used to help flatten the light curves. We discuss this choice more in \S \ref{activity:k2-199} and the note below Table \ref{tab:199_phot_model}.}

\subsection{Improvement on the precision of planet ephemerides and implications for future transit observations} \label{sub:photImprove}

\oneEightTwo and \oneNineNine are particularly attractive systems for follow-up investigation because they each have two \ktwo campaigns of photometry. The existence of multi-campaign photometry allows us to characterize stellar variability and improve upon the orbital ephemerides. Until now the literature for these systems has lacked a joint analysis of the multi-campaign \ktwo observations. Furthermore, for \oneNineNine b \cite{crossfield18} note that the ephemerides they derive from the C17 data disagree with those derived from the C6 data by \livingston at the 2 to 3-$\sigma$ level. With \jwst primed to usher in a new era in atmospheric studies for the sub-Neptune population, precise planet ephemerides are crucial for scheduling spectroscopic transit observations. This is especially important for \ktwo planets compared to those from \tess, which will have been observed much more recently relative to the planned \jwst launch \revision{in 2021 Nov or later}.

\begin{figure}
    \centering
    \includegraphics[width=\columnwidth]{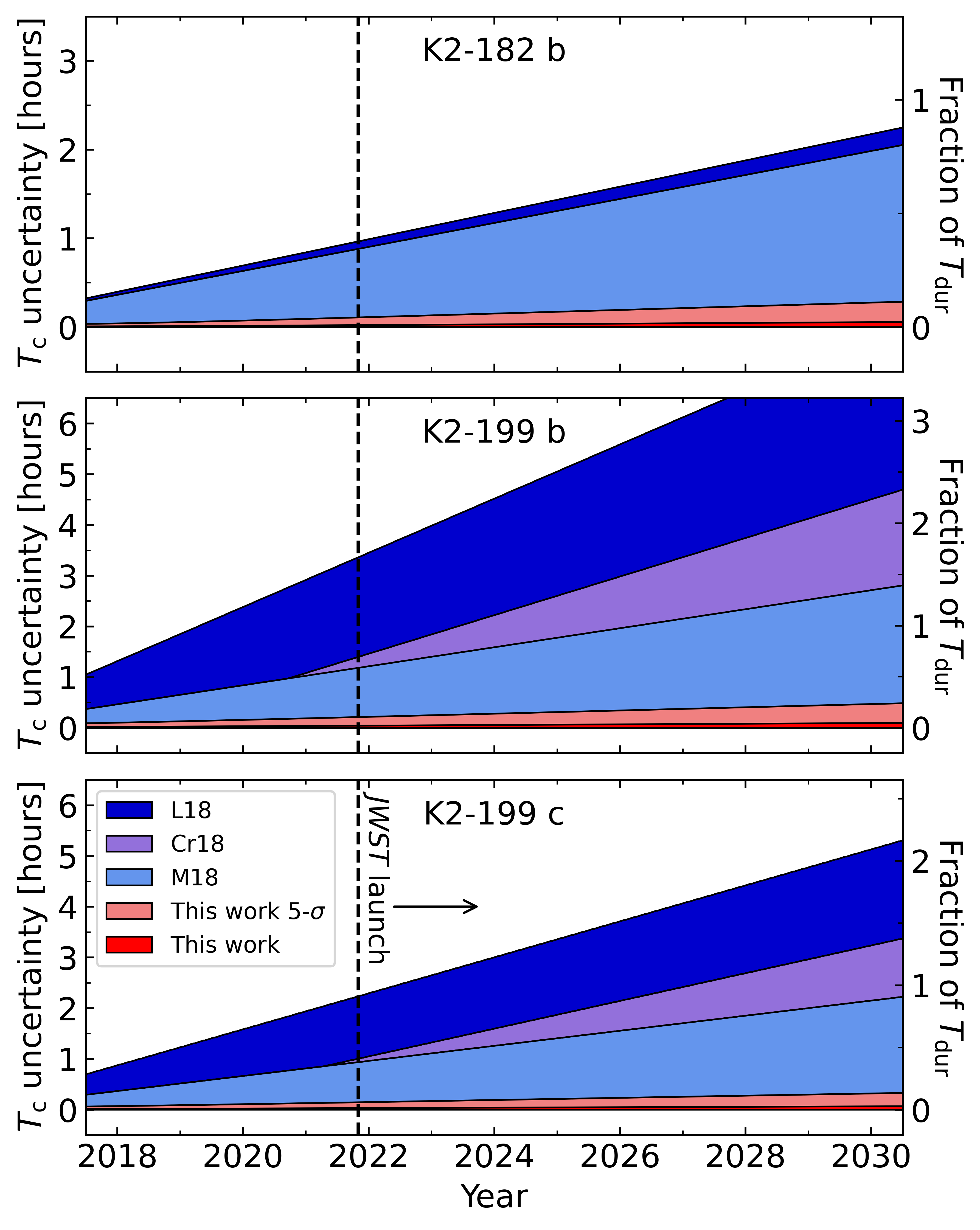}
    \caption{Transit time uncertainty for the three planets over the next decade. The left and right axes show the uncertainty in units of hours and as a fraction of the transit duration, respectively. Regions for \livingston, Cr18, and \mayo refer to the 1-$\sigma$ uncertainty in \transitTime as propagated from the values in \cite{livingston18}, \cite{crossfield18}, and \cite{mayo18}, respectively. Note that \cite{crossfield18} analyzed \ktwo C17 and therefore does not include \oneEightTwo b. 1-$\sigma$ and 5-$\sigma$ uncertainties in \transitTime over time are also plotted using the refined ephemerides from this work. \revision{The dashed vertical line and arrow mark \jwst's launch in 2021 Nov or later.} With our measurements of $P$ and \transitTime, the 5-$\sigma$ uncertainty in the transit time of each planet will be less than 20 minutes by the year 2025.}
    \label{fig:t0_errors}
\end{figure}

We compared the uncertainties on our orbital period ($P$) and \transitTime measurements as reported in Table \ref{tab:star_planet_combined} to those in the planet validation papers, \mayo and \livingston. For \oneNineNine b and c we also compared our uncertainties to \cite{crossfield18}. For \oneEightTwo b, our uncertainties on $P$ and \transitTime are improvements by a factor of about 30 and 2, respectively, over the values reported in both \mayo and \livingston. For \oneNineNine b we find improvements by a factor of about 30 and 2 over the $P$ and \transitTime values in \mayo and improvement factors of $\gtrsim60$ and $3$ compared to both \livingston and \cite{crossfield18}. We find similar improvement factors in each case for \oneNineNine c. Figure \ref{fig:t0_errors} visualizes how our refinement of $P$ and \transitTime greatly reduce the transit time uncertainty for each planet over the next decade compared to the literature values. Our improvements to the ephemerides are key in making these planets viable targets for future space-based missions like \emph{JWST}.

\section{Stellar characterization} \label{sec:stellar}

\subsection{High-resolution spectroscopy} \label{stellar:spec}
We used the High Resolution Echelle Spectrometer (HIRES; \citealt{vogt94}) on the 10-m Keck I telescope at the W. M. Keck Observatory on Mauna Kea to obtain an iodine-free (see \S\ref{rvs:obs}) spectrum of \oneEightTwo at high resolution and high signal-to-noise (a ``template" spectrum). For \oneNineNine we used the \keckhires template from \nasakeyproj. These spectra were collected with the B3 decker (14\arcsec$\times$0\arcsec.574, $R = 67,000$), the length of which allows for effective sky-subtraction (important for these $V > 12$ mag stars). The templates were obtained on UTC 2019 May 7 and 2017 Feb 3 for \oneEightTwo and \oneNineNine, respectively, with exposure times of 2700 s. \oneEightTwo was observed at an airmass of 1.33 and \oneNineNine was observed at an airmass of 1.14. The \oneEightTwo and \oneNineNine templates have a signal-to-noise ratio (SNR) of 173/pix and 163/pix, respectively, \revision{as measured at 5500 \AA}. Both templates were collected while the moon was below the horizon. Triple-shot exposures of rapidly-rotating B stars were taken with the iodine cell in the light path immediately before and after the high-resolution spectra were collected to precisely constrain the instrumental point spread function (PSF). The data collection and reduction followed the methods of the California Planet Search as described in \cite{howard10}.

We used \specMatchSynth\footnote{\url{https://github.com/petigura/specmatch-syn}} (\citealt{specmatchsynth_thesis} and applied by \citealt{specmatchsynth}) to constrain the stellar effective temperature (\teff), surface gravity (\logg), and metallicity ([Fe/H]) of \oneEightTwo directly from our high-SNR, iodine-free \keckhires template. \specMatchSynth fits segments of optical spectra (380 \AA\ between 5200 and 6260 \AA) by interpolating on a three-dimensional grid (in \teff, \logg, and [Fe/H]) of synthetic spectral models  \citep{coelho05}. The interpolated spectrum is then convovled with the kernel from \cite{hirano11} to account for rotational line broadening, as well as with a Gaussian that represents the \keckhires PSF during typical seeing conditions. For \oneEightTwo, \specMatchSynth finds $T_\text{eff} = 5170 \pm 100$ K, \logg $= 4.61 \pm 0.10$ dex, and $\text{[Fe/H]} = 0.12 \pm 0.06$ dex.

\oneNineNine is a K5V-dwarf \citep{dressing17a}. We used \specMatchEmp\footnote{\url{https://github.com/samuelyeewl/specmatch-emp}} \citep{yee17} to constrain \oneNineNine's \teff, stellar radius (\rstar), and [Fe/H] from the template spectra collected by \nasakeyproj. \specMatchEmp fits stellar spectra between 5000 and 5800 \AA\ in 100 \AA\ segments using linear combinations of spectral templates from a library of over 400 precisely-characterized FGKM stars. For our initial characterization of \oneNineNine we chose to use \specMatchEmp rather \specMatchSynth because the former is robust to cooler stars ($\sim$K4 and later) whose molecular spectral features may not be accounted for in synthetic models\footnote{\revision{This is not to say that \specMatchEmp fails when applied to hotter stars, however (i.e. F, G, and early-K dwarfs). We also ran \specMatchEmp on our \oneEightTwo template spectra and found the resulting \teff and [Fe/H] were 1-$\sigma$ consistent with the \specMatchSynth results. Using \specMatchSynth to characterize the \oneEightTwo template spectra rather than \specMatchEmp follows the \teff threshold from \cite{petigura17}.}}. For \oneNineNine, \specMatchEmp finds $T_\mathrm{eff} = 4491 \pm 70$ K, [Fe/H] $= -0.01 \pm 0.09$ dex, and $R_* = 0.70 \pm 0.10$ \rsun. 

To derive the final stellar parameters, we compiled the results from \specMatchSynth and \specMatchEmp with parallaxes from the second \emph{Gaia} data release (\gaiadrtwo; \citealt{gaia, gaiadr2}; \citealt{luri18}) and multiband photometry (\emph{Gaia} $G$ and Two Micron All Sky Survey [2MASS] $JHK$, \citealt{2mass}). We then input these data to \isoclassify\footnote{\url{https://github.com/danxhuber/isoclassify}} (\citealt{huber17} and \citealt{berger20}), using \texttt{grid} mode. \isoclassify infers marginal posteriors for stellar parameters by integrating over a grid of MIST isochrones \citep{choi16}. The stellar parameters are summarized in Table \ref{tab:star_planet_combined}. We find that our derived stellar parameters are generally consistent to within 1 to 2-$\sigma$ of those reported in the validation papers. We note that \mayo was published in March 2018, before \gaiadrtwo became public. \livingston was published in November 2018 and incorporated \gaiadrtwo parallaxes in the determination of stellar properties.

While our methodology for the stellar characterization closely follows the procedure in \livingston{}—who use iodine-free \keckhires spectra ($\text{SNR} \approx 40\text{/pixel}$) obtained by \cite{petigura17}—we chose to re-derive the stellar parameters ourselves because the templates we feed to \specMatchSynth and \specMatchEmp have much higher SNR. Additionally, we note that in their derivation of stellar parameters, \livingston use \specMatchSynth for \oneNineNine rather than \specMatchEmp. This is in contrast to \cite{petigura17}, who used the latter—\livingston only use \specMatchEmp for stars cooler than 4200 K, while \cite{petigura17} use it for stars cooler than 4600 K. For completeness, we tried using the \oneNineNine \specMatchSynth outputs as inputs to \isoclassify (along with the \gaiadrtwo parallax and multiband photometry) and found all stellar parameters to be 1-$\sigma$ consistent with the values in Table \ref{tab:star_planet_combined}. Moving forward we choose to quote the \isoclassify results that take input from \specMatchSynth for \oneEightTwo and \specMatchEmp for \oneNineNine because this follows the \teff threshold from \cite{petigura17}.

While we do not report a precise spectral classification for \oneEightTwo, we suggest it is an early-K dwarf based on \specMatchEmp, which models its spectrum as a linear combination of GL 144, HD 8553, HD 80367, HD 170657, and HD 189733, all of which are K0V to K2V. 

\begin{deluxetable*}{lccccc}
\tablecaption{System parameters \label{tab:star_planet_combined}}
\tabletypesize{\footnotesize}
\startdata
\tablehead{
    \vspace{0.01cm} \\ 
    \multicolumn{6}{c}{\textbf{Stellar Parameters}} \\
    \colhead{Parameter} & \colhead{Symbol} & \colhead{Units} & \colhead{\oneEightTwo} & \colhead{\oneNineNine} & \colhead{Provenance}
} \\
\sidehead{\emph{Identifying information}}
EPIC ID & & & 211359660 & 212779596 & EPIC \\
Right ascension & RA & deg (J2000) & $130.18004$ & $208.90145$ & \gaiadrtwo \\
Declination & DEC & deg (J2000) & $10.98295$ & $-6.13608$ & \gaiadrtwo \\
\emph{Kepler} magnitude & $K_\mathrm{p}$ & mag & 11.74 & 11.93 & EPIC \\
\sidehead{\emph{Spectroscopy}}
Effective temperature & \teff & K & $5170 \pm 100$ &  $4491 \pm 100$ & \texttt{SpecMatch}$^\mathrm{a}$ \\
Surface gravity & \logg & cgs & $4.61 \pm 0.10$  & \nodata & \texttt{SpecMatch}$^\mathrm{a}$ \\
Metallicity & $\text{[Fe/H]}$ & dex & $0.12 \pm 0.06$  & $-0.01 \pm 0.09$ & \texttt{SpecMatch}$^\mathrm{a}$ \\
\sidehead{\emph{Isochrone modeling}}
Mass & \mstar & \msun & \oneEightTwoIsoMstar &  \oneNineNineIsoMstar & \isoclassify \\
Radius & \rstar & \rsun & \oneEightTwoIsoRstar &  \oneNineNineIsoRstar & \isoclassify \\
Luminosity & $L$ & $L_\odot$ & \oneEightTwoIsoLum & \oneNineNineIsoLum & \isoclassify \\
Age &  & Gyr & $2.07^{+2.14}_{-1.36}$ & $5.04^{+6.12}_{-3.59}$ & \isoclassify \\
\vspace{0.01cm} \\ 
\multicolumn{6}{c}{\textbf{Planet Parameters}} \\
\colhead{Parameter} & \colhead{Symbol} & \colhead{Units} & \colhead{\oneEightTwo b} & \colhead{\oneNineNine b} & \colhead{\oneNineNine c} \\
\hline \\
\sidehead{\emph{Measured quantities}}
Orbital period & $P$ & Days & \oneEightTwoPeriod & \oneNineNineBPeriod & \oneNineNineCPeriod \\ 
Time of inferior conjunction & \transitTime & $\mathrm{BJD} - 2454833$ & \oneEightTwoEpoch & \oneNineNineBEpoch & \oneNineNineCEpoch \\
Occultation fraction & $\frac{R_\mathrm{p}}{R_*}$ & \% & \oneEightTwoRor & \oneNineNineBRor& \oneNineNineCRor \\
Impact parameter & $b$ & & \oneEightTwoImpact & \oneNineNineBImpact & \oneNineNineCImpact \\
Orbital eccentricity\textsuperscript{b} & $e$ & & \oneEightTwoEcc & \oneNineNineBEcc & \oneNineNineCEcc \\
Argument of periastron\textsuperscript{b} & $\omega$ & rad & \oneEightTwoOmega & \oneNineNineBOmega & \oneNineNineCOmega \\
RV semi-amplitude & $K$ & m s$^{-1}$ & \oneEightTwoModelAKamp & \oneNineNineBKamp & \oneNineNineCKamp \\
\sidehead{\emph{Derived quantities}}
Transit duration & $T_\mathrm{dur}$ & Hours & \oneEightTwoTdur & \oneNineNineBTdur & \oneNineNineCTdur \\
\revision{Orbital inclination} & $i$ & Rad & \oneEightTwoI & \oneNineNineBI & \oneNineNineCI \\
Orbital separation & $\frac{a}{R_*}$ &  & \oneEightTwoAor & \oneNineNineBAor &  \oneNineNineCAor\\
Orbital semi-major axis & $a$ & AU & \oneEightTwoSemiMaj & \oneNineNineBSemiMaj & \oneNineNineCSemiMaj \\
Radius & \rplanet & \rearth & \oneEightTwoRad & \oneNineNineBRad & \oneNineNineCRad \\
Mass & \mplanet & \mearth & \oneEightTwoMass & \oneNineNineBMass & \oneNineNineCMass \\
Bulk density & $\rho$ & g cm$^{-3}$ & \oneEightTwoRho & \oneNineNineBRho & \oneNineNineCRho \\
Equilibrium temperature\textsuperscript{c} & $T_\mathrm{eq}$ & K & \oneEightTwoTeq & \oneNineNineBTeq & \oneNineNineCTeq \\
Instellation flux & $S_\mathrm{p}$ & $S_\oplus$ & \oneEightTwoSinc & \oneNineNineBSinc & \oneNineNineCSinc \\
Core water mass fraction\textsuperscript{d} & \waterfrac & \% &  \oneEightTwoWater & \oneNineNineBWater & \oneNineNineCWater \\
H$_2$/He envelope mass fraction\textsuperscript{e} & \hhefrac & \% & \nodata & \oneNineNineBFhhe & \oneNineNineCFhhe \\
TSM\textsuperscript{f} & TSM & & \oneEightTwoTSM & \oneNineNineBTSM & \oneNineNineCTSM \\
\enddata
\tablecomments{\revision{Values reported in this table represent medians of the marginal posterior, while errors reflect the the upper and lower bounds of the 68\% confidence interval about the median. Errors on derived planetary parameters include errors on stellar mass and radius by way of Gaussian priors placed on the stellar properties (informed by our stellar characterization) during the modeling of the photometry and radial velocities.}\\
a: \specMatchSynth was used for \oneEightTwo and \specMatchEmp was used for \oneNineNine following the \teff threshold of \cite{petigura17}.\\
b: \revision{$e$ and $\omega$ were allowed to vary during the light curve modeling. Finding that the orbits were consistent with circular, they were held fixed at zero in models of the RVs.}\\
c: Equilibrium temperature calculated assuming zero Bond albedo.\\
d: Core water mass fraction calculated assuming a bulk composition of water and rock, using the grid from \cite{zeng16}.\\
e: H$_2$/He envelope mass fraction calculated assuming a $1\times$ solar metallicity H$_2$/He envelope on top of an Earth-like rock/iron core, using the grids from \cite{lopezforney14}. We do not infer \hhefrac for \oneEightTwo b because its mass lies at the edge of the grid (20 \mearth).\\
f: The transmission spectroscopy metric from \cite{kempton18}.\\ 
References for the provenance values in the order in which they appear in the table: EPIC (Ecliptic Plane Input Catalog; \citealt{epic}), \gaiadrtwo (\citealt{gaia, gaiadr2}), \isoclassify (\citealt{huber17} and \citealt{berger20}).}
\end{deluxetable*}

\subsection{High-resolution imaging} \label{stellar:img}

High-resolution imaging is a key component of the planet validation process because it can rule out astrophysical false positive scenarios and put limits on dilution levels from nearby stellar contaminants. While all three planets were previously validated by \mayo and \livingston, we acquired new (in the case of \oneEightTwo) and reduced publicly available (for \oneNineNine) high-resolution, high-contrast imaging observations in an attempt to further improve contrast limits.

\subsubsection{\oneEightTwo}

Neither \mayo or \livingston included imaging in their \vespa FPP calculations for \oneEightTwo b. While \gaiadrtwo does not detect any bright sources within the \everest aperture, we observed \oneEightTwo with the second-generation Near-infrared Camera (NIRC2; \citealt{nirc2:wizinowich14}) on the Keck II telescope to confidently rule out nearby stellar contaminants.

\oneEightTwo was observed with \kecknirctwo on UT 2020 May 28 at an airmass of 1.58. Observations were taken in narrow camera mode with a 1024” x 1024” FOV. A three-point dither pattern was used to avoid the noisy fourth quadrant of the detector. Observations were taken in the $K$ filter for a total integration time of 4 seconds.  

All of the \kecknirctwo AO data were processed and analyzed with a custom set of IDL tools. The science frames were flat-fielded and sky-subtracted. The flat fields were generated from a median average of dark subtracted flats taken on-sky. The flats were normalized such that the median value of the flats is unity. The sky frames were generated for each observation set individually; the dithered flat-fielded science frames were median averaged to produce a single sky frame that was then subtracted from all of the flat-field science frames.  The reduced science frames were combined into a single combined image using a intra-pixel interpolation that conserves flux, shifts the individual dithered frames by the appropriate fractional pixels, and coadds the frames. The final resolution of the combined dither was determined from the full-width half-maximum of the point spread function.

The sensitivities of the final combined \kecknirctwo image were determined by injecting simulated sources azimuthally around the primary target every $45\degree $ at separations of integer multiples of the central source's full width at half maximum (FWHM; \citealt{furlan17}). The brightness of each injected source was scaled until standard aperture photometry detected it with 5-$\sigma$ significance. The resulting brightness of the injected sources relative to the target set the contrast limits at that injection location. The final 5-$\sigma $ limit at each separation was determined from the average of all of the determined limits at that separation and the uncertainty on the limit was set by the RMS dispersion of the azimuthal slices at a given radial distance.

From our \kecknirctwo imaging of \oneEightTwo we find no companions to 5-$\sigma$ confidence for $\Delta K = 5.1$ mag at a minimum separation of 0\arcsec.2 and for $\Delta K = 7.0$ mag at a minimum separation of 0\arcsec.5. We find no companions to 5-$\sigma$ confidence for $\Delta K = 7.4$ mag at a minimum separation of 1\arcsec. The \kecknirctwo image is shown on the left in Figure \ref{fig:imgaing_combined}.

\revision{All of our observations suggest that \oneEightTwo is unambiguously single. If there was an undetected source within $\sim$1\arcsec, we would expect to see it in our spectroscopic observations. However, using the methodology from \cite{kolbl15}, we find no indication that \oneEightTwo is a double-lined spectroscopic binary (and the \citealt{kolbl15} algorithm is sensitive to companions down to $\sim$1\% the brightness of the primary in $V$-band). Furthermore, if there did happen to be an undetected source that our imaging is unable to rule out, even in the ``worst case scenario" (i.e. a $\Delta K = 7.4$ mag neighbor at a separation of 1\arcsec) the dilution level would be negligible at $<1\%$.}

\subsubsection{\oneNineNine}

\revision{As explained in \S\ref{sub:182_199_phot}, although a fainter neighbor falls in our \everest aperture for \oneNineNine ($\Delta K_\mathrm{p} = 5.6$ mag, separation $\approx 8$\arcsec) any contamination from this source would be $<1\%$ and is therefore negligible compared to our other sources of error (e.g. stellar properties).}

\livingston and \mayo use observations of \oneNineNine from the Differential Speckle Survey Instrument (DSSI; \citealt{dssi}) on both the WIYN telescope at KPNO and the 8.1-m Gemini-North telescope at the Gemini North Observatory on Mauna Kea to inform their \vespa FPP calculations. Two images, one centered at 6920 \AA\ and one centered at 8800 \AA, were taken with both WIYN and Gemini-North. The WIYN images were taken on UT 2016 Apr 21 and have an estimated contrast of $\Delta$3.5 mag at a separation of 0\arcsec.2. The Gemini-North images were taken on UT 2016 Jun 21 and have an estimated contrast of $\Delta$5.0 mag at a separation of 0\arcsec.2.

\begin{figure*}
\gridline{\fig{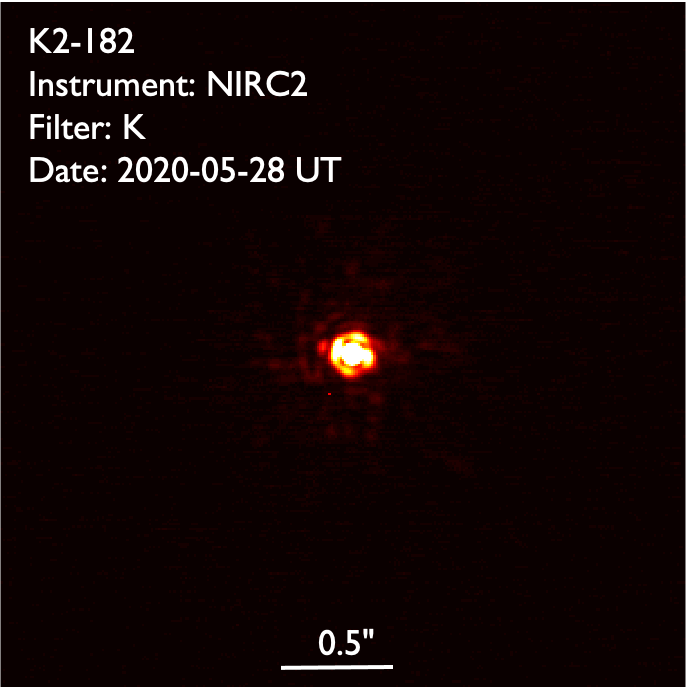}{0.5\textwidth}{(a) \oneEightTwo \kecknirctwo}
          \fig{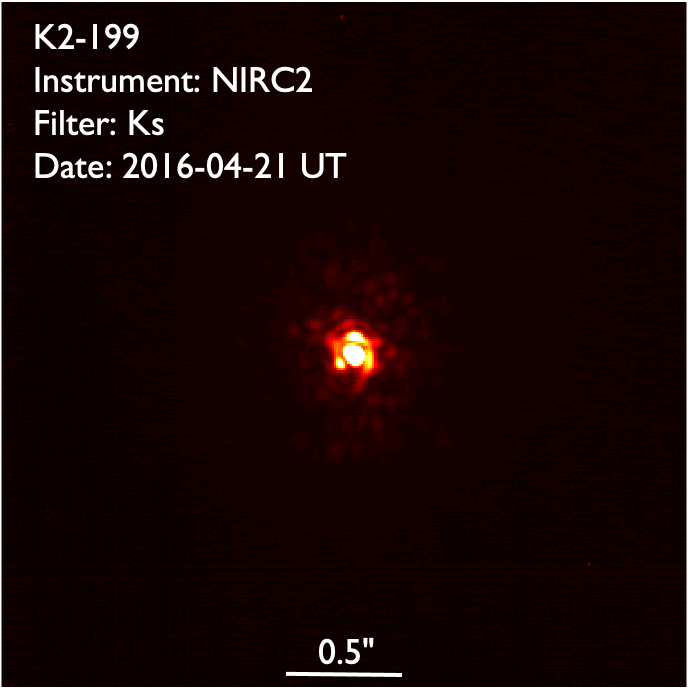}{0.5\textwidth}{(b) \oneNineNine \kecknirctwo}}
  \caption{Our high-resolution, high-contrast images of \oneEightTwo and \oneNineNine. \emph{Left:} The high-resolution image of \oneEightTwo with \kecknirctwo. The \kecknirctwo image rules out companions at 5-$\sigma$ confidence with a maximum contrast of $\Delta K = 7.4$ mag at a minimum separation of 1\arcsec. \emph{Right:} The \kecknirctwo image of \oneNineNine that we reduced from archival observations. We rule out nearby companions at 5-$\sigma$ confidence down to $\Delta K_\mathrm{s} = 8.8$ mag at a minimum separation of 1\arcsec.} \label{fig:imgaing_combined}
\end{figure*}

To improve contrast limits beyond those from DSSI, we identified and reduced a publicly available, archival AO image of \oneNineNine from \kecknirctwo. The archival \kecknirctwo image was acquired on UT 2016 Apr 21 at an airmass of 1.12. The observations were taken in the $K_\mathrm{s}$ filter with a total integration time of 288 seconds. Our data reduction for the archival image mimicked our method for the \kecknirctwo image of \oneEightTwo. We find no companions at 5-$\sigma$ confidence for $\Delta K_\mathrm{s} = 4.4$ mag at 0\arcsec.2 separation, $\Delta K_\mathrm{s} = 6.9$ mag at 0.\arcsec5 separation, and $\Delta K_\mathrm{s} = 8.8$ mag at 1\arcsec\ separation. The high-resolution image from \kecknirctwo is shown on the right in Figure \ref{fig:imgaing_combined}.

\revision{Similar to the case for \oneEightTwo, our observations of \oneNineNine suggest that the star is unambiguously single. We find no evidence that \oneNineNine is a double-lined spectroscopic binary \citep{kolbl15}. Any dilution from a neighbor that was not detected by the imaging observations would be $<1\%$, making its impact on the derived planetary radii negligible.}

\section{Radial velocities} \label{sec:rvs}

\subsection{Observations} \label{rvs:obs}

We obtained 12 high-resolution spectra of both \oneEightTwo and \oneNineNine with \keckhires to measure precise radial velocities (RVs). The observations of \oneEightTwo were taken between UTC 2019 Mar 27 and 2019 May 29 with a median exposure time of 1208 seconds and a median SNR of 100/pixel. The observations of \oneNineNine were taken between UTC 2019 Feb 18 and 2019 May 30 with a median exposure time of 1402 seconds and a median SNR of 100/pixel. All spectra were collected with the C2 decker (14\arcsec$\times$0\arcsec.86, $R =$ 60,000) to allow for sky subtraction. 

Radial velocities were determined following the procedures of \cite{howard10}. In brief, a warm ($50\degree$ C) cell of molecular iodine was placed at the entrance slit during the RV observations \citep{butler96}. The superposition of the iodine absorption lines on the stellar spectrum provides both a fiducial wavelength solution and a precise, observation-specific characterization of the instrument's PSF. Each RV spectrum was then modeled as the sum of the deconvolved template spectrum (see \S\ref{stellar:spec}) and the instrumental PSF convolved with an ``infinite resolution" laboratory spectrum of the molecular iodine. Our RV measurements of \oneEightTwo and \oneNineNine are listed in Table \ref{tab:k2-182_rvs} and Table \ref{tab:k2-199_rvs}, respectively. Before applying any models, the \oneEightTwo RV data had an RMS of 6.88 \mps and a median pointwise error of 1.55 \mps. The \oneNineNine RVs had an RMS of 5.21 \mps and a median pointwise error of 1.73 \mps.

For our analysis of \oneNineNine we include 33 \keckhires \revision{iodine-in spectra} from \nasakeyproj, which were obtained between UTC 2016 Jul 23 and 2018 Mar 17. \revision{These observations were also collected with the C2 decker and RVs of the combined ($N = 45$) data set were derived simultaneously.} Before applying any models, the combined \oneNineNine RV data had an RMS of 6.31 \mps and a median pointwise error of 1.69 \mps.

\subsection{Stellar activity} \label{rvs:activity}

Failure to account for just one coherent signal in an RV time series (e.g. stellar activity, additional planets, instrumental systematics) can produce discrepant planet mass measurements, even in the presence of large data sets \citep{rajpaul17}. As RV observations enter the era of sub-\mps measurement precision with next-generation instruments like \geminimaroon \citep{maroonx} and \wiynneid \citep{neid}, confounding signals in RV data, astrophysical or otherwise, will become the limiting factor in the pursuit of precise mass measurements for small planets.

Stellar activity can produce RV signals across a range of timescales by distorting spectral line profiles. As summarized by \cite{vanderburg16}, asteroseismic oscillations, granulation, starspots, and stellar magnetic activity can produce RV amplitudes on the order of 1 to several \mps for timescales of a few minutes, to hours and days, to months, and even years for stellar magnetic cycles (\citealt{butler04}, \citealt{dumusque11a}, \citealt{queloz01}, \citealt{gomesDaSilva12}). When the dominant stellar activity timescale is sufficiently short (typically minutes to a few days), it is common to treat the RV-signature of the activity as additional white noise about the planetary signal(s), added in quadrature with the pointwise RV measurement errors \citep{gregory05}. However, adding a simple ``RV jitter" term to the likelihood function can prove insufficient when correlated stellar activity signals extend to longer timescales ($\sim$tens of days). \cite{vanderburg16} show that the RV signals of rotating starspots can confuse planet searches, especially when the stellar rotation period and its first two harmonics fall near planet orbital periods.

As seen from their \ktwo light curves, both \oneEightTwo and \oneNineNine exhibit clear signs of starspots and spot modulation as well as significant spot evolution from one campaign to another. To obtain a rough estimate of the amplitude of the systematic RV contributions due to star spots ($\sigma_\mathrm{RV,\: spot}$), following \cite{dragomir19} we adapted Equation 2 from \cite{vanderburg16} as $\sigma_\mathrm{RV,\: spot} \approx \sigma_\mathrm{phot} \times v\sin i$, where $\sigma_\mathrm{phot}$ is the standard deviation of the light curve flux and $v\sin i$ is the sky-project stellar rotational velocity. 

For \oneEightTwo, $\sigma_\mathrm{phot} = 0.0026$ in units of relative flux from the \everest C5 and C18 light curves and \specMatchSynth finds $v\sin i = 0.25 \pm 1.0$ km s$^{-1}$ from our \keckhires template spectra. However, \specMatchSynth $v\sin i$ estimates below 2.0 km s$^{-1}$ should be interpreted as less than 2 km s$^{-1}$. Using $2$ km s$^{-1}$ as an upper limit on $v\sin i$ we calculate $\sigma_\mathrm{RV,\: spot} \lesssim 5.2$ m s$^{-1}$. Since \specMatchSynth only provides an upper limit on $v\sin i$ for \oneEightTwo, we also calculated $\sigma_\mathrm{RV,\: spot}$ using $v\sin i = 2 \pi R_* \sin i /P_\mathrm{rot}$ and $\sin i \approx 1$, where $R_* = 0.79$ \rsun (Table \ref{tab:star_planet_combined}) and $P_\mathrm{rot} \approx 25$ days (see \ref{activity:k2-182}). Estimating $v$ with $R_*$ and $P_\mathrm{rot}$, we find $\sigma_\mathrm{RV,\: spot} \approx 4.2$ \mps.

For \oneNineNine, the \everest C6 and C17 light curves have $\sigma_\mathrm{phot} = 0.0027$. While \specMatchEmp does not calculate an absolute $v \sin i$, \specMatchSynth finds $v \sin i < 2$ km s$^{-1}$ from the \keckhires template. Given our discussion in \S\ref{stellar:spec}, though, \oneNineNine is too cool for \specMatchSynth. Modulo this concern, $\sigma_\mathrm{RV,\: spot}$ would be $\lesssim 5.4$ m s$^{-1}$ for \oneNineNine using the \specMatchSynth upper limit. Extracting $P_\mathrm{rot}$ from the \ktwo photometry was nontrivial for \oneNineNine (see \ref{activity:k2-199}), so we do not try to estimate $\sigma_\mathrm{RV,\: spot}$ via $R_*$ and $P_\mathrm{rot}$ for \oneNineNine as we did for \oneEightTwo.

We employed a GP to model the RV-signatures of stellar activity in an attempt to mitigate their impact on the planet mass measurements. GPs are a popular tool for modeling correlated stellar activity and can help inform RV models when trained on complementary data such as photometry \citep{haywood14, grunblatt15} or activity indicators like Ca II H and K emission (\shk; \citealt{isaacson10, kosiarek21}) and H-$\alpha$ indices \citep{robertson13}. Recently, \cite{kosiarek20} used over 70 years of solar observations to demonstrate that photometry can act as a proxy for stellar activity in GP-enabled analyses. Below we discuss the results of training GPs on the \oneEightTwo and \oneNineNine photometry.

\subsubsection{\oneEightTwo} \label{activity:k2-182}

K2-182 is an early-K dwarf. Ca II H and K emission indicates it is moderately active—using the iodine-free \keckhires template we find $\log R'_\text{HK} = -4.68$ dex \citep{middelkoop82, noyes84}. For reference, over its magnetic cycle the Sun floats between $\log R'_\text{HK} = -5.05$ and $-4.84$ dex \citep{meunier10}. We used a GP with a quasi-periodic (QP) kernel to model the stellar rotation and spot modulation seen in the \ktwo C5 and C18 photometry. The kernel, 
\begin{equation}
\label{eq:kernel}
    k(t,t') = \eta_1^2 \ \mathrm{exp} \left[-\frac{(t-t')^2}{\eta_2^2}-\frac{\sin^2(\frac{\pi(t-t')}{\eta_3})}{2 \eta_4^2}\right],
\end{equation}
quantifies the covariance between data observed at times $t$ and $t'$. $\eta_{\text{1-4}}$ are the hyperparameters: $\eta_1$ represents the amplitude of the covariance, $\eta_2$ is interpreted as the evolutionary timescale of active stellar regions, $\eta_3$ is interpreted as the stellar rotation period, and $\eta_4$ is the length scale of the covariance's periodicity. The GP ``training" consists of performing a MAP fit to the training data (the out-of-transit \ktwo photometry) and posterior estimation for the GP hyperparameters with MCMC. The posteriors of $\eta_2$, $\eta_3$, and $\eta_4$ resulting from the training step are then used as numerical priors for these hyperparameters in a Keplerian $+$ GP fit to the RVs.

\revision{Although the QP kernel in Equation \ref{eq:kernel} is not the same kernel we use to simultaneously model stellar activity and planetary transits (see \S \ref{sub:phot_model_and_fit}), the QP kernel has been shown to be an effective model of stellar light curves (e.g. \citealt{angus18}) and both kernels have the capacity to explain exponentially decaying and periodic signals. We employ the QP kernel for the GP $+$ Keplerian modeling of the RVs because it is a familiar choice for stellar activity mitigation in RV time series \citep[e.g.][]{grunblatt15, kosiarek21}. In addition, it uses fewer free hyperparameters than the sum of SHOs we used in \S \ref{sub:phot_model_and_fit} (four versus seven), which is important for our smaller RV data sets. Given a covariance kernel that produces a smooth process (e.g. both the QP kernel and the sum of SHOs), even if hyperparameters differ, the GP's posterior predictions should be similar.}

The GP training on the out-of-transit \ktwo C5 and C18 photometry resulted in tight constraints on \oneEightTwo's active region evolutionary timescale, stellar rotation period, and periodic length scale ($\eta_2 = 37.4 \pm 5.2$ days, $\eta_3 = 24.92^{+0.26}_{-0.21}$ days, and $\eta_4 = 0.350^{+0.019}_{-0.018}$, respectively). This estimate of the stellar rotation period is consistent with the rotation period derived by the GP used to model the stellar activity signal in the \oneEightTwo photometry in \S\ref{sub:phot_model_and_fit} ($P_\mathrm{rot} = $ \oneEightTwoGProtP\xspace days). The posteriors on $\eta_2$, $\eta_3$, and $\eta_4$ were used as priors for an RV model that included a GP component, as described in \S\ref{rvs:k2-182}.

While \keckhires \shk values were measured simultaneously with RV observations, we forgo using them to inform models of correlated stellar activity given the small size of the data set. Our \oneEightTwo data set is also too small for a meaningful periodogram analysis so we reserve a discussion of the RV, RV residuals, and \shk periodograms for \oneNineNine.

\subsubsection{\oneNineNine} \label{activity:k2-199}
\oneNineNine is slightly cooler than \oneEightTwo and has a spectral classification of K5V \citep{dressing17a}. Ca II H and K emission indicates that it is also moderately active, with $\log R'_\text{HK} = -4.65$ dex. The \ktwo photometry shows stark differences in spot behavior from C6 to C17 (top panel of Figure \ref{fig:k2-199_phot}). While the C6 photometry appears to show a quasi-periodic signal, the amplitude of the modulation is damped significantly in C17. Using the kernel in Equation \ref{eq:kernel}, training a GP on the out-of-transit C6 and C17 photometry resulted in a bimodal posterior on $\eta_3$, with one peak near 14.15 days and another at 17.8 days. Examining a Lomb-Scargle periodogram \citep{lomb76, scargle82} of the \ktwo photometry, we find that the signal at 14.15 days is the first harmonic of a larger peak at 28.3 days. With the slight offset between the first harmonic of the peak at 28.3 days and the peak at 17.8 days, the GP had trouble modeling both signals with only a single hyperparameter to explain the periodicity ($\eta_3$). The kernel used to model the stellar activity signal in the photometry in \S\ref{sub:phot_model_and_fit}, which included a periodic term at $P_\mathrm{rot}$ and $P_\mathrm{rot}/2$, also had difficulty with the bimodality (see the note for Table \ref{tab:199_phot_model}). In the end, training a GP on the \oneNineNine photometry could be a moot point because of the star's apparent change in spot behavior over time—it is unclear which paradigm applies to the time period over which our RVs were collected.

In addition to experimenting with training a GP on the \oneNineNine photometry to search for signs of correlated stellar activity, we also examined a Lomb-Scargle periodogram of the RV times series, the RV residuals about our MAP solution from \S\ref{rvs:k2-199} (seen in Figure \ref{fig:k2-199_rv}), and the \shk indices we measured simultaneously with the RVs. In a periodogram of the RVs, the only peak to rise above the $1\%$ false alarm probability (FAP) level is the signal of \oneNineNine c at $P = 7.37$ days. A periodogram of the RV time series with both planets removed shows no peaks above the 10\% FAP level. A periodogram of the \shk values reveals a peak at 359.8 days that rises above the 0.01\% FAP level. However, nothing significant (above the 1\% FAP level) is detected shortward of the the 359.8-day peak's first harmonic at 179.9 days, so while there is power at these longer periods we do not believe it should affect the planetary signals ($P = 3.23$ and $7.37$ days).

We choose not to pursue an RV model of \oneNineNine that includes a GP due to the change in \oneNineNine's spot behavior between campaigns and the lack of Lomb-Scargle power in the \keckhires \shk indices on timescales comparable to the planet orbital periods. Moving forward we suggest that additional photometric monitoring, contemporaneous with RV and \shk observations, may help better inform the correlated RV signatures of stellar activity (if any) for \oneNineNine \citep{grunblatt15}. On the other hand, experimentation with a wider variety of GP kernels may help model the existing \ktwo photometry.

\subsection{Radial velocity analysis} \label{rvs:analysis}
We used \radvel\footnote{\url{https://github.com/California-Planet-Search/radvel}} \citep{radvel} to model the RVs. \radvel uses MAP estimation to fit Keplerian orbits to an RV time series. The orbit of each planet is described using five parameters: orbital period ($P$), time of inferior conjunction (\transitTime), orbital eccentricity ($e$), argument of periastron passage ($\omega$), and RV semi-amplitude ($K$). \revision{Modeling the velocities as a sum of Keplerian orbits and noise produces a measurement of \mplanet$\sin i$ for each planet, where $i$ is the planet's orbital inclination. Using Equation 7 in \cite{winn10}, we derive $i$ for each planet from our models of the photometry (see Table 1), enabling the conversion from \mplanet$\sin i$ to \mplanet.}

\radvel estimates posterior distributions with an MCMC scheme that uses an Affine Invariant sampler \citep{goodman10} from \emcee \citep{emcee}. For the MCMC uncertainty estimation we followed the default prescription for burn-in criteria, the number of walkers, the number of steps, and convergence testing as described by \cite{radvel}. \revision{In our analysis of the radial velocities we opted to use \emcee for the posterior estimation rather than employ an HMC sampler as we did for the models of the photometry (see \S\ref{sub:phot_post_est}). The biggest benefit of using HMC over other sampling methods is to enable the efficient exploration of high-dimensional posteriors. Since our models of the radial velocities have many fewer free parameters than the models of the photometry—\oneNineNine's photometric model has 25 free parameters (Table \ref{tab:199_phot_model}) while our adopted model of the system's radial velocities has seven (Table \ref{tab:199_rv_model})—we found that posterior estimation with \emcee for the models of the radial velocities was computationally tractable. Since \emcee is already integrated with \radvel (and \texttt{pymc3}, which enabled the HMC sampling for our models of the \ktwo photometry, is not), we chose to use \emcee for posterior estimation in our radial velocity analysis.}

\revision{We used the AICc, the small sample-corrected version of the Akaike Information Criterion (AIC; \citealt{akaike74}) to compare models of the radial velocities:
\begin{equation}
    \mathrm{AICc} = \mathrm{AIC} + \frac{2k^2 + 2k}{n - k - 1},
\end{equation}
where 
\begin{equation}
    \mathrm{AIC} = 2 k - 2 \log \mathcal{\hat{L}},
\end{equation}
$k$ is the number of free parameters in the model, $n$ is the sample size, $\log$ is the natural logarithm, and $\mathcal{\hat{L}}$ is the maximum of the likelihood function with respect to the model parameters. The AICc is essentially the same as the AIC, but with an additional penalty term for model complexity, since the AIC has a tendency to favor models that overfit when $n$ is small. \cite{burnham02} recommend using the AICc in place of the AIC when $n \lesssim 40 \times k$. Note that as $n \rightarrow +\infty$ the AICc reduces to the AIC. We elect to use the AICc over other popular model comparison metrics, namely the Bayesian Information Criterion (BIC; \citealt{schwarz78}), because simulation studies suggest that for finite $n$, the BIC may be at risk of selecting very poor models \citep{burnham04, vrieze12}.}

\revision{Let $\Delta \mathrm{AICc}_i \equiv \mathrm{AICc}_i - \mathrm{AICc}_\mathrm{min}$, where AICc$_i$ is the AICc of the $i$th model under consideration and AICc$_\mathrm{min}$ is the lowest AICc of all models considered. When comparing models with the AICc, \cite{burnham04} provide the following guidelines:
\begin{itemize}
    \item If $\Delta \mathrm{AICc}_i < 2$, the two models are nearly indistinguishable.
    \item If $2 < \Delta \mathrm{AICc}_i < 10$, the $i$th model is disfavored.
    \item If $\Delta \mathrm{AICc}_i > 10$, the $i$th model is essentially ruled out.
\end{itemize}
}

\subsubsection{\oneEightTwo} \label{rvs:k2-182}

Our models of the \oneEightTwo RVs hold a few common assumptions: Each model assumes there is a single planet orbiting \oneEightTwo. Additional planets would have to have an orbital period $>80$ days (the length of a normal \ktwo campaign) and/or be non-transiting. Our RV data set is not large enough ($N = 12$) to comment on the system's multiplicity in the absence of constraints from the \ktwo photometry. Each of our \oneEightTwo models also assume a circular orbit for \oneEightTwo b. Our RV data set is not large enough to comment on the orbit's eccentricity, though the \ktwo photometry suggests it is nominally low, albeit the posterior is skewed towards higher values ($e = $ \oneEightTwoEcc), perhaps due to Lucy-Sweeney bias \citep{lucy-sweeney71}. In addition, allowing the orbital eccentricity to vary introduces two free parameters ($e$ and $\omega$), which we preferred to avoid given our data set's modest size. Finally, we fix \oneEightTwo b's orbital period and time of inferior conjunction unless otherwise specified, since the RVs are unlikely to add appreciable constraints to the planet ephemerides compared to the two campaigns of \ktwo photometry. 

\begin{figure*}
\gridline{\fig{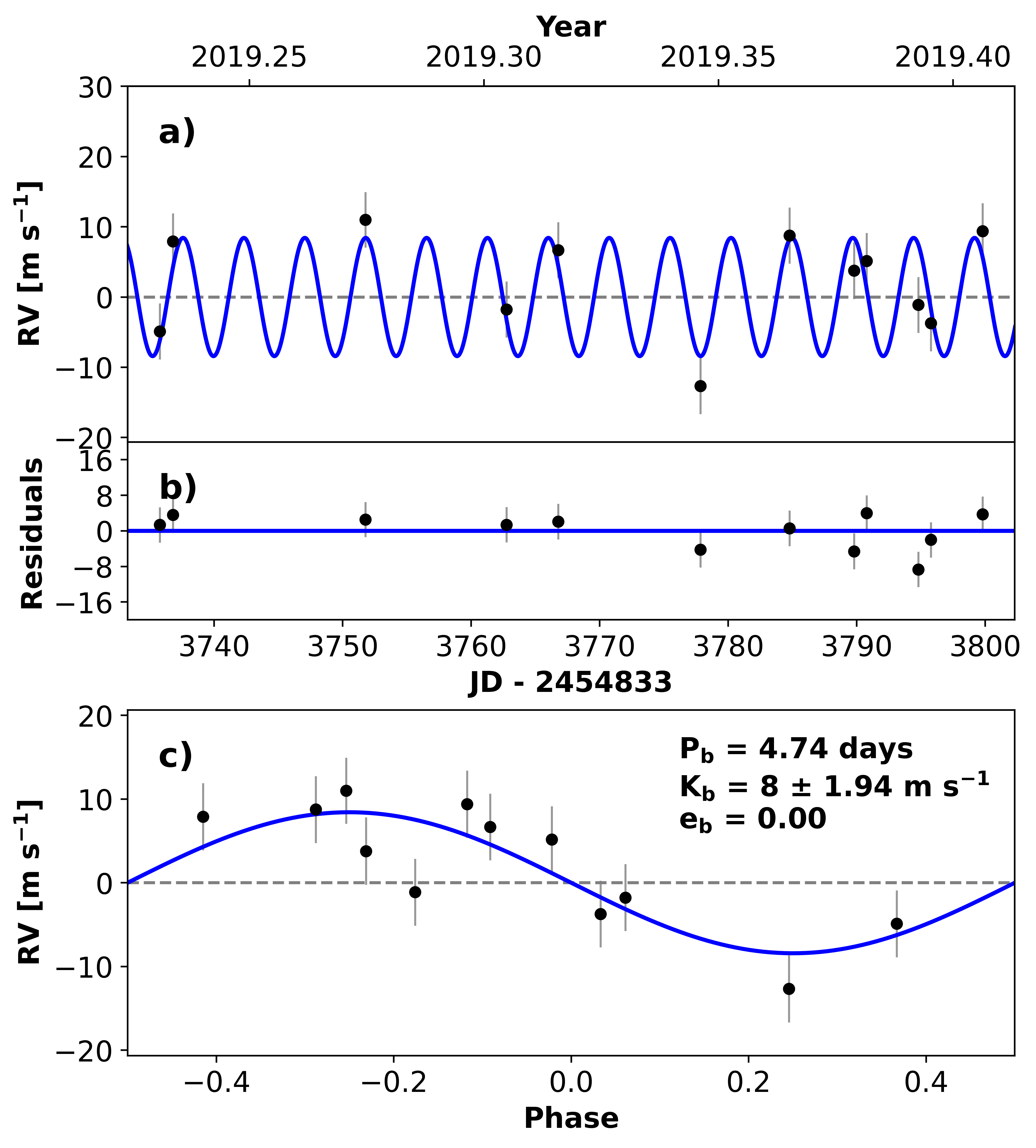}{0.5\textwidth}{(a) \oneEightTwo b RV Model A (adopted)}
          \fig{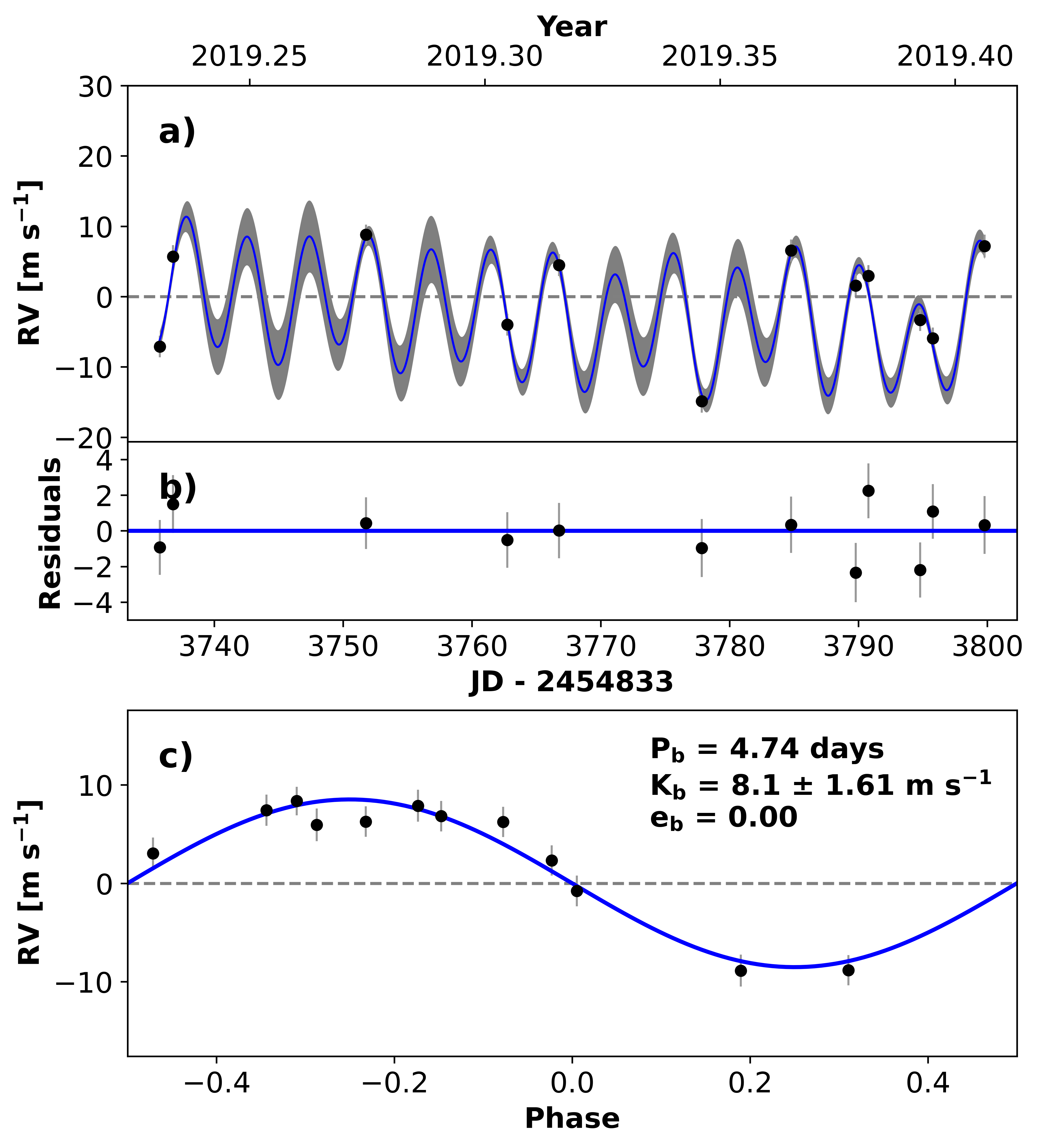}{0.5\textwidth}{(b) \oneEightTwo b RV Model B}}
  \caption{\emph{Left:} The best-fitting 1-planet Keplerian model for \oneEightTwo from \radvel, referred to as \oneEightTwoModelA. \emph{Right:} The best-fitting 1-planet Keplerian $+$ GP model, referred to as \oneEightTwoModelB. For both Model A and B, the MAP solution is shown in panel \emph{a)} as the blue line and \keckhires RVs are shown as the black points. For Model B, the 1-$\sigma$ GP error envelope is shown in gray about the MAP solution. For Model A, data error bars represent the 1-$\sigma$ pointwise RV measurement uncertainty added in quadrature with the RV jitter parameter, $\sigma_j$. For Model B, error bars represent the pointwise RV measurement uncertainty and GP posterior prediction uncertainty added in quadrature. Note that because GPs tend to ``snap" to the data, the typical GP posterior prediction uncertainty in Model B is much smaller than the RV jitter term from Model A. (This is why the error bars appear much smaller in panel \emph{c)} on the right than they do on the left.) \emph{b)} Residuals about the MAP solution. \emph{c)} The phase-folded orbital solution. In the case of Model B, panel \emph{c)} is solely the Keplerian component of the MAP solution, with the GP having been subtracted out. While both Model A and B produce a consistent, $\sim$4-$\sigma$ mass measurement for \oneEightTwo b, we emphasize that additional RVs are required to rule out scenarios of model misspecification e.g. additional planetary signals or yet unmitigated activity signatures.
  }\label{fig:k2-182_rvs_combined}
\end{figure*}

First, we modeled the RVs with a circular orbit where the only free parameter for \oneEightTwo b was the RV semi-amplitude, $K$. We also fit two global parameters, an RV zero-point offset ($\gamma_j$) and RV jitter ($\sigma_j$), to absorb instrumental systematics and RV contributions from stellar variability. $\gamma_j$ was calculated analytically to center the data about zero and held fixed during MCMC—we found that allowing it to vary made virtually no difference in the derived parameters \revision{and resulted in a nearly indistinguishable model with a slightly higher AICc ($\Delta \mathrm{AICc} < 2$)}. $\sigma_j$ was allowed to vary during the MCMC and is added in quadrature to the pointwise RV measurement errors when evaluating the likelihood function. For clarity, we will refer to this RV model as \oneEightTwoModelA. The MAP solution for \oneEightTwoModelA is shown on the left in Figure \ref{fig:k2-182_rvs_combined}. The model parameters, priors, and posterior estimates are summarized in Table \ref{tab:182_rv_model}. For \oneEightTwoModelA we imposed the prior that $K$ be strictly non-negative, since a negative RV semi-amplitude is unphysical. However, in reference to our discussion in \S\ref{sub:phot_post_est}, to ensure the prior did not bias the model towards larger values of $K$ we fit the same model without the prior and found no difference in the MAP solution or posteriors. \revision{We also tested a model that included a linear trend ($\dot{\gamma}$) in the RVs. The best-fitting trend was not convincing, though ($\dot{\gamma} = -0.07 \pm 0.06$ \mps d$^{-1}$). We also find $\Delta \mathrm{AICc} = 2$ compared to \oneEightTwoModelA, suggesting that a model including a trend is nearly indistinguishable from one without.}

Using the posteriors of $\eta_2$, $\eta_3$, and $\eta_4$ from the training on the \ktwo photometry as priors for a fit to the RVs (see \S\ref{activity:k2-182}), we produce the MAP solution shown on the right in Figure \ref{fig:k2-182_rvs_combined}. We refer to this model as \oneEightTwoModelB. \oneEightTwoModelB differs from Model A in the inclusion of the GP which introduces four additional parameters to the \radvel model. Also, in \oneEightTwoModelB we exclude the parameter for RV jitter, $\sigma_j$ (i.e. held  $\sigma_j$ fixed at 0). We did this for two reasons: First, the GP is meant to represent a more sophisticated treatment of the RV-signatures of stellar activity than $\sigma_j$. However, the RV jitter term is also meant to absorb instrumental systematics, while the GP is not, so the error bars on the data points shown on the right in Figure \ref{fig:k2-182_rvs_combined} are probably underestimated. The second reason we excluded the RV jitter parameter is because a fit that included both a GP and $\sigma_j$ resulted in a MAP value for $\sigma_j$ of more than 9 m s$^{-1}$ (which would be unexpectedly large for \keckhires) while the GP amplitude, $\eta_1$, was forced to zero. We believe this is likely a symptom of our small data set wherein an unusually large RV jitter term is able to produce a similar value for the maximum likelihood compared to the 4-parameter GP by inflating the data error bars. More RV data will help disambiguate this degeneracy.

As another check, we fit a \radvel model identical to \oneEightTwoModelB but which allowed $P$ and \transitTime\ to vary with tight Gaussian priors stemming from the posteriors listed in Table \ref{tab:star_planet_combined}. This model had a total of \emph{seven} free parameters ($P$, \transitTime, $K$, and $\eta_\mathrm{1-4}$) fit to just 12 data points. This model returned a mass measurement of $M = 14.4^{+3.4}_{-3.8}$ \mearth, which disagrees with those from \oneEightTwoModelA and B and would make \oneEightTwo b a more typical sub-Neptune for its given radius (see Figure \ref{fig:mass_radius}). However, we interpret this result as the GP sliding the Keplerian very slightly about the $P$ and \transitTime priors so that it can overfit the RVs to achieve a higher maximum likelihood value. We do not consider this a viable model and only mention it to illustrate the point (which we do not claim be novel) that stellar activity signatures and the GPs used to account for them can have significant impacts on the mass determinations of small planets. This fact may be particularly relevant for spotted K-dwarfs like \oneEightTwo. In \S\ref{discuss:spur} we explore this idea further and discuss how it might relate to other super-dense sub-Neptunes.

In summary, we explored a variety of models to explain the RVs of \oneEightTwo. We hone in on two: \oneEightTwoModelA, a circular fit with all parameters held fixed save for $K$ and $\sigma_j$, and \oneEightTwoModelB, which is the same as Model A but $\sigma_j$ has been replaced with a GP trained on the \ktwo photometry. The measured semi-amplitudes for \oneEightTwo b are consistent within 1-$\sigma$ between \oneEightTwoModelA and B. The parameters, priors, and posterior estimates for the two models are summarized in Table \ref{tab:182_rv_model}.

Moving forward we adopt the results of \oneEightTwoModelA, which has an AICc of $70$ compared to $95$ for \oneEightTwoModelB. While we believe stellar activity may be contributing a correlated signal to the RVs based on the clear modulation in the \ktwo photometry, our data set is too small to justify the inclusion of the additional GP hyperparameters. We note, however, that the RV jitter term from \oneEightTwoModelA, $\sigma_j = 4.28^{+1.42}_{-1.00}$ m s$^{-1}$, is seemingly consistent with our estimate of the RV contribution from star spots, $\sigma_\mathrm{RV,\: phot} \approx 4.2$ m s$^{-1}$, stemming from our discussion in \S\ref{rvs:activity}. This may suggest that our uncertainties on $K$ in Model A are not severely underestimated. Overall, we regard the mass measurement from \oneEightTwoModelA, quoted in Table \ref{tab:star_planet_combined}, with caution and emphasize the need for additional RV monitoring.

\subsubsection{\oneNineNine} \label{rvs:k2-199}

\begin{figure}
    \centering
    \includegraphics[width=\columnwidth]{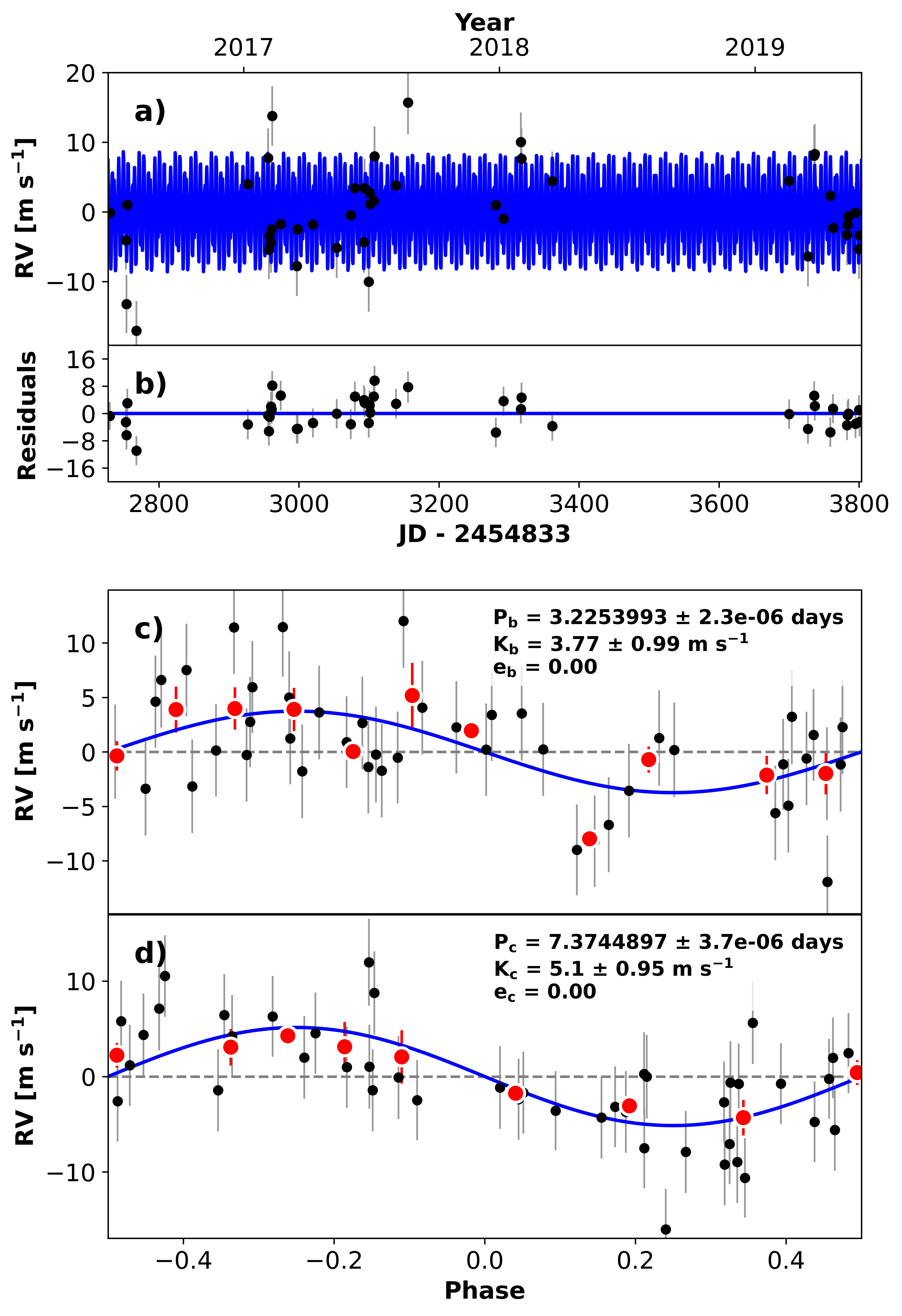}
    \caption{The best-fitting 2-planet Keplerian model for the \oneNineNine RVs. \emph{a)} The MAP solution is the blue line, with \keckhires RVs shown as the black points. Error bars represent pointwise RV measurement errors added in quadrature with an RV jitter term. \emph{b)} Residuals about the best-fitting two-planet model. \emph{c)} The phase-folded orbital solution for \oneNineNine b. Red points are data binned in units of 0.08 orbital phase. \emph{d)} The same for \oneNineNine c.}
    \label{fig:k2-199_rv}
\end{figure}

\oneNineNine b and c were recently confirmed by \nasakeyproj who modeled the \ktwo C6 photometry to derive planet ephemerides and used 33 \keckhires RVs to measure $M_\mathrm{b} = 7.8 \pm 2.2$ \mearth and $M_\mathrm{c} = 11.0^{+2.7}_{-2.9}$ \mearth for \oneEightTwo b and c, respectively. \nasakeyproj adopt a 2-planet, circular Keplerian model for the \oneNineNine RVs and suggest that extending the RV baseline will help determine the validity of a potential linear trend.

We explored several \radvel models to fit the \oneNineNine RVs, all of which were 2-planet, circular Keplerian fits. In the end we choose a model that varies $P$, \transitTime, and $K$ for each planet, fits a global RV jitter term, $\sigma_j$, and analytically calculates the RV instrumental offset, $\gamma_j$. We found that allowing $\gamma_j$ to vary during the MCMC produced near-identical results but slightly increased the model AICc \revision{($\Delta \mathrm{AICc} < 2$)}. Similar to \oneEightTwoModelA, we impose a prior to keep $K > 0$ for both planets, however a fit without the prior produced entirely similar results and did not bias the planet masses towards higher values. We elect to include the prior in our final solution because it is physically motivated.  We allowed the planet ephemerides to vary within the tight priors from the \ktwo photometry because of the time difference between the end of \ktwo C17 and the start of our new RV observations (about 10 months). We would have done the same for \oneEightTwoModelA if not for the small size of the data set. Model parameters, priors, and posterior estimates are summarized in Table \ref{tab:199_rv_model}. We improve the RV detection significance of both planets, especially \oneNineNine c, whose mass is now constrained to better than 5-$\sigma$ precision.

As we discussed in \S\ref{activity:k2-199}, we forgo a model of the \oneNineNine RVs that includes a GP trained on either the \ktwo photometry or the \keckhires \shk values because of the significant change in spot behavior between the two \ktwo campaigns and the lack of periodicity in the \shk time series. Though by eye it seems like there may be slight correlation in the residuals in Figure \ref{fig:k2-199_rv}, a periodogram of the RVs with the two Keplerians removed shows no peaks above the 0.01\% false alarm probability level.

In addition to improving the planet mass determinations, our other main contribution to this system is in extending the RV observing baseline—our observations extend the baseline by more than a year when added to the \nasakeyproj data set. We checked for a potential linear trend as mentioned by \nasakeyproj but found no evidence for one. Our final mass measurements for \oneNineNine b and c are listed in Table \ref{tab:star_planet_combined}.

\begin{figure*}[htb]
    \centering
    \includegraphics[width=\textwidth]{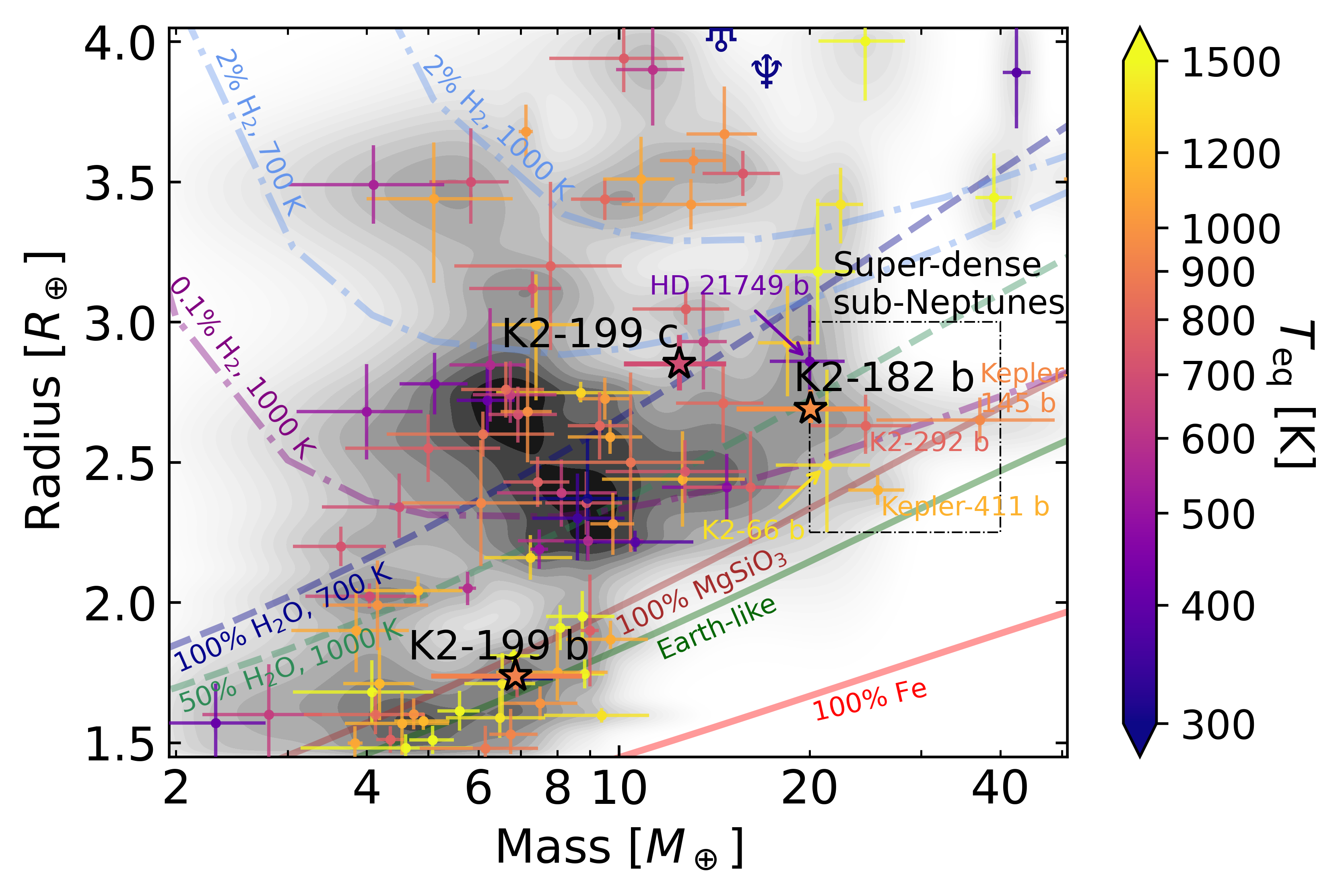}
    \caption{The mass-radius diagram in the sub-Neptune regime. Data was compiled from the NASA Exoplanet Archive on 2020 Nov 20. We only include planets with better than 15\% fractional measurement precision in radius and 33\% fractional precision in mass. We note that the sample of planets shown here may lack completeness for low-density sub-Neptunes due to our fractional precision requirements, but they should not impact completeness at high-density \citep{burt18, montet18, teske21}. Planets and their 1-$\sigma$ measurement uncertainties are colored by equilibrium temperature assuming zero Bond albedo. Underlying contours from Gaussian kernel density estimation correspond to equal levels of fractional population in the mass-radius plane. We plot various density profiles to demonstrate that the sub-Neptune regime is host to numerous degeneracies in planet bulk composition. Dash-dot curves represent planets with Earth-like rock/iron cores surrounded by either a 0.1\% or 2\% H$_2$ envelope by mass \citep{zeng19}. The green dashed curve refers to planets with Earth-like rock/iron cores surround by a 50\% layer of \water by mass \citep{zeng16}. Other composition curves are taken from \cite{zeng16}. We draw attention to a small but growing sample of ``super-dense" sub-Neptunes (\rplanet $< 3$ \rearth and \mplanet $> 20$ \mearth), of which \oneEightTwo b is a member. Others include two planets with masses from transit timing variations, Kepler-145 b \citep{xie14} and Kepler-411 b \citep{sun19}, and three radial velocity confirmations, K2-66 b \citep{sinukoff17}, K2-292 b \citep{luque19}, and \hdTwoOneSevenFourNine b \citep{gan21}. We discuss these super-dense sub-Neptunes further in \S\ref{discuss:spur}.}
    \label{fig:mass_radius}
\end{figure*}

\section{Discussion}
\label{sec:discuss}

Our discussion is broken into three parts: First, in \S\ref{discuss:comp} we place \oneEightTwo b and \oneNineNine b and c on the mass-radius diagram and compare their physical parameters to models of interior composition. In \S\ref{discuss:spur}, we use \oneEightTwo b as a launching point to take a closer look at a seemingly emerging group of super-dense sub-Neptunes (\rplanet $< 3$ \rearth, \mplanet $> 20$ \mearth). We review the literature for five of these planets to determine whether or not their high mass measurements could be explained by untreated signatures of correlated stellar variability in radial velocity data. We follow with a discussion of possible formation and evolution mechanisms for these unusually dense planets. Finally, in \S\ref{discuss:atmos} we return to \oneEightTwo b and \oneNineNine b and c to assess their viability as targets for space-based spectroscopic observations. 

\subsection{Bulk composition} \label{discuss:comp}
The growing sub-Neptune population spans a wide range of bulk densities on the mass-radius plane. Furthermore, it is located in a region where numerous theoretical models of planet interiors and volatile envelopes converge. Precise mass, radius, and instellation flux estimates are the first step in discriminating between these degenerate compositions. 

To infer the bulk compositions of \oneEightTwo b (\rhob$ = $ \oneEightTwoRho\ \gcc), \oneNineNine b (\rhob $ = $ \oneNineNineBRho\ \gcc), \oneNineNine c (\rhoc $ = $ \oneNineNineCRho\ \gcc), we first compared their locations on the mass-radius plane with bulk density profiles from \revision{\cite{zeng16, zeng19}}. Figure \ref{fig:mass_radius} shows the planets in the mass-radius diagram along with a sample of confirmed planets and various composition curves. By eye, the composition curves suggest that \oneEightTwo b may have a significant core \water mass fraction, \oneNineNine b is likely rocky, and \oneNineNine c may have a substantial ($\gtrsim 1.5-2\%$ by mass) H$_2$-dominated envelope.

\begin{figure*}
\gridline{\fig{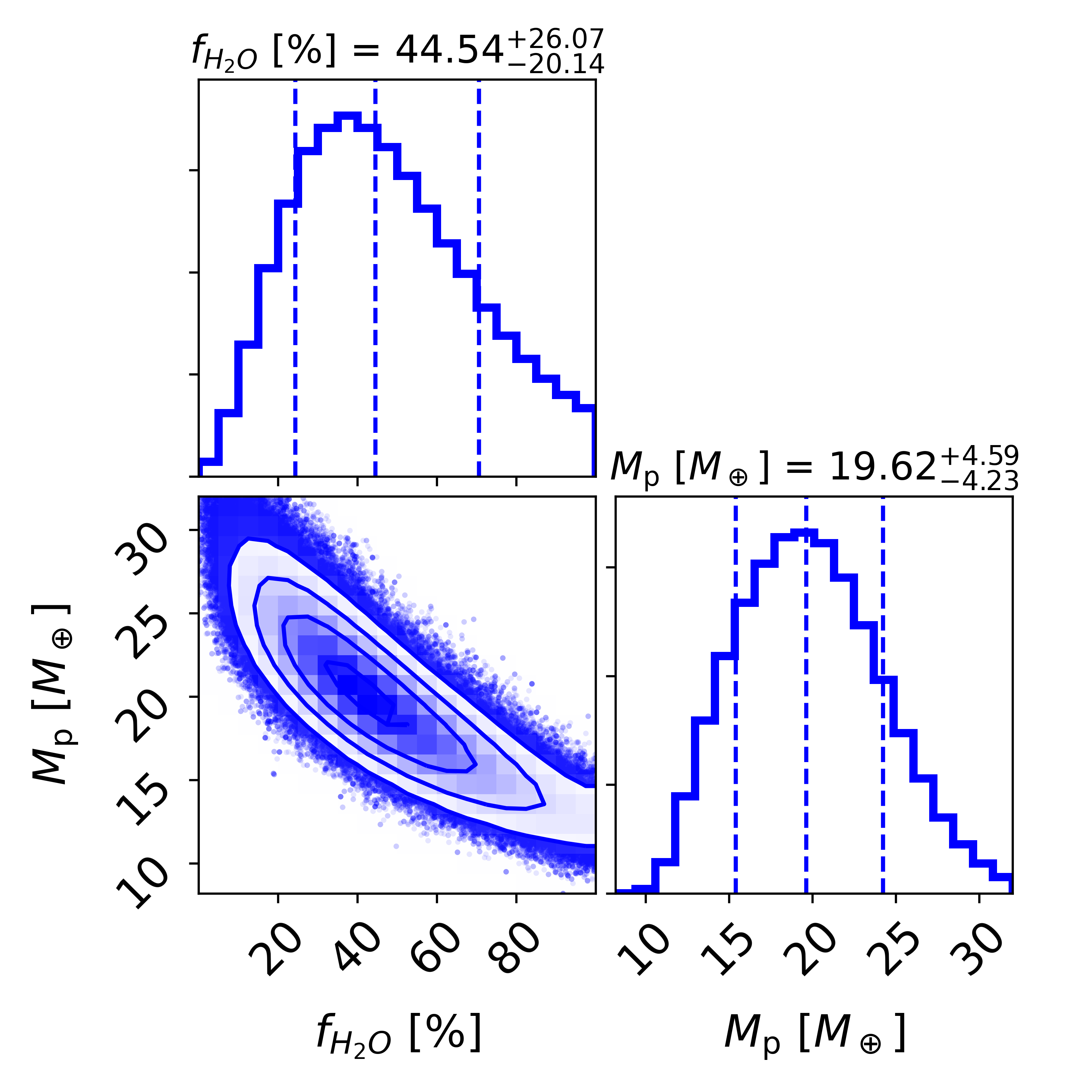}{0.5\textwidth}{(a) \oneEightTwo b}}
\gridline{\fig{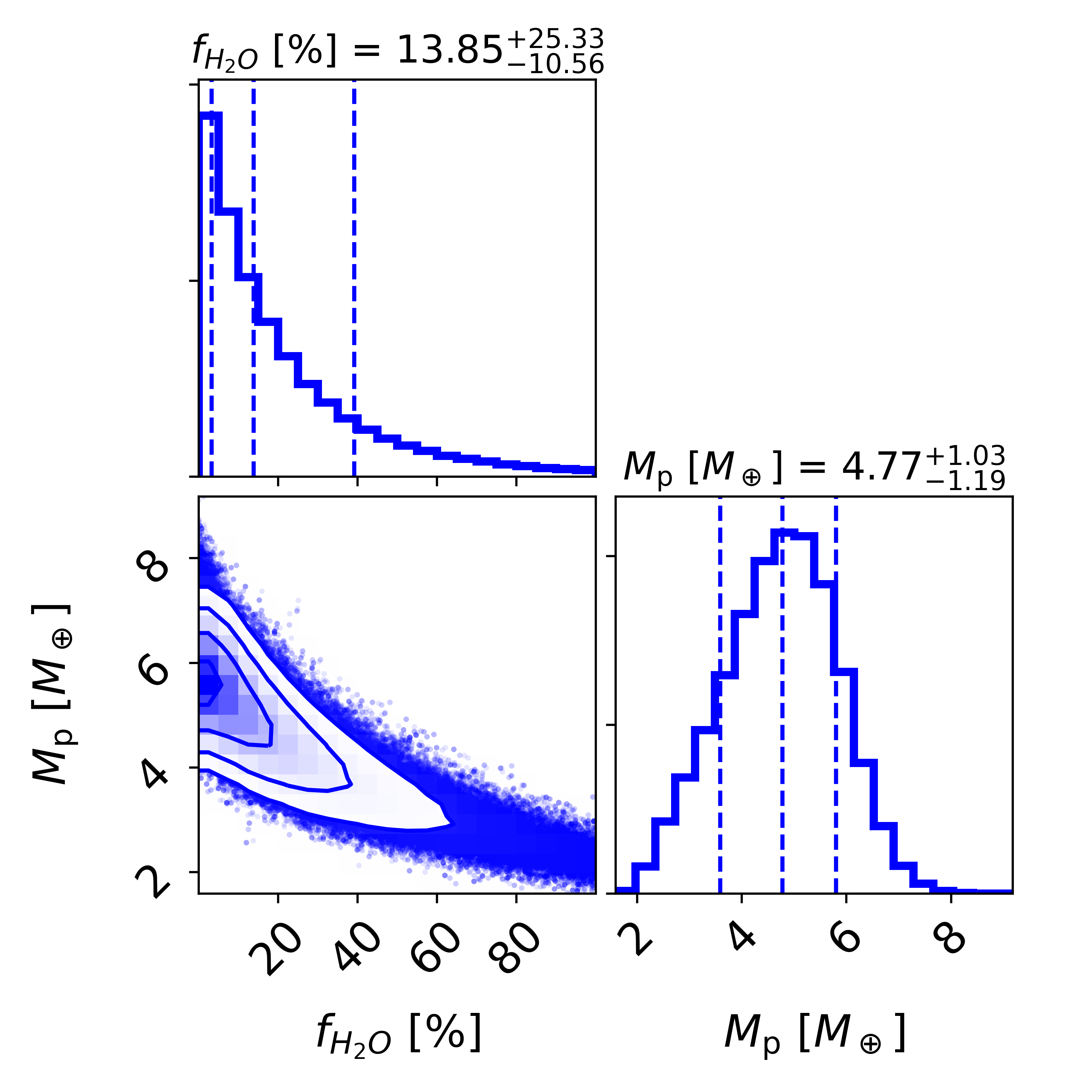}{0.5\textwidth}{(b) \oneNineNine b}
          \fig{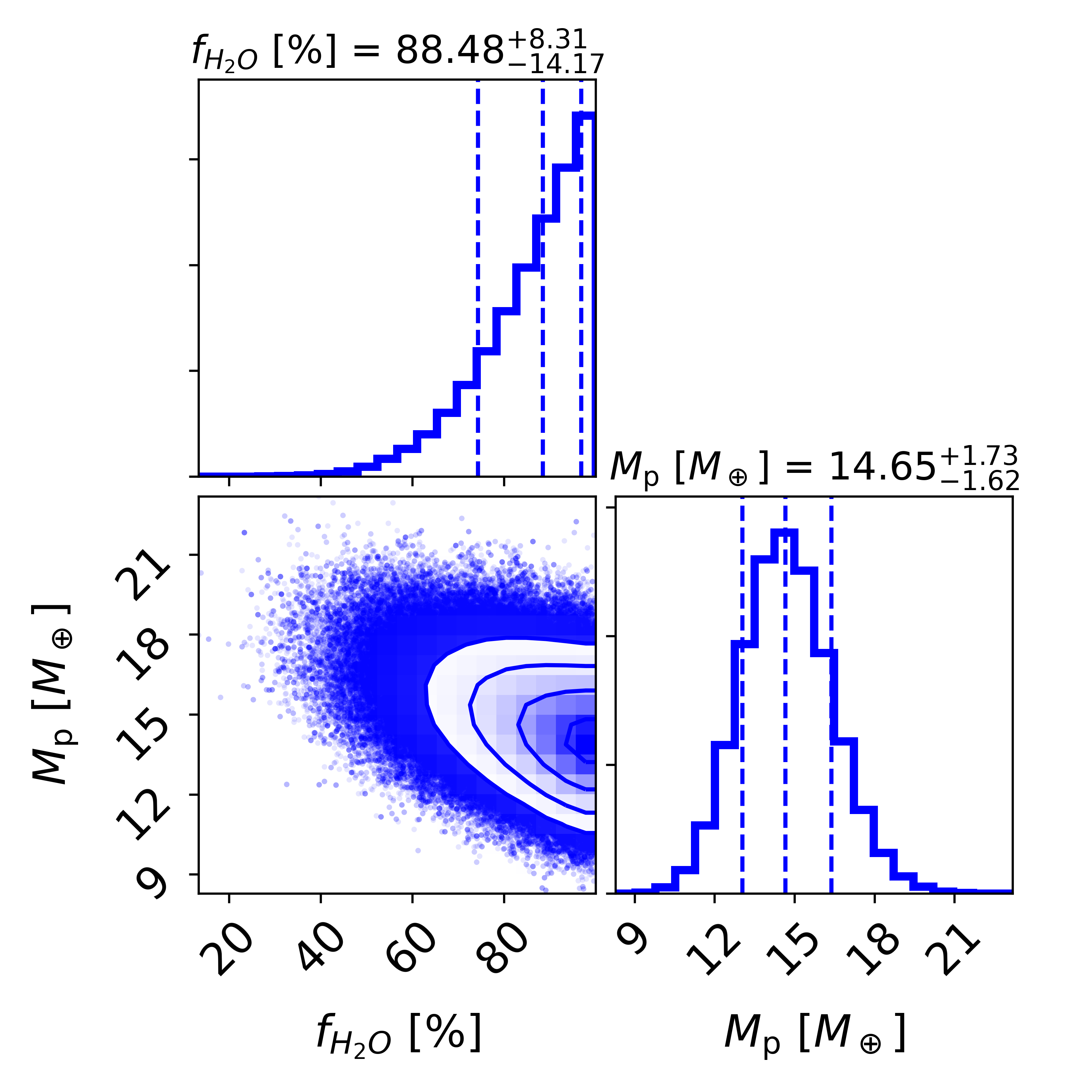}{0.5\textwidth}{(c) \oneNineNine c}}
  \caption{Joint and marginalized posteriors for our inference of \waterfrac for the three planets, assuming a bulk composition of water and rock. Posterior median values and 68\% confidence intervals are labeled at the top of each panel. Vertical dashed lines represent the median and bounds of the 68\% confidence interval. \oneEightTwo b's mass and radius are consistent with a substantial core \water mass fraction. This may have implications for its formation mechanism, as discussed at the end of \S\ref{spur:formEvo}.}\label{fig:fh2o_combined}
\end{figure*}

To make more quantitative statements about the possible bulk compositions of these planets, we compared our derived planet parameters to theoretical grids of interior composition. First, we used the grid from \cite{zeng16} to infer planet core water mass fractions (\waterfrac) assuming a two-component bulk composition of water and rock. We did this using the the Structure Model INTerpolator tool \revision{(\smint; \citealt{piaulet21})}, which performs linear interpolation on the grid of possible \waterfrac and planet mass (\mplanet) values to find the combination that best fits the measured planet radius. 

We explored the posteriors of the inferred \waterfrac and \mplanet values for each planet with \emcee. We used an informed Gaussian prior on \mplanet stemming from our results in Table \ref{tab:star_planet_combined} and a uniform prior on \waterfrac between 0 and 100\%. We ran the MCMC with 100 chains for at least 20,000 steps each, discarding the first 60\% of steps in each chain as burn-in. To ensure convergence, we continued sampling until each chain had run for at least 50 times the maximum autocorrelation time ($\tau$) across all parameters. In addition, we enforced that the maximum relative change in $\tau$ between convergence checks (every 1000 steps) was less than 1\%. Finally, we visually inspected the chains to confirm a stationary and common distribution for each parameter. For each planet, the \smint results for our interpolation on the \cite{zeng16} grid are shown in Figure \ref{fig:fh2o_combined} and summarized in Table \ref{tab:star_planet_combined}. We find that \oneEightTwo b's mass and radius is consistent with a substantial water mass fraction of \waterfrac $=$ \oneEightTwoWater\%. We infer small (\oneNineNineBWater\%) and large (\oneNineNineCWater\%) core water mass fractions for \oneNineNine b and c, respectively. 

\begin{figure*}
\gridline{\fig{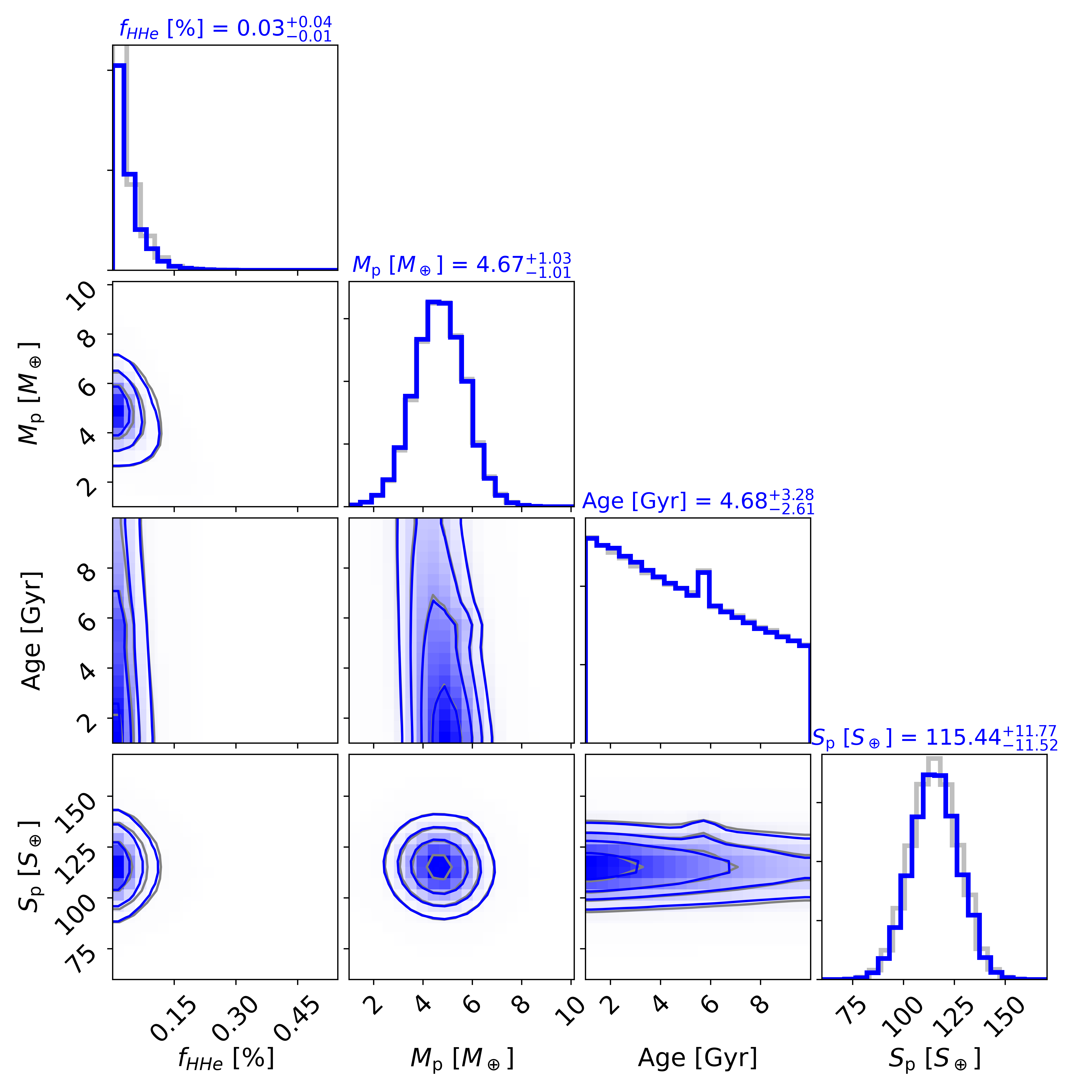}{0.5\textwidth}{(a) \oneNineNine b}
          \fig{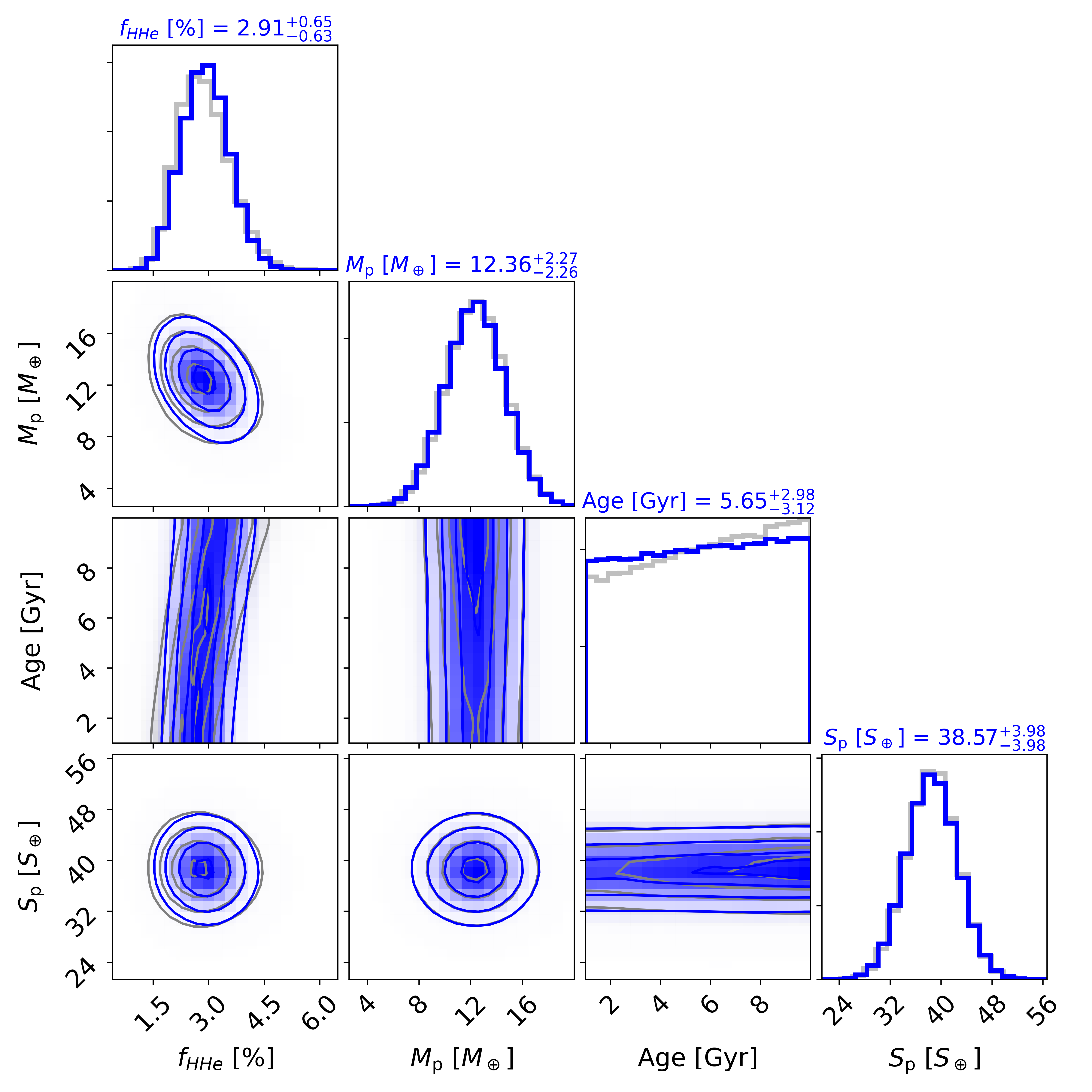}{0.5\textwidth}{(b) \oneNineNine c}}
  \caption{Joint and marginalized posteriors from our inference of \hhefrac for \oneNineNine b and c are shown in blue (gray) assuming a 1$\times$ (50$\times$) solar metallicity H$_2$/He envelope on top of an Earth-like rock/iron core. Posterior median values and 68\% confidence intervals for the $1\times$ solar metallicity case are labeled at the top of each panel. }\label{fig:fhhe_combined}
\end{figure*}

For \oneNineNine b and c we also used \smint to infer the fraction of their mass that might be contained in a H$_2$/He envelope, assuming an Earth-like core of rock/iron, according to the grids of thermal evolution from \cite{lopezforney14}. We do not infer \hhefrac for \oneEightTwo b because at 20 \mearth it lies at the edge of the grid—though this is not to say that an ice-rich core is more likely for this planet than an Earth-like one \citep{owen17}. The analysis operated in an analogous way to our interpolation on the \cite{zeng16} grid, though the parameters in the fit were now the H$_2$/He envelope mass fraction (\hhefrac), \mplanet, system age, and planet instellation flux (\sincplanet). \smint interpolated on the \cite{lopezforney14} grids to find the values of these parameters that best matched the planet radius. The posterior estimation was similar to our method for \waterfrac. We used Gaussian priors on \mplanet and \sincplanet for each planet using the values in Table \ref{tab:star_planet_combined}. We placed uniform priors on \hhefrac from 0.1 to 20\% (the bounds of the \citealt{lopezforney14} grid) and on the system age from 1 to 10 Gyr. Our results for \oneNineNine b and c are shown in Figure \ref{fig:fhhe_combined} and summarized in Table \ref{tab:star_planet_combined}. \oneNineNine b's mass, radius, and instellation flux are consistent with a very small H$_2$/He envelope (\hhefrac $\leq 0.15$\% at 3-$\sigma$ confidence), potentially indicative of photoevaporation. \oneNineNine c's physical parameters are consistent with a larger H$_2$/He envelope mass fraction, \hhefrac $=$ \oneNineNineCFhhe\%.

\revision{Figure \ref{fig:mass_radius} also shows models of bulk interior composition from \cite{zeng19} corresponding to Earth-like rock/iron cores surrounded by an H$_2$-dominated envelope. We show the \cite{zeng19} models in Figure \ref{fig:mass_radius} for consistency with the \cite{zeng16} models. However, we choose to use the \cite{lopezforney14} grid to infer \hhefrac rather than \cite{zeng19} because the former results from a study of thermal evolution, allowing us to marginalize over the age of the system.}

\subsection{Super-dense sub-Neptunes} \label{discuss:spur}
\oneEightTwo b's mass measurement makes it one of the densest sub-Neptunes known to date. However, RV models of small planets around spotted stars can be greatly influenced by the in- or exclusion of GPs meant to account for stellar activity \citep{rajpaul15, faria16, dumusque17, jones17, rajpaul17}. While we found that our mass measurement for \oneEightTwo b was consistent between RV models with and without a GP trained on the \ktwo photometry, we suggest that additional monitoring will better inform the influence of stellar activity on the spectroscopic observations. Furthermore, cases where star spots contribute to planet RV amplitude may be subject to publication bias because the inflated signal can more easily overcome standard fractional precision thresholds, \revision{like \mplanet$/\sigma_{M_\mathrm{p}} \ge 5$ \citep{burt18, montet18, batalha19, teske21}}.

Here we review the literature for the handful of other ``super-dense" sub-Neptunes (\rplanet $< 3$ \rearth, \mplanet $> 20$ \mearth) and question whether or not their high masses can be explained by unmitigated stellar activity. For a summary and general takeaways, see \S\ref{spur:upshot}.

\subsubsection{Sample and caveats}

From our sample of confirmed planets shown in Figure \ref{fig:mass_radius}, there are five sub-Neptunes other than \oneEightTwo b that we consider ``super-dense": Kepler-145 b \citep{xie14}, Kepler-411 b \citep{sun19}, K2-66 b \citep{sinukoff17}, K2-292 b \citep{luque19}, and \hdTwoOneSevenFourNine b (GJ 143 b; TOI-186.01; \citealt{trifonov19, dragomir19, gan21}). The masses of Kepler-145 b and Kepler-411 b were measured using transit timing variations (TTVs) and the remaining three are radial velocity confirmations. While our RV/stellar activity discussion does not apply to TTV measurements, we briefly summarize the physical parameters of Kepler-145 b and Kepler-411 b to add context to the super-dense sub-Neptunes as a whole. 

For the purpose of our discussion, we chose to highlight these five planets because they are similar in mass and radius to \oneEightTwo b. We acknowledge that they may not comprise a complete sample and the \rplanet $< 3$ \rearth, \mplanet $> 20$ \mearth limits are a bit arbitrary. \revision{For example, we do not discuss the ultra-short period (USP) exposed planetary core orbiting the late-G dwarf, TOI-849 ($P = 0.76$ days, \rplanet $= 3.45$ \rearth, \mplanet $= 40.8$ \mearth, $\rho_\mathrm{p} = 5.5$ \gcc; \citealt{armstrong20})—stellar activity signals due to star spots (timescales of several to tens of days for F, G, and K dwarfs) are unlikely to confuse the mass measurements of USPs ($P < 1$ day), so we do not expect the RV mass of TOI-849 b to be biased due to stellar activity.} Similarly, we do not include K2-110 b ($P = 13.9$ days, \rplanet $= 2.59$ \rearth, \mplanet $= 16.7$ \mearth, $\rho = 5.2$ \gcc), a dense sub-Neptune around a metal-poor ([Fe/H] $= -0.34 \pm 0.03$ dex) K3V dwarf \citep{osborn17}. The K2-110 b authors note spot variations on the order of weeks are seen in the \ktwo light curve, meaning that the stellar rotation period (or one of its harmonics) could be in the neighborhood of the planet's longer orbital period. Their $v\sin i$ measurement implies $\mathrm{Min}(P_\mathrm{rot}) > 9.2$ days, so further investigation may be warranted.

\subsubsection{Kepler-145 b}

Kepler-145 (\teff $= 6110 \pm 122$ K, \citealt{berger18}) hosts two transiting planets confirmed by \cite{xie14}. Kepler-145 b ($P = 22.9$ days) has \rplanet $= 2.65 \pm 0.08$ \rearth, \mplanet $= 37.1 \pm 11.6$ \mearth, and $\rho = 10.9 \pm 3.6$ \gcc, potentially making it the densest of all the super-dense sub-Neptunes we consider. However, when \cite{otegi20a} curated a high-fidelity sample of exoplanet mass and radius measurements from the NASA Exoplanet Archive, they excluded all planets from \cite{xie14} because the masses disagree significantly with those from \cite{haddenLithwick14, haddenLithwick17}. Furthermore, Kepler-145 b was not included in the broad sample of planets with secure TTV signals in either \cite{haddenLithwick14} or \cite{haddenLithwick17}, making its mass even more suspect. 

\subsubsection{Kepler-411 b}

Kepler-411 is an active K2V dwarf hosting four planets whose masses were measured by \cite{sun19} with TTVs. Kepler-411 b ($P = 3.0$ days) is a hot ($T_\mathrm{eq} = 1138$ K) sub-Neptune with \rplanet $=2.40 \pm 0.05$ \rearth, \mplanet $= 25.6 \pm 2.6$ \mearth, and $\rho = 10.3 \pm 1.3$ \gcc. For reference, \cite{otegi20a} do include masses from \cite{sun19} in their revised exoplanet mass and radius catalog.

\subsubsection{K2-66 b}

K2-66 (EPIC 206153219) is a relatively quiet ($\log R'_\mathrm{HK}$ $= -5.27$ dex) G1 subgiant (\mstar $= 1.11$ \msun, \rstar $= 1.67$ \rsun) with a transiting planet in the hot sub-Neptune desert (K2-66 b; $P = 5.0$ days, \rplanet $= 2.49^{+0.34}_{-0.24}$ \rearth, $T_\mathrm{eq} = 1372 \pm 51$ K; \citealt{sinukoff17}; \citealt{lundkvist16}). \cite{sinukoff17} confirm the planetary nature of K2-66 b using 38 precision RV measurements from \keckhires, finding the planet is \mplanet $= 21.3 \pm 3.6$ \mearth and $\rho = 7.8 \pm 2.7$ \gcc. 

After extracting the \ktwo C3 \everest light curve in a similar way to our procedure in \S\ref{phot:lightcurve}, we find that K2-66 shows long-term spot evolution ($\sim 50$ days) but no obvious rotation signatures closer to the time scale of K2-66 b's orbital period. Using our adaptation of Equation 2 from \cite{vanderburg16} as in \S\ref{rvs:activity}, we find that $\sigma_\mathrm{phot} = 0.00075$ in units of relative flux (about an order of magnitude smaller than for \oneEightTwo and \oneNineNine). Combined with $v\sin i = 3.7$ km s$^{-1}$ from the confirmation paper, $\sigma_\mathrm{RV,\: spot} \approx 2.8$ \mps for K2-66 (a little more than 0.5$\times$ our upper limits on $\sigma_\mathrm{RV,\: spot}$ for \oneEightTwo and \oneNineNine).

\cite{sinukoff17} use a variety of \radvel models to explain their RV measurements, settling on a 1-planet circular Keplerian fit. They do not find any significant peaks in a periodogram of the RV residuals after removing the signal for K2-66 b. With the star's low magnetic activity and without clear signs of a stellar rotation period in the \ktwo C3 light curve, it seems that including a GP in a model of the RVs would not greatly affect the mass measurement. \keckhires RV monitoring for this system in ongoing (Howard et al. in prep).

\subsubsection{K2-292 b}
\cite{luque19} confirm K2-292 b ($P = 17.0$ days) using 18 precision RVs from the Calar Alto high-Resolution search for M dwarfs with Exoearths with Near-infrared and optical Échelle Spectrographs instrument (CARMENES; \citealt{carmenes14, carmenes18}) on the 3.5-m telescope at the Calar Alto Observatory in Spain. K2-292 (EPIC 212628254, HD 119130) is a G3V dwarf \citep{houk99} observed in \ktwo C17. Through a joint analysis of the \ktwo photometry and CARMENES RVs, \cite{luque19} find \rplanet $=2.63^{+0.11}_{-0.10}$ \rearth, \mplanet $= 24.5 \pm 4.4$ \mearth, and $\rho = 7.4^{+1.6}_{-1.5}$ \gcc. They also detect a linear trend in the RVs that could represent an outer companion with a minimum mass of $\sim 33$ \mearth. The authors suggest that additional data is required to rule out the possibilities that the trend is caused by stellar rotation or instrumental systematics. 

The \ktwo C17 \everest light curve does not show significant spot modulation like those of \oneEightTwo and \oneNineNine. Repeating the process we carried out for K2-66, we find that $\sigma_\mathrm{phot} = 0.00079$ in units of relative flux. The confirmation paper finds $v\sin i = 4.6$ km s$^{-1}$, yielding $\sigma_\mathrm{RV,\: spot} \approx 3.6$ \mps. \cite{luque19} check for periodic signals in various stellar indices measured simultaneously with their RVs but find no evidence for correlated stellar activity, save for a peak in the H-$\alpha$ indices and the cross-correlation function FWHM at $P \sim 3.5$ days. Though additional RV observations could help constrain the nature of the potentially non-transiting outer companion, without clear evidence of spot modulation from the \ktwo photometry it does not seem like K2-292 warrants a GP-based analysis of its RVs.

\subsubsection{\hdTwoOneSevenFourNine b (GJ 143 b; TOI-186.01)}
\hdTwoOneSevenFourNine is a bright ($V = 8.1$ mag) K4.5 dwarf. With multi-sector \tess photometry, \citet[hereafter \dragomir]{dragomir19} discovered \hdTwoOneSevenFourNine hosts a sub-Neptune (HD 21749 b; $P = 35.6$ days, \rplanet $= 2.61^{+0.17}_{-0.16}$ \rearth) and an Earth-size planet (HD 21749 c; $P = 7.8$ days, \rplanet $= 0.892^{+0.064}_{-0.058}$ \rearth). 

Previously, \cite{trifonov19} had identified \hdTwoOneSevenFourNine b as a single-transit planet candidate from the \tess Sector 1 and 2 photometry, and used a total of 58 publicly-available archival RVs from the High-Accuracy Radial-velocity Planet Searcher (HARPS; \citealt{mayor03}) instrument, mounted on the European Southern Observatory 3.6-m telescope at La Silla Observatory in Chile, to constrain the orbital period and measure the mass of planet b. Using the HARPS data, \cite{trifonov19} recover the orbital period and measure \mplanet $= 30.63^{+2.63}_{-2.67}$ \mearth by modeling the RVs with a moderately eccentric Keplerian orbit ($e = 0.325^{+0.079}_{-0.079}$). They do not include \hdTwoOneSevenFourNine c in a model of the RVs.

A few months after \cite{trifonov19} was published, \dragomir combined the HARPS RVs with archival ($N = 48$) and newly-acquired ($N = 34$) velocities from the Planet Finder Spectrograph (PFS; \citealt{crane10}) on the 6.5-m Magellan II Telescope at Las Campanas Observatory in Chile. This was in addition to two more sectors of \tess photometry which confirmed the orbital period of planet b at $P = 35.6$ days. \dragomir jointly model the \tess photometry with the HARPS and PFS RVs, finding \hdTwoOneSevenFourNine b is \mplanet $= 22.7^{+2.2}_{-1.9}$ \mearth and $\rho = 7.0^{+1.6}_{-1.3}$ \gcc, and placing an upper limit on the mass of planet c. The \cite{trifonov19} mass is in near-3-$\sigma$ disagreement with the measurement from \dragomir. The \dragomir orbital solution also employs slightly smaller eccentricity for planet b, with $e = 0.188^{+0.076}_{-0.078}$, and prefers a linear trend over the span of the $\sim15$-year RV baseline. In the following, we discuss the results and methods from \dragomir rather than \cite{trifonov19} because the former was able to confirm the orbital period of planet b with \tess photometry before fitting their (larger) RV time series.

The \emph{TESS} photometry of \hdTwoOneSevenFourNine shows clear spot modulation with a peak-to-peak amplitude of about four parts per thousand in relative flux. \dragomir use stellar activity indices derived from their HARPS and PFS spectra along with long-term photometric monitoring from the Kilodegree Extremely Little Telescope (KELT; \citealt{kelt04}) to determine that the stellar rotation period is around 37 to 39 days, just longward of the period for planet b. Notably, while periodograms of the activity indices and photometry have peaks in this range, the same peaks do not manifest themselves in a periodogram of the RVs above the 0.01\% FAP level (\dragomir Figure 5). The authors calculate $\sigma_\mathrm{RV,\: spot} \approx 1.3$ \mps and suggest that uncertainties for their mass measurement may be slightly underestimated.

\hdTwoOneSevenFourNine represents an interesting case where the stellar rotation period is likely very close (delta of a $\sim$few days) to the orbital period of the super-dense sub-Neptune. Though the rotation signals from the activity indicators and KELT photometry do not appear in the RV periodogram, a model of the RVs that includes a GP trained on the complementary data would provide a useful sanity check for the mass measurement of \hdTwoOneSevenFourNine b. As a bright system with ample archival photometry and activity indicator measurements, \hdTwoOneSevenFourNine is particularly amenable to a GP-based analysis.

\revision{Recently, \citet[hereafter \gan]{gan21} performed a more detailed accounting of stellar activity when they re-analyzed the \hdTwoOneSevenFourNine system using the multi-sector \tess photometry and the HARPS $+$ PFS velocities from \dragomir, as well as 147 additional PFS observations. Using an updated \tess dilution factor, \gan find the radii of the two planets are slightly larger than reported in \dragomir. They also use a (slightly eccentric, $e_\mathrm{b} = 0.164^{+0.062}_{-0.058}$) Keplerian $+$ GP model of the RVs to account for stellar rotation (the periodic hyperparameter of their GP kernel finds $P_\mathrm{rot} = 34.1^{+2.4}_{-2.7}$ days). The resulting parameters they report for \hdTwoOneSevenFourNine b are \rplanet $= 2.86 \pm 0.20$ \rearth, \mplanet $= 20.0 \pm 2.7$ \mearth, and $\rho = 4.7^{+1.9}_{-1.4}$ \gcc. The mass measurements for \hdTwoOneSevenFourNine b from \dragomir and \gan overlap within 1-$\sigma$, and the more intermediate bulk density of the planet owes primarily to the increase in its radius measurement—had the radius remained the same, when combined with the \mplanet $= 20$ \mearth measurement from \gan the bulk density would be $\sim$6.2 \gcc.}

\revision{Comparing the \dragomir and \gan results, it does not appear that a stellar activity signal in the RV data set (though present) greatly affected the mass measurement of \hdTwoOneSevenFourNine b in the \dragomir analysis.} To conduct our own test of the hypothesis that unmitigated stellar activity may be artificially inflating the mass measurements of the super-dense sub-Neptunes, we re-analyzed the RVs from \dragomir with our own \radvel models. The details of the analysis are described in Appendix \ref{app:hd21749_rv_modeling}, but the GP training process largely follows the methodology in \S\ref{rvs:activity}. \revision{For the sake of comparison with the Keplerian-only model of the RVs from \dragomir, we restricted our analysis to use only the velocities included in \dragomir rather than use the larger data set from \gan.}

We compared a variety of models of the \hdTwoOneSevenFourNine RVs using the AICc, finding strong preference for models that included a GP component over ones without (generally $\Delta$AICc $> 30$). Figure \ref{fig:hd21749_rv_combined} compares a model of the RVs that attempts to replicate the solution in \dragomir (\hdTwoOneSevenFourNine RV Model A) and one that differs only in the inclusion of a GP trained on the HARPS and PFS H-$\alpha$ indices (\hdTwoOneSevenFourNine RV Model B). The \hdTwoOneSevenFourNine b mass measurement from \hdTwoOneSevenFourNine RV Model B is in good agreement with the one from Model A, and both agree with the results from \dragomir, further ameliorating concerns that the measurement is artificially inflated due to unmitigated stellar activity. \revision{Our mass measurements, though slightly larger, also overlap within 1-$\sigma$ with the result from \gan. While not explicitly the same, the quasi-periodic GP kernel we employ is similar in effect to the kernel used in \gan (both represent a combination of exponentially-decaying and periodic signals).}

\subsubsection{Literature review summary} \label{spur:upshot}
We briefly reviewed the confirmation papers for five super-dense sub-Neptunes, Kepler-145 b, Kepler-411 b, K2-66 b, K2-292 b, and HD 21749 b, to determine if unmitigated stellar activity signatures in RVs could have inflated the mass measurements of these planets. Though our activity discussion does not apply to TTV measurements, based on the concerns raised by \cite{otegi20a} Kepler-145 b's TTV mass measurement from \cite{xie14} should be handled with caution—we do not include it in the discussion below. Kepler-411 is an active K2V dwarf but masses in the system were also measured with TTVs so, again, our RV/activity discussion does not apply. \everest light curves do not show clear signs of spot modulation for the G1 subgiant K2-66 or the solar-like G3V dwarf K2-292. These systems do not seem to warrant an RV model that includes a GP. 

\hdTwoOneSevenFourNine is a spotted K4.5 dwarf with a rotation period that is likely close to the orbital period of planet b. \dragomir offer a thorough investigation into possible signatures of stellar activity in the RVs, finding little sign of the stellar rotation period in their extensive ($N = 141$) RV time series. \revision{\gan provide a more detailed accounting of stellar activity in the \hdTwoOneSevenFourNine RVs, adding 147 PFS observations to the \dragomir data set and including a GP in a model of the velocities. Their mass measurement of \hdTwoOneSevenFourNine b, though slightly smaller, overlaps with the result from \dragomir within 1-$\sigma$. The more intermediate density \gan report for planet b is primarily due to their larger planet radius measurement, which results from an updated \tess dilution factor. As another sanity check, we modeled the \hdTwoOneSevenFourNine RVs as a combination of Keplerians and a GP, restricting ourselves to the \dragomir data set to create a more direct comparison between their Keplerian-only model.} We find that while a model of the RVs which includes a GP trained on activity indicators is strongly preferred by the AICc and produces smaller scatter in the residuals (see Figure \ref{fig:hd21749_rv_combined}), its mass measurement for \hdTwoOneSevenFourNine b is entirely consistent with a model that does not include the GP (and both models are consistent with the \dragomir and \gan results). \revision{Though the mass measurement of \hdTwoOneSevenFourNine b from \gan is slightly smaller than reported in \dragomir and this work, it does not seem to indicate that untreated stellar activity was greatly impacting the planet mass measurement.} Having found no compelling evidence that stellar activity is producing inflated RV mass measurements for these planets, we are confident that this is a growing, bonafide population of super-dense sub-Neptunes. 

\subsubsection{Super-dense sub-Neptune formation and evolution} \label{spur:formEvo}

Several explanations have been posited as the formation and/or evolution mechanism(s) behind the super-dense sub-Neptunes. These include photoevaporation, migration, and giant impacts. Here we briefly discuss how these scenarios relate to our sample and revisit \oneEightTwo b.

In the case of the highly-irradiated K2-66 b ($S_\mathrm{p} \approx 840$ $S_\oplus$), the planet was probably stripped of its primordial atmosphere through photoevaporation as K2-66 evolved off of the main sequence. With $S_\mathrm{p} \approx 220$ $S_\oplus$, it seems that photoevaporation could also have a hand in the density of Kepler-411 b. However the system is relatively young with a gyrochronological age of $212 \pm 31$ Myr \citep{sun19}. Using Equation 15 from \cite{lecavelier07}, we find that with $a = 0.038$ AU, Kepler-411 b would only lose 0.01 \mearth due to extreme ultraviolet radiation (EUV) from its K2V host over the system's gyrochronological age. \cite{otegi20a} suggest that Kepler-411 b could be ice-rich and represent the maximum allowed core mass, in line with models of Saturn and Jupiter's interiors, which infer core masses up to 20 and 25 \mearth, respectively \citep{saumonGuillot04, wahl17}.

\cite{luque19} show that K2-292 b receives too little instellation ($S_\mathrm{p} \approx 67$ $S_\oplus$) for its density to be explained by EUV-driven photoevaporation. Instead, following an explanation similar to the one for K2-110 b \citep{osborn17}, \cite{luque19} propose that K2-292 b was formed at high density. In situ formation is ruled out, as the required disk mass enhancement factor to the minimum-mass solar nebula (MMSN) is greater than 40 \citep{schlichting14}. Rather, \cite{luque19} suggest K2-292 b could have formed beyond the snow line and migrated inward due to Kozai-Lidov oscillations \citep{kozai62, lidov62, dawsonChiang14, mustill17} excited by an outer companion (possibly the linear trend they detect in the RVs).

\hdTwoOneSevenFourNine b is the coolest super-dense sub-Neptune ($T_\mathrm{eq} = 407^{+22}_{-19}$ K from \gan) meaning that photoevaporation is not a viable explanation. Given its moderate orbital eccentricity ($e = 0.164^{+0.062}_{-0.058}$ from \gan), the presence of the close-in Earth-size \hdTwoOneSevenFourNine c, and the possibility of an outer companion due to the linear trend in the \dragomir RVs \revision{(though no trend is found in the \gan data set),} \hdTwoOneSevenFourNine b could have also formed beyond the snow line and migrated to its present location ($a = 0.21$ AU from \gan) due to dynamical interactions.

Returning to \oneEightTwo b, we first tried to place an upper limit on its mass-loss rate due to EUV radiation. Using Equation 15 from \cite{lecavelier07}, we find \oneEightTwo b could lose up to 0.5 \mearth over 10 Gyr from EUV instellation, i.e. $2.5$\% of its current mass. According to the grids from \cite{lopezforney14}, if a planet with \oneEightTwo b's mass (20 \mearth) and radius (2.7 \rearth) and $S_\mathrm{p} = 10$ $S_\oplus$ (instead of \oneEightTwo b's $\approx 150$ $S_\oplus$) were to lose all of a primordial 2\% H$_2$/He envelope over 10 Gyr, this would correspond to a reduction in radius of $\sim 0.8$ \rearth. The radius reduction would be even more dramatic in the case of maximum EUV-driven mass-loss at \oneEightTwo b's actual instellation. It seems suspect that \oneEightTwo ($L \approx 0.4$ $L_\odot$) could strip away 1 \rearth from \oneEightTwo b, so while photoevaporation may play a role in \oneEightTwo b's density, it does not appear to be the dominating factor.

To rule out formation in situ, using Equation 7 from \cite{schlichting14} we calculated the required disk mass enhancement factor ($\mathcal{F}$) to the MMSN needed in order to form our 20 \mearth planet at 0.05 AU. We find $\mathcal{F} > 100$, implying that like K2-292 b, \oneEightTwo b must have migrated to its current location from farther out ($> 2$ AU) in the protoplanetary disk. However, the delivery of 50-100 \mearth of rocky material to the inner disk combined with the constructive collisions of primordial icy cores could potentially provide an out for the in situ formation scenario \citep{hansenMurray12, inamdar15, zeng19}. In situ formation via giant impacts is not viable for typical sub-Neptunes with \hhefrac of a few percent because the impacts drive atmospheric mass-loss, but it could help explain the small envelope mass for planets like \oneEightTwo b. Furthermore, our results from interpolation on the \cite{zeng16} grids show that \oneEightTwo b's mass and radius are consistent with a substantial core \water mass fraction (see Figure \ref{fig:fh2o_combined}). 

Similar to Kepler-411 b, if \oneEightTwo b is the product of constructive collisions of icy cores, then it could represent the maximum allowed core mass. \cite{otegi20a} point out that models of giant planet formation via pebble accretion and migration predict pebble isolation masses on the order of 10 to 20 \mearth \citep{andersMichiel17, bitsch19}, which is seemingly consistent with these $\sim$20-25 \mearth super-dense sub-Neptunes. Therefore some combination of pebble accretion, giant impacts, and migration caused by interactions with a gaseous disk could explain \oneEightTwo b's formation.

Additional RV observations will shed light on \oneEightTwo b's dynamical history, helping to distinguish between the scenarios we have already mentioned and migration due to an interaction with an outer companion. RV follow-up could do this by checking for a linear trend in the velocities and/or better-constraining the orbital eccentricity, which is nominally small but slightly skewed (perhaps in-part due to Lucy-Sweeney bias) towards moderate values as measured from the \ktwo transits ($e = $ \oneEightTwoEcc).

In summary, save for K2-66 b, photoevaporation is unable to explain the super-dense sub-Neptune population—this should be unsurprising, given that the Fulton gap \citep{fulton17} does not extend to \rplanet $\gtrsim 2$ \rearth. Migration probably played a roll in the formation and evolution of Kepler-411 b, K2-292 b, \hdTwoOneSevenFourNine b, and \oneEightTwo b. Extending \oneEightTwo b's RV baseline to check for an outer companion and moderate eccentricity could help distinguish between scenarios of in situ formation, pebble accretion plus migration due to planet-disk interactions, and migration from a dynamical interaction.

\subsection{Prospects for the atmospheric characterization of \oneEightTwo b and \oneNineNine b and c} \label{discuss:atmos}

For sub-Neptunes that lie between the Earth-like and pure \water composition curves (e.g. the super-dense sub-Neptunes), improvements to mass and radius measurement precision alone cannot break degeneracies between various planet core, mantle, water, and envelope mass fractions \citep{valencia07, otegi20b}. Instead, measurements of atmospheric metallicity via space-based transmission spectroscopy are needed to constrain models of interior structure. While massive planets with small volatile envelopes are unattractive targets for these studies because of their meager atmospheric scale heights, a better understanding of their composition could inform important questions in planet formation \citep{kite20}.

We used the transmission spectroscopy metric (TSM; \citealt{kempton18}) to quantify the expected SNR of a 10-hour observing program with \jwst-NIRISS for \oneEightTwo b, assuming a cloud-free, solar metallicity atmosphere:
\begin{equation} \label{eq:tsm}
\centering
\mathrm{TSM} = (\textrm{scale factor}) \times \frac{R_\mathrm{p}^3   T_\mathrm{eq}}{M_\mathrm{p}  R_*^2} \times 10^{-m_J/5}.
\end{equation}
The ``scale factor" is a dimensionless normalization constant equal to 1.26 for planets $1.5 < R_p < 2.75$ \rearth (i.e. \oneEightTwo b and \oneNineNine b) and 1.28 for planets $2.75 < R_p < 4.0$ \rearth (i.e. \oneNineNine c).

\oneEightTwo b's TSM (\oneEightTwoTSM) is very low—\cite{kempton18} propose a cutoff of TSM $> 92$ for planets between $1.5 < R_p < 2.75$ \rearth to warrant space-based follow-up. \oneEightTwo b's TSM scales to an expected single-transit SNR of \oneEightTwoSingleTransitSNR. \revision{\hdTwoOneSevenFourNine b might serve as the best candidate for atmospheric observations of a super-dense sub-Neptune (see \S5.1 in \gan), though at $J = 6.0$ mag \hdTwoOneSevenFourNine is too bright for observations with \jwst-NIRISS.}

On the other hand, \oneNineNine is better-suited for atmospheric studies given \oneNineNine c's intermediate density ($\rho = $ \oneNineNineCRho\ \gcc) and 5-$\sigma$-precision mass measurement \citep{batalha19}. Using Equation \ref{eq:tsm} we find \oneNineNine c has a TSM value of \oneNineNineCTSM, which scales to a single-transit SNR of \oneNineNineCSingleTransitSNR. \oneNineNine b's TSM value is low (\oneNineNineBTSM; single-transit SNR of \oneNineNineBSingleTransitSNR), though as a multi-planet system, atmospheric observations of \oneNineNine b and c would be especially valuable to test theories of planet formation.

\section{Conclusion}
\label{sec:conclusion}

Using multi-campaign \ktwo photometry and precise RV measurements from \keckhires, we measured the physical parameters of \oneEightTwo b ($P = 4.7$ days, \rplanet $=$ \oneEightTwoRad\ \rearth, \mplanet $=$ \oneEightTwoMass\ \mearth), \oneNineNine b ($P = 3.2$ days, \rplanet $=$ \oneNineNineBRad\ \rearth, \mplanet $=$ \oneNineNineBMass\ \mearth), and \oneNineNine c ($P = 7.4$ days, \rplanet $=$ \oneNineNineCRad\ \rearth, \mplanet $=$ \oneNineNineCMass\ \mearth). In \S\ref{sec:phot} we extracted the \everest light curves and modeled the two campaigns of \ktwo photometry for each system. Importantly, this resulted in significant improvements to the orbital ephemerides (Figure \ref{fig:t0_errors}). In \S\ref{sec:stellar} we characterized \oneEightTwo (an early-K dwarf) and \oneNineNine (a K5V dwarf) with high-resolution spectroscopy and imaging. In \S\ref{sec:rvs} we reported our RV observations, searched for signatures of stellar activity, and measured the planet masses. Our adopted RV solution for \oneEightTwo is shown on the left in Figure \ref{fig:k2-182_rvs_combined}. Our RV solution for \oneNineNine is shown in Figure \ref{fig:k2-199_rv}. Measured and derived planet parameters are summarized in Table \ref{tab:star_planet_combined}. We find that \oneEightTwo b may be a ``super-dense" sub-Neptune, with \rplanet $=$ \oneEightTwoRad\ \rearth and $\rho =$ \oneEightTwoRho\ \gcc, though additional RV monitoring is needed to more confidently place it among this group of unusually dense planets. For \oneNineNine, we surpass the 5-$\sigma$ detection level for \oneNineNine c. 

In \S\ref{discuss:comp} we inferred the bulk compositions of our three planets using theoretical models of planet interiors and thermal evolution. We find that \oneEightTwo b's mass and radius are consistent with an ice-rich core (\waterfrac $ = $ \oneEightTwoWater\%). Using the grids from \cite{lopezforney14}, we find that \oneNineNine b has a very small H$_2$/He envelope mass fraction (\hhefrac $=$ \oneNineNineBFhhe\%) meaning it is probably rocky, while \oneNineNine c has a more substantial envelope (\hhefrac $=$ \oneNineNineCFhhe\%).

In \S\ref{discuss:spur} we reviewed the literature of the ``super-dense" sub-Neptunes (\rplanet $< 3$ \rearth, \mplanet $> 20$ \mearth; see Figure \ref{fig:mass_radius}) to examine whether or not unmitigated stellar activity could have artificially inflated their high mass measurements. We conducted our own analysis of the \hdTwoOneSevenFourNine RVs as reported by \cite{dragomir19} and found agreement in the mass measurement for the super-dense sub-Neptune \hdTwoOneSevenFourNine b between a model of the RVs with and without a Gaussian process trained on stellar activity indicators. \revision{We also found agreement between our measurements and the results of \cite{gan21}, which uses a larger RV data set in addition to a GP model of the stellar activity signal.} Details and results of our analysis are found in Appendix \ref{app:hd21749_rv_modeling}. While the mass measurements for \hdTwoOneSevenFourNine b agree between our two models (and both agree with the results in \citealt{dragomir19} and \citealt{gan21}) the GP-enabled model is heavily favored by the Akaike Information Criterion ($\Delta$AICc $> 30$) and produces a lower RMS scatter in the residuals ($\Delta$RMS $> 2.5$ \mps). Overall, we find no evidence that unmitigated stellar activity can explain the high mass measurements of the super-dense sub-Neptunes.

We discussed formation and evolution scenarios for these unusually dense planets, including \oneEightTwo b, in \S\ref{spur:formEvo}. Save for K2-66 b, which lies in the hot sub-Neptune desert, it seems that photoevaporation cannot explain the masses and radii of the super-dense sub-Neptunes. Instead, formation via giant impacts or migration due to dynamical interactions with the gaseous protoplanetary disk or an outer companion likely occurred (or some combination thereof).

\oneEightTwo b and \oneNineNine b and c are not enticing targets for atmospheric characterization with \jwst according to the transmission spectroscopy metric from \cite{kempton18}. However, we note that due to degeneracies in models of planet bulk composition, measurements of atmospheric metallicity may be the only way to shed light on the interiors and formation histories of super-dense sub-Neptunes like \oneEightTwo b. In addition, as a multi-planet system with a precise ($>5$-$\sigma$) mass measurement for planet c, \oneNineNine is a valuable opportunity for atmospheric studies to probe models of planet formation. 

\facilities{Keck I Telescope (\emph{HIRES}), Keck II Telescope (\emph{NIRC2}).}
\software{\texttt{astropy} \citep{exoplanet:astropy13, exoplanet:astropy18}, \xspace\celerite \citep{celerite}, \exoplanet \citep{exoplanet:exoplanet}, \isoclassify \citep{huber17, berger20}, \texttt{matplotlib} \citep{matplotlib}, \texttt{numpy} \citep{numpy}, \texttt{pandas} \citep{pandas},  \texttt{pymc3} \citep{exoplanet:pymc3}, \texttt{python} 3 \citep{python3}, \texttt{RadVel} \citep{radvel}, \texttt{scipy} \citep{scipy}, \smint \revision{\citep{piaulet21}}, \specMatchSynth \citep{specmatchsynth}, \specMatchEmp \citep{yee17}, \texttt{starry} \citep{starry}, \xspace\texttt{theano} \citep{exoplanet:theano}.}

\begin{center}
ACKNOWLEDGMENTS
\end{center}
\revision{We thank the anonymous referee for their thorough reading of this work—their detailed comments improved the quality of the manuscript.} J.M.A.M. is supported by the National Science Foundation Graduate Research Fellowship Program under Grant No. DGE-1842400. J.M.A.M. acknowledges the LSSTC Data Science Fellowship Program, which is funded by LSSTC, NSF Cybertraining Grant No. 1829740, the Brinson Foundation, and the Moore Foundation; his participation in the program has benefited this work. M.R.K. is supported by the NSF Graduate Research Fellowship, grant No. DGE 1339067. A.B. is supported by the NSF Graduate Research Fellowship, grant No. DGE 1745301. L.M.W. is supported by the Beatrice Watson Parrent Fellowship and NASA ADAP Grant 80NSSC19K0597.

We acknowledge use of the \texttt{lux} supercomputer at UC Santa Cruz, funded by NSF MRI grant AST 1828315. J.M.A.M. would like to thank Brant Robertson, Josh Sonstroem, and Ryan Hausen for their help with \texttt{lux} access and setup. J.M.A.M. would also like to thank Jonathan Fortney for sharing computational resources and Aarynn Carter for helpful conversations.

The authors wish to recognize and acknowledge the very significant cultural role and reverence that the summit of Mauna Kea has always had within the indigenous Hawaiian community. We are most fortunate to have the opportunity to conduct observations from this sacred mountain which is now colonized land.

\bibliography{main}{}
\bibliographystyle{aasjournal}
\appendix

\section{\oneEightTwo and \oneNineNine photometric models} \label{app:phot_modeling}

Here we include tables describing the parameters, priors, and posteriors for our photometric models of the \oneEightTwo and \oneNineNine \everest light curves. In many cases, the actual values for the model parameters may not be of physical interest (most of the relevant physical planet parameters can be found in Table \ref{tab:star_planet_combined}) but are listed here for completeness.

\begin{deluxetable}{lccccc}[H]
\tablecaption{\oneEightTwo photometric model \label{tab:182_phot_model}}
\centering
\tabletypesize{\footnotesize}
\tablehead{\colhead{Parameter} & \colhead{Symbol} & \colhead{Units} & \colhead{Prior} & \colhead{Posterior Median $\pm$ 1-$\sigma$} & \colhead{Notes}}
\startdata
\sidehead{\emph{Light curve parameters}}
Light curve mean offset & $\mu$ & ppt & $\mathcal{N}$(0, 10) & \oneEightTwoMean & \\
Log photometric variance & $\log$ \photjitter & $\log$ ppt$^2$ & $\mathcal{N}$($\log s^2_\mathrm{phot}$, 5) & \oneEightTwoLogsTwo & A \\
\sidehead{\emph{Stellar parameters}}
Limb darkening parameter 1 & $q_1$ & & $\mathcal{U}$[0, 1] & \oneEightTwoqOne & B \\
Limb darkening parameter 2 & $q_2$ & & $\mathcal{U}$[0, 1] & \oneEightTwoqTwo & B \\
Stellar mass & \mstar & \msun & $\mathcal{N}$(0.86, 0.1)[0, 1.5] & \oneEightTwoMStar & C \\
Stellar radius & \rstar & \rsun & $\mathcal{N}$(0.79, 0.1)[0, 1.5] & \oneEightTwoRStar & C \\
\sidehead{\emph{Planet parameters}}
Log orbital period & $\log P$ & $\log$ days & $\mathcal{N}$($\log P_\mathrm{\livingston}$, 1) & \oneEightTwoLogP & D \\ 
Time of inferior conjunction & \transitTime & $\mathrm{BJD} - 2454833$ & $\mathcal{N}$($T_{\mathrm{c},\:\mathrm{\livingston}}, 1)$ & \oneEightTwoEpoch & D \\
Log planet radius & $\log$ \rplanet & $\log$ \rearth & $\mathcal{N}$($\log R_{\mathrm{p}0}$, 1) & \oneEightTwoLogR & E \\
Impact parameter & $b$ & & $\mathcal{U}$[0, $1 + \frac{R_\mathrm{p}}{R_*}$] & \oneEightTwoImpact & F \\ 
$\sqrt{e}\cos({\omega})$ & $\xi_1$ & & $\mathcal{D}$($\xi_1$, $\xi_2$)[0, 1], VE($e | \bm{\theta}$)& \oneEightTwoSecosw & \\ 
$\sqrt{e}\sin({\omega})$ & $\xi_2$ & & $\mathcal{D}$($\xi_1$, $\xi_2$)[0, 1], VE($e | \bm{\theta}$) & \oneEightTwoSesinw & \\ 
\sidehead{\emph{GP hyperparameters}}
GP$_\mathrm{dec}$ log amplitude parameter & $\log S_{0\mathrm{, dec}}\omega_{0\mathrm{, dec}}^4$ & log ppt$^2$ rad$^3$ days$^{-3}$ & $\mathcal{N}$($\log s^2_\mathrm{phot} (\frac{2 \pi}{10})^4$, 5) & \oneEightTwoGPdecLogSZerowFour & G \\
GP$_\mathrm{dec}$ log angular frequency & $\log \omega_{0\mathrm{, dec}}$ & $\log$ rad days$^{-1}$ & $\mathcal{N}$($\log \frac{2\pi}{10}$, 5) & \oneEightTwoGPdecLogw & \\
GP$_\mathrm{dec}$ quality factor & $Q_{\mathrm{dec}}$ & & fixed & $\equiv \frac{1}{\sqrt{2}}$ & H \\ 
GP$_\mathrm{rot}$ log amplitude & $\log S_{0\mathrm{, rot}}$ & $\log$ ppt$^2$ rad$^{-1}$ days & $\mathcal{N}$($\log s^2_\mathrm{phot}$, 5) & \oneEightTwoGProtLogAmp & \\
GP$_\mathrm{rot}$ log period & $\log P_\mathrm{rot}$ & $\log$ days & $\mathcal{N}$($\log$ 24.82, 5)[0, 50] & \oneEightTwoGProtLogP & I \\
GP$_\mathrm{rot}$ log quality factor & $\log Q_{0\mathrm{, rot}}$ & & $\mathcal{N}$(1, 2) & \oneEightTwoGProtLogQ & J \\
GP$_\mathrm{rot}$ log quality factor separation & $\log \delta Q_{0\mathrm{, rot}}$ & & $\mathcal{N}$(1, 2) & \oneEightTwoGProtLogdQ & K \\
GP$_\mathrm{rot}$ mixture fraction & $f$ & & $\mathcal{U}$[0, 1] & \oneEightTwoGProtf & L
\enddata
\tablecomments{``Log" refers to the natural logarithm. $\mathcal{N}$(X, Y) refers to a Gaussian distribution with mean X and standard deviation Y. $\mathcal{N}$(X, Y)[A, B] refers to a bounded Gaussian with mean X, standard deviation Y, and hard bounds at A and B. $\mathcal{U}$[X, Y] refers to a uniform distribution inclusive on the interval X and Y. $\mathcal{D}$($\xi_1$, $\xi_2$)[0, 1] refers to a uniform distribution over the unit disk (i.e. $\sqrt{\xi_1^2 + \xi_2^2} \leq 1$). VE($e | \bm{\theta}$) refers to the mixture distribution from \cite{exoplanet:vaneylen19} which is used as a prior on $e$ and whose hyperparameters, $\bm{\theta}$, are marginalized over. GP$_\mathrm{dec}$ and GP$_\mathrm{rot}$ refer to the exponential decay and rotation terms of the GP kernel, respectively.\\
\textbf{A}: $\sigma_\mathrm{phot}$ is treated as a uniform point-wise flux measurement error. $s^2_\mathrm{phot}$ indicates the sample variance of the flux data. \\
\textbf{B}: The parameterization $q_1 \equiv (u_1 + u_2)^2$ and $q_2 \equiv 0.5 u_1(u_1 + u_2)^{-1}$, where $u_1$ and $u_2$ are the usual quadratic limb darkening coefficients, follows the prescription by \cite{exoplanet:kipping13} for efficient, uninformative sampling of $u_1$ and $u_2$. \\
\textbf{C}: The stellar mass and radius were given Gaussian priors according to the results from our analysis in \S\ref{stellar:spec}, with hard bounds at 0 and 1.5 \msun and \rsun, respectively.\\
\textbf{D}: $P_\mathrm{\livingston} = 4.7368$ days and $T_{\mathrm{c},\:\mathrm{\livingston}} = 2308.20626$ BJD - 245833 are the orbital period and time of inferior conjunction for \oneEightTwo b as reported by \livingston. \\
\textbf{E}: $\log R_{\mathrm{p}0} = \log\big((\frac{R_\mathrm{p}}{R_*}_{\livingston})^2 \times R_* \big)$ where $\frac{R_\mathrm{p}}{R_*}_{\livingston} = 0.00317$ is the occultation fraction for \oneEightTwo b as reported by \livingston and $R_* = 0.86$ \rsun is the stellar radius resulting from our analysis in \S\ref{stellar:spec}.\\
\textbf{F}: The upper bound of the uniform prior on $b$ is conditional on the ratio of the (free) model parameters for planet and stellar radius. \\ 
\textbf{G}: The power of the GP$_\mathrm{dec}$ term (in units of ppt$^2$) at $\omega_\mathrm{GP} = 0$ is $\sqrt{\frac{2}{\pi}} S_{0\mathrm{, dec}}$. Our model fits the logarithm of $S_{0\mathrm{, dec}} \omega_{0\mathrm{, dec}}^4$ and calculates $S_{0\mathrm{, dec}}$ deterministically because $S_{0\mathrm{, dec}}$ and $\omega_{0\mathrm{, dec}}$, the angular frequency of the undamped oscillation, can show strong covariance. \\
\textbf{H}: A quality factor of $\frac{1}{\sqrt{2}}$ gives this stochastic damped harmonic oscillator the same power spectral density as stellar granulation \citep{harvey85, kallinger14}.\\
\textbf{I}: The center of the prior comes from a Lomb-Scargle periodogram \citep{lomb76, scargle82} of the light curve.\\
\textbf{J}: $Q_{0\mathrm{, rot}}$ is the quality factor minus $\frac{1}{2}$ for the GP rotation term's oscillator at $\frac{P_\mathrm{rot}}{2}$. \\
\textbf{K}: The difference between the quality factors of the oscillators at $P_\mathrm{rot}$ and  $\frac{P_\mathrm{rot}}{2}$. \\
\textbf{L}: The fractional amplitude of the oscillator at $\frac{P_\mathrm{rot}}{2}$ relative to the one at $P_\mathrm{rot}$.
}
\end{deluxetable}

\begin{deluxetable}{lccccc}[H]
\tablecaption{\oneNineNine photometric model \label{tab:199_phot_model}}
\centering
\tabletypesize{\footnotesize}
\tablehead{\colhead{Parameter} & \colhead{Symbol} & \colhead{Units} & \colhead{Prior} & \colhead{Posterior Median $\pm$ 1-$\sigma$}}
\startdata
\sidehead{\emph{Light curve parameters}}
Light curve mean offset & $\mu$ & ppt & $\mathcal{N}$(0, 10) & \oneNineNineMean\\
Log photometric variance & $\log$ \photjitter & $\log$ ppt$^2$ & $\mathcal{N}$($\log s^2_\mathrm{phot}$, 5) & \oneNineNineLogsTwo\\
\sidehead{\emph{Stellar parameters}}
Limb darkening parameter 1 & $q_1$ & & $\mathcal{U}$[0, 1] & \oneNineNineqOne\\
Limb darkening parameter 2 & $q_2$ & & $\mathcal{U}$[0, 1] & \oneNineNineqTwo\\
Stellar mass & \mstar & \msun & $\mathcal{N}$(0.86, 0.1)[0, 1.5] & \oneNineNineMStar \\
Stellar radius & \rstar & \rsun & $\mathcal{N}$(0.79, 0.1)[0, 1.5] & \oneNineNineRStar \\
\sidehead{\emph{Planet b parameters}}
Log orbital period & $\log P$ & $\log$ days & $\mathcal{N}$($\log P_\mathrm{\livingston}$, 1) & \oneNineNineBLogP \\ 
Time of inferior conjunction & \transitTime & $\mathrm{BJD} - 2454833$ & $\mathcal{N}$($T_{\mathrm{c},\:\mathrm{\livingston}}$, 1) & \oneNineNineBEpoch \\
Log planet radius & $\log$ \rplanet & $\log$ \rearth & $\mathcal{N}$($\log R_{\mathrm{p}0}$, 1) & \oneNineNineBLogR \\
Impact parameter & $b$ & & $\mathcal{U}$[0, $1 + \frac{R_\mathrm{p}}{R_*}$] & \oneNineNineBImpact \\ 
$\sqrt{e}\cos({\omega})$ & $\xi_1$ & & $\mathcal{D}$($\xi_1$, $\xi_2$)[0, 1], VE($e | \bm{\theta}$)& \oneNineNineBSecosw\\ 
$\sqrt{e}\sin({\omega})$ & $\xi_2$ & & $\mathcal{D}$($\xi_1$, $\xi_2$)[0, 1], VE($e | \bm{\theta}$) & \oneNineNineBSesinw \\ 
\sidehead{\emph{Planet c parameters}}
Log orbital period & $\log P$ & $\log$ days & $\mathcal{N}$($\log P_\mathrm{\livingston}$, 1) & \oneNineNineCLogP \\ 
Time of inferior conjunction & \transitTime & $\mathrm{BJD} - 2454833$ & $\mathcal{N}$($T_{\mathrm{c},\:\mathrm{\livingston}}$, 1) & \oneNineNineCEpoch \\
Log planet radius & $\log$ \rplanet & $\log$ \rearth & $\mathcal{N}$($\log R_{\mathrm{p}0}$, 1) & \oneNineNineCLogR \\
Impact parameter & $b$ & & $\mathcal{U}$[0, $1 + \frac{R_\mathrm{p}}{R_*}$] & \oneNineNineBImpact \\ 
$\sqrt{e}\cos({\omega})$ & $\xi_1$ & & $\mathcal{D}$($\xi_1$, $\xi_2$)[0, 1], VE($e | \bm{\theta}$)& \oneNineNineCSecosw\\ 
$\sqrt{e}\sin({\omega})$ & $\xi_2$ & & $\mathcal{D}$($\xi_1$, $\xi_2$)[0, 1], VE($e | \bm{\theta}$) & \oneNineNineCSesinw \\ 
\sidehead{\emph{GP hyperparameters}}
GP$_\mathrm{dec}$ log amplitude parameter & $\log S_{0\mathrm{, dec}}\omega_{0\mathrm{, dec}}^4$ & log ppt$^2$ rad$^3$ days$^{-3}$ & $\mathcal{N}$($\log s^2_\mathrm{phot} (\frac{2 \pi}{10})^4$, 5) & \oneNineNineGPdecLogSZerowFour \\
GP$_\mathrm{dec}$ log angular frequency & $\log \omega_{0\mathrm{, dec}}$ & $\log$ rad days$^{-1}$ & $\mathcal{N}$($\log \frac{2\pi}{10}$, 5) & \oneNineNineGPdecLogw\\
GP$_\mathrm{dec}$ quality factor & $Q_{\mathrm{dec}}$ & & fixed & $\equiv \frac{1}{\sqrt{2}}$ \\ 
GP$_\mathrm{rot}$ log amplitude & $\log S_{0\mathrm{, rot}}$ & $\log$ ppt$^2$ rad$^{-1}$ days & $\mathcal{N}$($\log s^2_\mathrm{phot}$, 5) & \oneNineNineGProtLogAmp \\
GP$_\mathrm{rot}$ log period & $\log P_\mathrm{rot}$ & $\log$ days & $\mathcal{N}$($\log$ 15.36, $\log$ 5)[$\log$ 1, $\log$ 21] & \oneNineNineGProtLogP \\
GP$_\mathrm{rot}$ log quality factor & $\log Q_{0\mathrm{, rot}}$ & & $\mathcal{N}$(1, 2) & \oneNineNineGProtLogQ  \\
GP$_\mathrm{rot}$ log quality factor separation & $\log \delta Q_{0\mathrm{, rot}}$ & & $\mathcal{N}$(1, 2) & \oneNineNineGProtLogdQ \\
GP$_\mathrm{rot}$ mixture fraction & $f$ & & $\mathcal{U}$[0, 1] & \oneNineNineGProtf
\enddata
\tablecomments{Notation and parameters are analogous to Table \ref{tab:182_phot_model}. We include a tight prior on $\log P_\mathrm{rot}$ because in using a wide prior similar to the one in our model of \oneEightTwo, we found the marginal posterior was multimodal and caused sampling performance to suffer. The multimodality is probably a symptom of the stark difference in behavior between the \ktwo C6 and C17 light curves of \oneNineNine, seen in the top panel of Figure \ref{fig:k2-199_phot}. Therefore, for \oneNineNine $\log P_\mathrm{rot}$ should not be thought of as a robust estimate of the stellar rotation period and only as a non-physical model parameter used to flatten the light curve.
}
\end{deluxetable}

\section{\oneEightTwo and \oneNineNine radial velocity models} \label{app:rv_modeling}

Here we include tables describing the parameters, priors, and posteriors for our models of the \oneEightTwo and \oneNineNine \keckhires RVs. Derived parameters are summarized in Table \ref{tab:star_planet_combined}. We also include tables with our \keckhires RV measurements.

\begin{deluxetable}{lccccc}[H]
\tablecaption{\oneEightTwo RV models\label{tab:182_rv_model}}
\centering
\tabletypesize{\normalsize}
\tablehead{\colhead{Parameter} & \colhead{Symbol} & \colhead{Units} & \colhead{Prior} & \colhead{Posterior Median $\pm$ 1-$\sigma$}}
\startdata
\rule{0pt}{2ex} & & & & \\ 
\multicolumn{5}{l}{\bf{\oneEightTwoModelA (1-planet circular Keplerian; adopted)}} \\
\hline
\sidehead{\emph{Planet b orbital parameters}}
Orbital period & $P$ & days & fixed & $\equiv 4.7369683$ \\
Time of inferior conjunction & \transitTime & BJD - 2454833 & fixed & $\equiv 2886.11517$ \\
Orbital eccentricity & $e$ & & fixed & $\equiv 0$ \\
Argument of periastron & $\omega$ & rad & fixed & $\equiv 0$ \\
RV semi-amplitude & $K$ & \mps & $K > 0$ & \oneEightTwoModelAKamp \\
\sidehead{\emph{Instrument parameters}}
HIRES RV offset & $\gamma_j$ & \mps & fixed & $\equiv -2.21$ \\
HIRES RV jitter & $\sigma_j$ & \mps & $\mathcal{U}$[0, 20] & \oneEightTwoModelAJit \\
\rule{0pt}{2ex} & & & & \\
\multicolumn{5}{l}{\bf{\oneEightTwoModelB (1-planet circular Keplerian $+$ GP)}} \\
\hline
\sidehead{\emph{Planet b orbital parameters}}
Orbital period & $P$ & days & fixed & $\equiv 4.7369683$ \\
Time of inferior conjunction & \transitTime & BJD - 2454833 & fixed & $\equiv 2886.11517$ \\
Orbital eccentricity & $e$ & & fixed & $\equiv 0$ \\
Argument of periastron & $\omega$ & rad & fixed & $\equiv 0$ \\
RV semi-amplitude & $K$ & \mps & $K > 0$ & \oneEightTwoModelBKamp \\
\sidehead{\emph{Instrument parameters}}
HIRES RV offset & $\gamma_j$ & \mps & fixed & $\equiv 0$ \\
HIRES RV jitter & $\sigma_j$ & \mps & fixed & $\equiv 0$ \\
\sidehead{\emph{GP hyperparameters}}
Amplitude & $\eta_1$ & \mps & $\mathcal{U}$[0, 20] & \oneEightTwoModelBEtaone \\
Evolutionary timescale & $\eta_2$ & days & \ktwo photometry & \oneEightTwoModelBEtatwo \\
Rotation period & $\eta_3$ & days & \ktwo photometry & \oneEightTwoModelBEtathree \\
Lengthscale & $\eta_4$ & & \ktwo photometry & \oneEightTwoModelBEtafour \\
\enddata
\tablecomments{Numerical priors on $\eta_\mathrm{2-4}$ come from Gaussian kernel density estimation of the posteriors from the training on the \ktwo C5 and C18 photometry. Identical models were tested which did not include the positive prior on $K$ and produced entirely similar results, suggesting the prior is not biasing $K$ towards larger values. We choose to include the prior because it is physically motivated. In Model A $\gamma_j$ was fit to center the data about 0 but held fixed during the MCMC. We found that allowing it vary returned near-identical results but slightly increased the model AIC.
}
\end{deluxetable}

\begin{deluxetable}{lccccc}[H]
\tablecaption{\oneNineNine RV model \label{tab:199_rv_model}}
\centering
\tabletypesize{\normalsize}
\tablehead{\colhead{Parameter} & \colhead{Symbol} & \colhead{Units} & \colhead{Prior} & \colhead{Posterior Median $\pm$ 1-$\sigma$}}
\startdata
\rule{0pt}{2ex} & & & & \\
\multicolumn{5}{l}{\bf{2-planet circular Keplerian}} \\
\hline
\sidehead{\emph{Planet b orbital parameters}}
Orbital period & $P$ & days & \ktwo photometry & $3.2253993^{+0.0000023}_{-0.0000024}$ \\
Time of inferior conjunction & \transitTime & BJD - 2454833 & \ktwo photometry & $2385.73733^{+0.00053}_{-0.00053}$ \\
Orbital eccentricity & $e$ & & fixed & $\equiv 0$ \\
Argument of periastron & $\omega$ & rad & fixed & $\equiv 0$ \\
RV semi-amplitude & $K$ & \mps & $K > 0$ & \oneNineNineBKamp \\
\sidehead{\emph{Planet c orbital parameters}}
Orbital period & $P$ & days & \ktwo photometry & $7.3744897^{+0.0000036}_{-0.0000037}$ \\
Time of inferior conjunction & \transitTime & BJD - 2454833 & \ktwo photometry & $2389.93033^{+0.00036}_{-0.00035}$ \\
Orbital eccentricity & $e$ & & fixed & $\equiv 0$ \\
Argument of periastron & $\omega$ & rad & fixed & $\equiv 0$ \\
RV semi-amplitude & $K$ & \mps & $K > 0$ & \oneNineNineCKamp \\
\sidehead{\emph{Instrument parameters}}
HIRES RV offset & $\gamma_j$ & \mps & fixed & $\equiv -2.05$ \\
HIRES RV jitter & $\sigma_j$ & \mps & $\mathcal{U}$[0, 20] & \oneNineNineRVJit \\
\enddata
\tablecomments{Priors on $P$ and \transitTime for each planet are a Gaussian whose mean is the median posterior value quoted in Table \ref{tab:star_planet_combined} and whose width is the median of corresponding upper and lower uncertainties. An identical model was tested which did not include the positive priors on $K$ for each planet and produced entirely similar results, suggesting the priors are not biasing $K$ towards larger values. We choose to include the priors because they are physically motivated. $\gamma_j$ was fit to center the data about 0 but held fixed during the MCMC. We found that allowing it vary returned near-identical results but slightly increased the model AIC.
}
\end{deluxetable}

\begin{deluxetable}{lrrc}
\tablecaption{K2-182 radial velocities \label{tab:k2-182_rvs}}
\tablehead{
\colhead{Time} & 
\colhead{RV} & 
\colhead{RV Unc.} & 
\colhead{Inst.} \\
\colhead{(BJD)} & 
\colhead{(m s$^{-1}$)} & 
\colhead{(m s$^{-1}$)} & 
\colhead{}}
\startdata
2458568.77000 & -7.10 & 1.54 & HIRES \\
2458569.80300 & 5.70 & 1.62 & HIRES \\
2458584.77900 & 8.79 & 1.45 & HIRES \\
2458595.74400 & -3.97 & 1.56 & HIRES \\
2458599.75800 & 4.47 & 1.55 & HIRES \\
2458610.82800 & -14.86 & 1.63 & HIRES \\
2458617.77600 & 6.54 & 1.58 & HIRES \\
2458622.78000 & 1.57 & 1.66 & HIRES \\
2458623.77300 & 2.95 & 1.54 & HIRES \\
2458627.77900 & -3.33 & 1.55 & HIRES \\
2458628.76900 & -5.93 & 1.54 & HIRES \\
2458632.79400 & 7.17 & 1.62 & HIRES \\
\enddata
\tablecomments{Uncertainties reported here are added in quadrature with an instrumental parameter for RV jitter, $\sigma_j = 4.3$ \mps, that was fit to the data. This table is available online in machine readable format.
}
\end{deluxetable}

\begin{deluxetable}{lrrrrc}
\tablecaption{K2-199 radial velocities \label{tab:k2-199_rvs}}
\tablehead{
\colhead{Time} & 
\colhead{RV} & 
\colhead{RV Unc.} &
\colhead{Inst.} \\
\colhead{(JD)} & 
\colhead{(m s$^{-1}$)} & 
\colhead{(m s$^{-1}$)} & 
\colhead{}
}
\startdata
2458533.11808 & 2.39 & 1.82 & HIRES \\
2458560.06257 & -8.44 & 1.73 & HIRES \\
2458568.90252 & 6.03 & 1.60 & HIRES \\
2458569.91236 & 6.26 & 1.72 & HIRES \\
2458591.97627 & 0.24 & 1.76 & HIRES \\
2458595.86178 & -4.35 & 1.60 & HIRES \\
2458615.86603 & -5.36 & 1.89 & HIRES \\
2458616.85860 & -3.86 & 1.82 & HIRES \\
2458617.85161 & -2.68 & 1.74 & HIRES \\
2458627.86391 & -2.16 & 1.53 & HIRES \\
2458632.84003 & -7.41 & 1.71 & HIRES \\
2458633.85508 & -5.40 & 1.47 & HIRES \\
\enddata
\tablecomments{\revision{The 12 RVs listed here come from the spectroscopic observations of \oneNineNine conducted for this work. Our analysis combined these observations with 33 spectra taken by \nasakeyproj. The RVs of the combined dataset ($N = 45$, including the 12 RVs listed here) were calculated simultaneously and are available online in machine readable format.} The RV uncertainties listed here are added in quadrature with an instrumental parameter for RV jitter, $\sigma_j =  4.14$ \mps, that was fit to the data.
}
\end{deluxetable}

\section{\hdTwoOneSevenFourNine radial velocity models} \label{app:hd21749_rv_modeling}

Here we include figures of our MAP solutions as well as tables describing the parameters, priors, and posteriors for our models of the \hdTwoOneSevenFourNine HARPS and PFS RVs, as reported by \citet[referred to as \dragomir]{dragomir19}. \revision{We restricted ourselves to the \dragomir data set rather than using the (larger) data set from \citet[referred to as \gan]{gan21} to enable a more direct comparison between our models and the Keplerian-only model from \dragomir.}

We downloaded the HARPS and PFS data used by \dragomir from ExoFOP-\tess\footnote{\url{https://exofop.ipac.caltech.edu/tess/target.php?id=279741379}} and the Data Analysis Center for Exoplanets (DACE) website\footnote{\url{https://dace.unige.ch/radialVelocities/?pattern=HD\%2021749\#}}. We did not find the four HARPS velocities from 2016 at either location, so they are not included in our analysis. We do not anticipate this to make a large difference, though, given the size of the remaining data set ($N = 137$). To reproduce the results from \dragomir, we first modeled the HARPS and PFS RVs using a 2-planet Keplerian fit, allowing the eccentricity of planet b's orbit to vary and holding it fixed at zero for planet c. The Keplerian model also included a linear trend ($\dot{\gamma}$). Gaussian priors were placed on the planet orbital period and time of inferior conjunction using the posteriors from \dragomir. We included a uniform prior between 0 and 0.99 on the orbital eccentricity of planet b. We also included wide, uniform priors on the instrument RV offset and jitter terms. We refer to this model as \hdTwoOneSevenFourNine RV Model A. Model parameters, priors, and posteriors are summarized in Table \ref{tab:21749_rv_model_A}.

\begin{figure*}[!ht]
\gridline{\fig{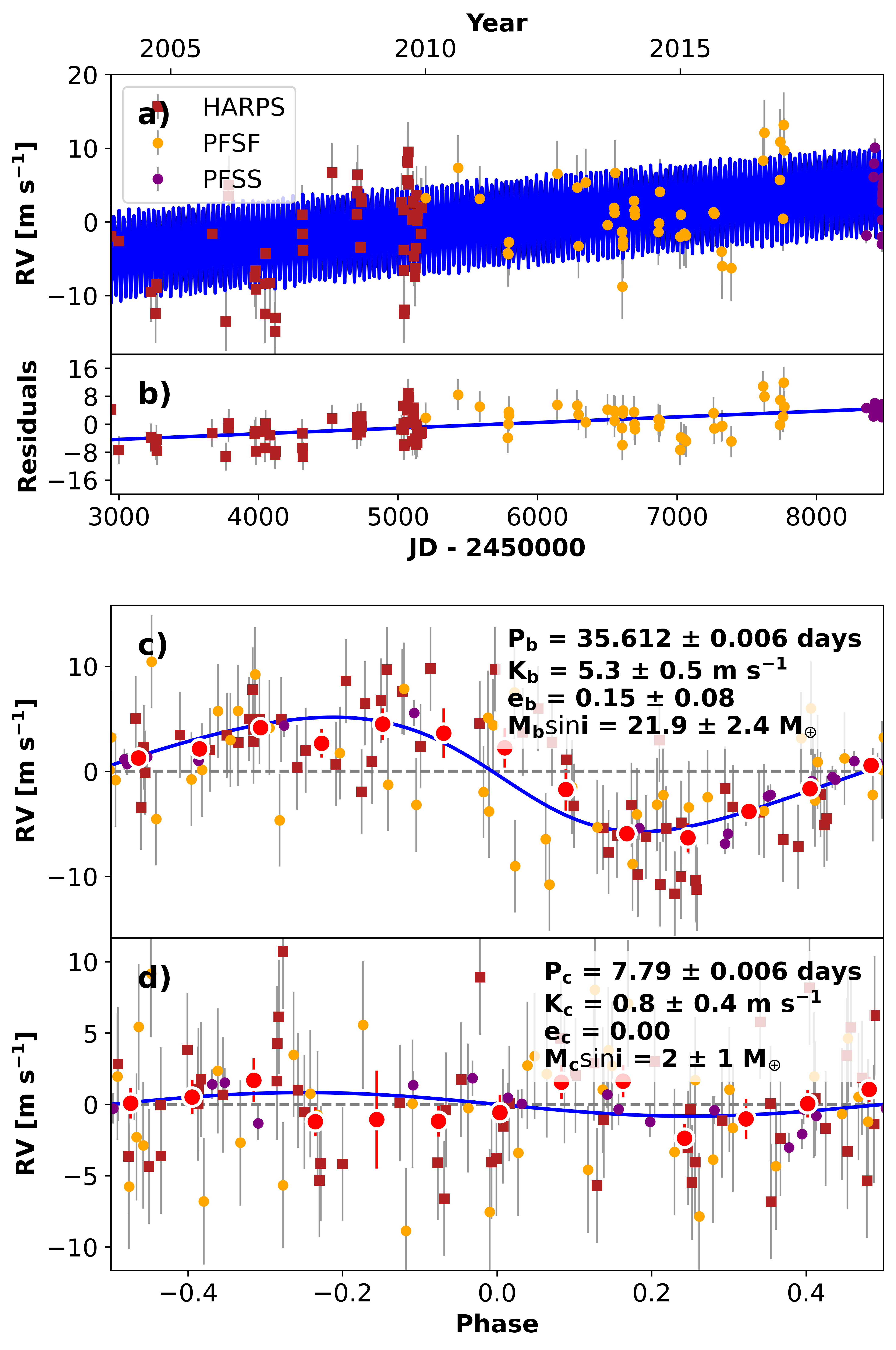}{0.5\textwidth}{(a) \hdTwoOneSevenFourNine RV Model A (mimic of \dragomir solution)}
      \fig{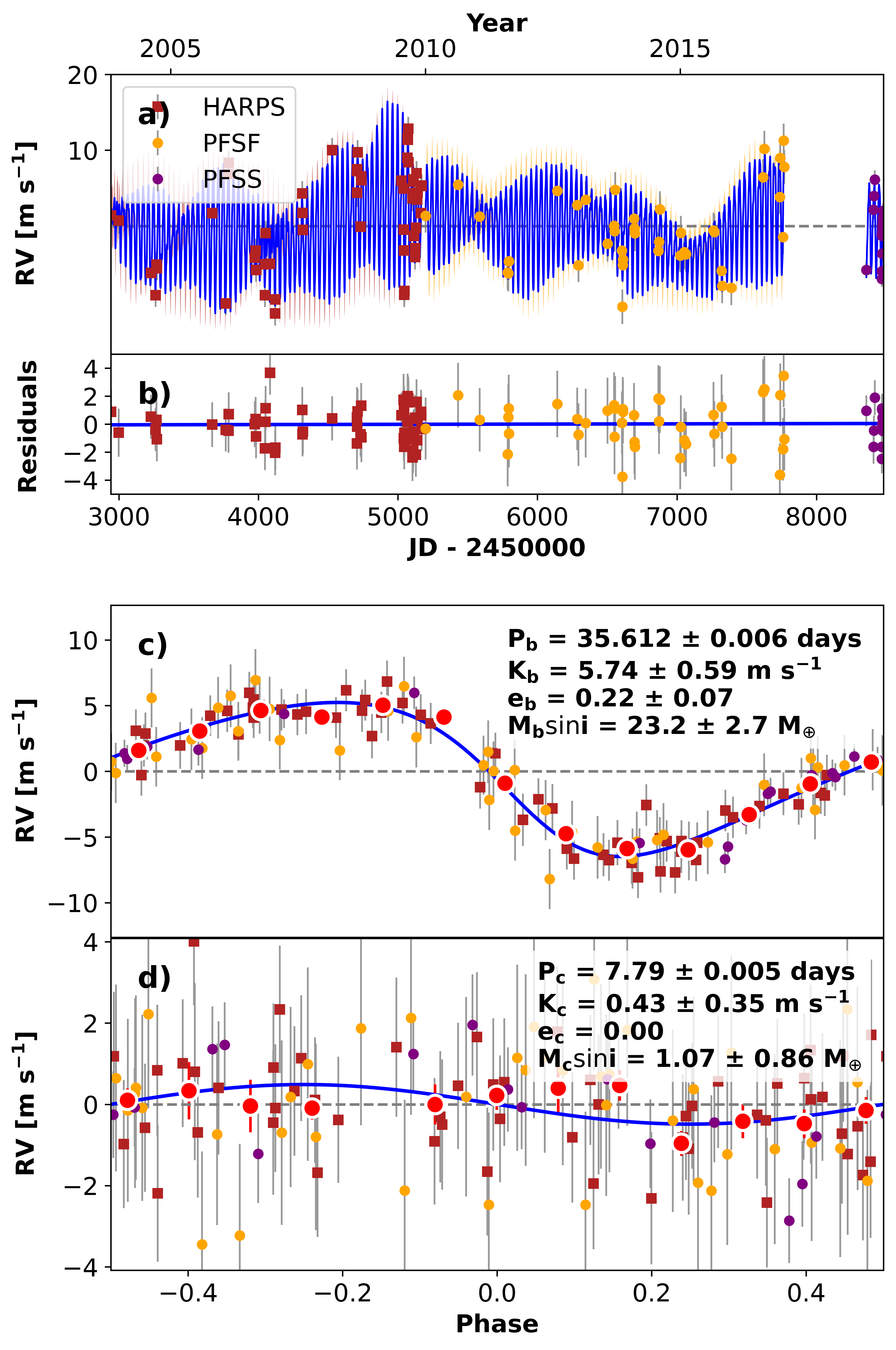}{0.5\textwidth}{(b) \hdTwoOneSevenFourNine RV Model B (Model A $+$ GP trained on H-$\alpha$ indices)}}
  \caption{\emph{Left:} The MAP solution for \hdTwoOneSevenFourNine RV Model A, which mimics the solution presented in \dragomir. \emph{Right:} The MAP solution for \hdTwoOneSevenFourNine RV Model B, which differs only in the inclusion of a GP trained on the H-$\alpha$ indices. For both models, panel \emph{a)} shows the best-fitting 2-planet solution. Dark red squares correspond to HARPS velocities, yellow circles are measurements from PFS prior to a detector upgrade in 2018 Feb (PFSF), and purple circles are measurements from PFS post-upgrade (PFSS). Data from the same instrument were binned in groups of 2.4 hours prior to fitting the models. We note that we did not find the additional four HARPS velocities from 2016 used by \dragomir on either ExoFOP-\tess or the Data Analysis Center for Exoplanets (DACE) website, so they are not included here. Measurements from HARPS, PFSF, and PFSS were all modeled with separate instrument RV offset and jitter terms. Additionally, the GP in Model B included a separate amplitude hyperparameter ($\eta_1$) for each instrument, while $\eta_{2-4}$ were shared between instruments. 1-$\sigma$ error bars on the data points reflect measurement errors added in quadrature with the corresponding instrument RV jitter term. For Model B, the GP posterior prediction uncertainty is also added in quadrature. MAP solutions are shown as the blue line. For Model B, the GP 1-$\sigma$ error envelope for each instrument is shown as the lightly shaded region about the MAP solution and is in the same color as the instrument data points. For both models, panel \emph{b)} shows the residuals about the MAP solution. Panel \emph{c)} shows the phase-folded MAP orbital solution for \hdTwoOneSevenFourNine b. Data (from all instruments) binned in units of 0.08 orbital phase are shown as the large red circles. Panel \emph{d)} shows the same for \hdTwoOneSevenFourNine c. Model B is strongly favored by the AICc ($\Delta$ AICc $\bm{>}$ 30) and produces residuals with smaller scatter (RMS = 1.4 \mps about the MAP solution for Model B, compared to RMS = 3.9 \mps for Model A). \revision{While \mplanet$\sin i$ is reported in the plots, this was converted to \mplanet for the values listed in Tables \ref{tab:21749_rv_model_A} and \ref{tab:21749_rv_model_B} using the orbital inclinations reported by \dragomir (the conversion did not change the values by enough to notice given the number of significant figures we use to report \mplanet).}}
  \label{fig:hd21749_rv_combined}
\end{figure*}

Our second model of the \hdTwoOneSevenFourNine RVs was entirely similar to the first, but it included a GP trained on the HARPS and PFS \halpha indices. The GP used the kernel shown in Equation \ref{eq:kernel}, and training produced tight constraints on the periodicity of the activity, finding $\eta_3 = 40.13^{+0.93}_{-0.36}$ days. Including the posteriors on $\eta_{2-4}$ from the GP training as priors in a fit to the RVs resulted in $\eta_3 = 37.25^{+0.35}_{-0.38}$ days, which is consistent with the estimates of the stellar rotation period from \dragomir \revision{and is 1.5-$\sigma$ consistent with the $P_\mathrm{rot}$ GP hyperparameter from the adopted RV model in \gan}. We refer to this model as \hdTwoOneSevenFourNine RV Model B. Model parameters, priors, and posteriors are summarized in Table \ref{tab:21749_rv_model_B}. MAP solutions for Model A and B are shown side-by-side in Figure \ref{fig:hd21749_rv_combined}.

While \hdTwoOneSevenFourNine RV Model A and B return similar mass measurements for planet b, Model B is strongly favored by the AICc ($\Delta$ AICc $> 30$) and produces residuals with smaller scatter (RMS $= 1.4$ \mps about the MAP solution for Model B, compared to RMS $= 3.9$ \mps for Model A). We note that the linear trend seems to disappear in Model B. We believe this is mainly due to the GP's ability to account for constant RV offsets itself (by tuning $\eta_1$), which, in combination with fitting for the instrumental offsets ($\gamma$), is able to explain away the long-term trend. The joint posteriors of $\dot{\gamma}$ and the instrumental offsets for HARPS ($\gamma_\mathrm{HARPS}$) and PFSS ($\gamma_\mathrm{PFSS}$) show evidence for this explanation as the parameters are strongly covariate. We mention this only to say that the linear trend likely vanishes in Model B due to degeneracy between $\dot{\gamma}$ and both $\gamma_\mathrm{HARPS}$ and $\gamma_\mathrm{PFSS}$, not because of the GP's modeling of the correlated activity signal at $\sim37$ days.\revision{ It should be noted that \gan's Keplerian $+$ GP model of their larger RV data set does not require a linear trend.}

\begin{deluxetable}{lccccc}[H]
\tablecaption{\hdTwoOneSevenFourNine RV Model A\label{tab:21749_rv_model_A}}
\centering
\tabletypesize{\normalsize}
\tablehead{\colhead{Parameter} & \colhead{Symbol} & \colhead{Units} & \colhead{Prior} & \colhead{Posterior Median $\pm$ 1-$\sigma$}}
\startdata
\rule{0pt}{2ex} & & & & \\
\multicolumn{5}{l}{\bf{Model parameters (2-planet Keplerian; mimic of \dragomir solution)}} \\
\hline
\sidehead{\emph{Planet b orbital parameters}}
Orbital period & $P$ & days & $\mathcal{N}$(\dragomir) & $35.61238^{+0.00062}_{-0.00061}$ \\
Time of inferior conjunction & \transitTime & BJD & $\mathcal{N}$(\dragomir) & $2458385.92503^{+0.00055}_{-0.00056}$ \\
$\sqrt{e}\cos(\omega)$ & $\xi_1$ & & $\mathcal{D}$($\xi_1$, $\xi_2$)[0, $\sqrt{0.99}$] & $-0.11^{+0.12}_{-0.12}$ \\
$\sqrt{e}\sin(\omega)$ & $\xi_2$ & & $\mathcal{D}$($\xi_1$, $\xi_2$)[0, $\sqrt{0.99}$] & $0.35^{+0.11}_{-0.17}$ \\
RV semi-amplitude & $K$ & \mps & $\mathcal{U}$[$-\infty$, $+\infty$] & $5.3 \pm 0.5$ \\
\sidehead{\emph{Planet c orbital parameters}}
Orbital period & $P$ & days & $\mathcal{N}$(\dragomir) & $7.79003\pm 0.00048$ \\
Time of inferior conjunction & \transitTime & BJD & $\mathcal{N}$(\dragomir) & $2458371.2287^{+0.0016}_{-0.0015}$ \\
Orbital eccentricity & $e$ & & fixed & $\equiv 0$ \\
Argument of periastron & $\omega$ & & fixed & $\equiv 0$ \\
RV semi-amplitude & $K$ & \mps & $\mathcal{U}$[$-\infty$, $+\infty$] &  $0.8 \pm 0.4$ \\
\sidehead{\emph{Instrument parameters}}
Linear trend & $\dot{\gamma}$ & \mps d$^{-1}$ & $\mathcal{U}$[$-\infty$, $+\infty$] & $0.00164^{+0.00069}_{-0.00071}$ \\
HARPS RV offset & $\gamma_\mathrm{HARPS}$ & \mps & $\mathcal{U}$[$-10$, 10] & $1.68^{+1.00}_{-1.03}$ \\
HARPS RV jitter & $\sigma_\mathrm{HARPS}$ & \mps & $\mathcal{U}$[0, 20] & $4.14^{+0.49}_{-0.40}$ \\
PFSF RV offset & $\gamma_\mathrm{PFSF}$ & \mps & $\mathcal{U}$[$-10$, 10] & $-1.41^{+1.02}_{-1.03}$ \\
PFSF RV jitter & $\sigma_\mathrm{PFSF}$ & \mps & $\mathcal{U}$[0, 20] & $4.57^{+0.62}_{-0.51}$ \\
PFSS RV offset & $\gamma_\mathrm{PFSS}$ & \mps & $\mathcal{U}$[$-10$, 10] &  $-2.37^{+1.95}_{-1.93}$ \\
PFSS RV jitter & $\sigma_\mathrm{PFSS}$ & \mps & $\mathcal{U}$[0, 20] & $1.15^{+0.33}_{-0.26}$ \\
\rule{0pt}{2ex} & & & & \\
\multicolumn{5}{l}{\bf{Derived parameters}} \\
\hline
\sidehead{\emph{Planet b}}
Eccentricity & $e$ & & & $0.15^{+0.08}_{-0.07}$ \\
Argument of periastron & $\omega$ & rad & & $1.88^{+0.55}_{-0.33}$ \\
Mass & \mplanet & \mearth & & $21.9^{+2.4}_{-2.3}$\\
Bulk density & $\rho$ & \gcc & & $6.7^{+1.7}_{-1.3}$\\ 
\sidehead{\emph{Planet c}}
Mass & \mplanet & \mearth & & $< 5.0$\\
Bulk density & $\rho$ & \gcc & & $< 42.8$\\ 
\enddata
\tablecomments{Gaussian priors on planet ephemerides come from \dragomir. Bulk densities calculated using $R_\mathrm{b} = 2.61^{+0.17}_{-0.16}$ \rearth and $R_\mathrm{c} = 0.0892^{+0.064}_{-0.058}$ \rearth from \dragomir. Upper limits on planet c's mass and density represent 99.7\% confidence.
}
\end{deluxetable}
\begin{deluxetable}{lccccc}[H]
\tablecaption{\hdTwoOneSevenFourNine RV Model B\label{tab:21749_rv_model_B}}
\centering
\tabletypesize{\footnotesize}
\tablehead{\colhead{Parameter} & \colhead{Symbol} & \colhead{Units} & \colhead{Prior} & \colhead{Posterior Median $\pm$ 1-$\sigma$}}
\startdata
\rule{0pt}{2ex} & & & & \\ 
\multicolumn{5}{l}{\bf{Model parameters (2-planet Keplerian $+$ GP)}} \\
\hline
\sidehead{\emph{Planet b orbital parameters}}
Orbital period & $P$ & days & $\mathcal{N}$(\dragomir) & $35.61238^{+0.00062}_{-0.00061}$ \\
Time of inferior conjunction & \transitTime & BJD & $\mathcal{N}$(\dragomir) & $2458385.92502\pm 0.00055$ \\
$\sqrt{e}\cos(\omega)$ & $\xi_1$ & & $\mathcal{D}$($\xi_1$, $\xi_2$)[0, $\sqrt{0.99}$] & $-0.21^{+0.11}_{-0.10}$ \\
$\sqrt{e}\sin(\omega)$ & $\xi_2$ & & $\mathcal{D}$($\xi_1$, $\xi_2$)[0, $\sqrt{0.99}$] & $0.41^{+0.09}_{-0.14}$ \\
RV semi-amplitude & $K$ & \mps & $\mathcal{U}$[$-\infty$, $+\infty$] & $5.7\pm 0.6$ \\
\sidehead{\emph{Planet c orbital parameters}}
Orbital period & $P$ & days & $\mathcal{N}$(\dragomir) & $7.78996\pm 0.00047$ \\
Time of inferior conjunction & \transitTime & BJD & $\mathcal{N}$(\dragomir) & $2458371.2287\pm 0.0016$ \\
Orbital eccentricity & $e$ & & fixed & $\equiv 0$ \\
Argument of periastron & $\omega$ & & fixed & $\equiv 0$ \\
RV semi-amplitude & $K$ & \mps & $\mathcal{U}$[$-\infty$, $+\infty$] & $0.4 \pm 0.3$ \\
\sidehead{\emph{Instrument parameters}}
Linear trend & $\dot{\gamma}$ & \mps d$^{-1}$ & $\mathcal{U}$[$-\infty$, $+\infty$] & $0.00004^{+0.0012}_{-0.0013}$ \\
HARPS RV offset & $\gamma_\mathrm{HARPS}$ & \mps & $\mathcal{U}$[$-10$, 10] & $-1.55^{+2.63}_{-2.75}$ \\
HARPS RV jitter & $\sigma_\mathrm{HARPS}$ & \mps & $\mathcal{U}$[0, 20] & $1.63^{+0.34}_{-0.28}$ \\
PFSF RV offset & $\gamma_\mathrm{PFSF}$ & \mps & $\mathcal{U}$[$-10$, 10] & $0.51^{+1.77}_{-1.70}$ \\
PFSF RV jitter & $\sigma_\mathrm{PFSF}$ & \mps & $\mathcal{U}$[0, 20] & $2.20^{+0.77}_{-0.88}$ \\
PFSS RV offset & $\gamma_\mathrm{PFSS}$ & \mps & $\mathcal{U}$[$-10$, 10] &  $1.60^{+3.56}_{-3.38}$ \\
PFSS RV jitter & $\sigma_\mathrm{PFSS}$ & \mps & $\mathcal{U}$[0, 20] & $1.16^{+0.36}_{-0.27}$ \\
\sidehead{\emph{GP hyperparameters}}
HARPS amplitude & $\eta_{1,\mathrm{HARPS}}$ & \mps & $\mathcal{U}$[0, 20] & $5.29^{+1.28}_{-0.98}$ \\
PFSF amplitude & $\eta_{1,\mathrm{PFSF}}$ & \mps & $\mathcal{U}$[0, 20] & $4.14^{+1.30}_{-1.05}$ \\
PFSS amplitude & $\eta_{1,\mathrm{PFSS}}$ & \mps & $\mathcal{U}$[0, 20] & 3.8e-8$^{+7\mathrm{e-}6}_{-3.8\mathrm{e-}8}$ \\
Evolutionary timescale & $\eta_2$ & days & \halpha indices & $320.19^{+102.33}_{-89.40}$ \\
Rotation period & $\eta_3$ & days & \halpha indices & $37.25^{+0.35}_{-0.38}$ \\
Lengthscale & $\eta_4$ & & \halpha indices & $0.49^{+0.05}_{-0.05}$ \\
\rule{0pt}{2ex} & & & & \\
\multicolumn{5}{l}{\bf{Derived parameters}} \\
\hline
\sidehead{\emph{Planet b}}
Eccentricity & $e$ & & & $0.221^{+0.07}_{-0.068}$ \\
Argument of periastron & $\omega$ & rad & & $2.04^{+0.35}_{-0.26}$ \\
Mass & \mplanet & \mearth & & $23.2 \pm 2.7$\\
Bulk density & $\rho$ & \gcc & & $7.2^{+1.8}_{-1.4}$\\ 
\sidehead{\emph{Planet c}}
Mass & \mplanet & \mearth & & $< 3.6$\\
Bulk density & $\rho$ & \gcc & & $< 29.6$\\ 
\enddata
\tablecomments{Gaussian priors on planet ephemerides come from \dragomir. Numerical priors on $\eta_\mathrm{2-4}$ come from Gaussian kernel density estimation of the posteriors from the training on the HARPS and PFS \halpha indices. Bulk densities calculated using $R_\mathrm{b} = 2.61^{+0.17}_{-0.16}$ \rearth and $R_\mathrm{c} = 0.0892^{+0.064}_{-0.058}$ \rearth from \dragomir. Upper limits on planet c's mass and density represent 99.7\% confidence.
}
\end{deluxetable}

\end{document}